\documentclass[traditabstract]{aa}
\pdfoutput=1
\usepackage{amsmath}
\usepackage{graphicx}
\usepackage{dcolumn}
\usepackage{amssymb}
\usepackage{color}
\usepackage{url}
\usepackage{natbib}

\definecolor{mypurple}{rgb}{0.255, 0.0, 0.255}

\bibpunct{(}{)}{;}{a}{}{,}
\newcommand{\changeRi}[1]{\textcolor{black}{#1}}
\newcommand{\changeRiII}[1]{\textcolor{black}{#1}}
\newcommand{\changeRiNo}[1]{\textcolor{black}{#1}}
\newcommand{\changeSu}[1]{\textcolor{black}{#1}}

\newcommand{\add}[1]{\textcolor{black}{#1}}
\newcommand{\addII}[1]{\textcolor{black}{#1}}

\begin{document}

\title{\changeRi{Probing the clumping structure of Giant Molecular 
Clouds 
through the spectrum, polarisation and morphology of  
X-ray Reflection Nebulae}}

\author{Margherita Molaro\inst{\ref{inst1}}
\and
Rishi Khatri\inst{\ref{inst1},\ref{inst2}}
\and
Rashid A. Sunyaev\inst{\ref{inst1}{,}\ref{inst3}}
}

\institute{ Max Planck Institut f\"{u}r Astrophysik, Karl-Schwarzschild-Str. 1,
  85741, Garching, Germany \email{molaro@mpa-garching.mpg.de}\label{inst1}
\and
\changeRiII{
Tata Institute of Fundamental Research, Homi Bhabha Road, Mumbai, 400005, India 
\label{inst2}}
\and
 Space Research Institute, Russian Academy of Sciences, Profsoyuznaya
 84/32, 117997 Moscow, Russia \label{inst3}}
\date{\today}
\authorrunning{Molaro,Khatri \& Sunyaev}
\titlerunning{Internal structure of XRNe}

\abstract{We suggest a new method for probing global properties 
of 
clump populations in Giant Molecular Clouds (GMCs) in the case where these act 
as X-ray 
reflection nebulae (XRNe), based on the study of the clumping's overall effect 
on the reflected X-ray signal\changeSu{, in particular on the Fe K-$\alpha$ line's shoulder}. We consider the 
particular case of Sgr B2, one of the brightest and most massive XRN in our 
Galaxy. \changeRi{We parametrise the gas distribution inside the cloud using a 
simple clumping model with
 the  }
slope of the clump mass function ($\alpha$), the minimum clump mass 
($m_{min}$), the fraction of the cloud's mass 
contained in clumps ($f_{\text{DGMF}}$), and the mass-size relation of 
individual 
clumps as free parameters, and investigate how these affect the 
reflected X-ray spectrum. 
In the case of very dense clumps, similar to those presently 
observed in Sgr B2, these occupy a small volume of the cloud and present a 
small projected area to the incoming X-ray radiation. We find that 
these contribute negligibly to the scattered X-rays.  
Clump populations with volume filling factors of $> 10^{-3}$, do leave observational signatures, that are sensitive to the clump model
parameters, in the reflected spectrum and polarisation. \changeRiII{Future  
high-resolution} X-ray observations could therefore complement the traditional 
optical and radio observations of these GMCs, and prove to be a powerful probe 
in the study of their internal structure. 
Finally, clumps in GMCs should be visible both as bright spots and regions
of heavy absorption in high resolution X-ray observations.
We therefore further study the time-evolution of the X-ray morphology, under illumination 
by a transient source, as a probe of the 3d
distribution and column density of individual clumps by future X-ray 
observatories.}

\maketitle

\section{Introduction}
\label{intro}
Understanding the \changeRi{internal structure of} giant molecular 
clouds (GMCs), which is driven by the interplay of turbulence, 
self-gravitation, and magnetic fields, is crucial when studying star 
formation processes in galaxies. It is, in fact, inside GMCs that 
dense, gravitationally unstable regions of gas, known as 
prestellar cores, form and collapse to give birth to stars 
\citep{Williams2000}.\\
Direct and exhaustive studies of the internal structure of GMCs are severely 
limited by 
issues of spatial and mass resolution when observing small-scale gas 
substructures. This is particularly true for GMCs located at a great distance, 
for example  
those found in the Central Molecular Zone (CMZ), the innermost region of the 
Galaxy (within $\sim$400 pc from the Galactic centre). Dense regions inside 
GMCs, often studied as discrete objects loosely classified as clumps and 
cores, span 
spatial ranges of 0.2-2 pc and 0.02-0.4 pc and mass ranges of  
$10-10^3M_{\sun}$ and $0.3-10^2 M_{\sun}$, respectively \citep{Draine}. At 
distances comparable to that from the Sun to the Galactic Center 
(GC), subarcsec angular resolution is therefore required for these structures 
to 
be studied in detail. Despite the 
challenge 
that such high-resolution observations pose, obtaining a clear and complete 
 picture of the overall properties of the clump and core populations 
in GMCs remains a vital effort in developing theoretical models of star 
formation \changeRi{\citep{Williams2000}}.  \newline
In this paper, we suggest a new method for probing global properties of the 
clump and core population in GMCs in the case where these act as X-ray 
reflection nebulae 
(XRNe), based on the study of their overall
effect on the 
reflected X-ray signal.  \\
X-ray emission from XRNe is composed both of a continuum, shaped by the 
interplay of scattering and absorption of the illuminating X-rays by atoms and 
molecules in
the GMC, and by 
characteristic 
spectral features in the keV regime. \changeRi{The latter are} caused by the 
emission of 
fluorescent photons by heavy elements following the photoionisation 
of tightly bound electrons by hard X-rays. \changeSu{The inelastic scattering of fluorescent photons down to lower energies results in a characteristic increase in the continuum 
at energies lower than the fluorescent features - the so called ``shoudler''. This feature is most easily visible in the case of bright fluorescent lines, such as the Fe K-$\alpha$ line.}  \newline
Fluorescent emission following illumination by X-rays was predicted by 
\citet{Sunyaev1993} in support of the claim that GMCs surrounding the Galactic 
Center (GC) should act as XRNe of past flares of Sgr 
A$^{*}$, the super-massive black-hole located at the center of the Galaxy. If 
this were the case, then part of the diffuse, hard, X-ray emission 
observed from these GMCs should be composed of a 
flux in the neutral Fe 
fluorescent line energy (6.4 keV), caused by the imprint of 
past, 
prequiescent activity of Sgr A$^{*}$
on the present-day (due to time delays) X-ray emission of 
GMCs located in its proximity. A high, time-varying flux 
at 
6.4 keV was indeed observed in the GMC Sgr B2 by Koyama et al 1996, 
\changeRiNo{and has 
been 
extensively studied} ever since 
\citep{sgr3,sgr5,sgr6,sgr7,sgr8,sgr10,sgr11,
sgr12,sgr13, Ponti2013, Zhang2015}. Scattered flux from Sgr B2 in hard X-rays was also detected using Integral \citep{Revnivtsev2004}. Similarly, other GMCs in the CMZ 
\citep{Murakami2001, Marin2015} have since been shown to act as XRNe. \newline
Both the continuum and fluorescent spectral 
features of the reflected X-ray spectrum are dependent 
on the structure and composition of the gas itself, as well as on the 
properties of the X-ray source illuminating the gas, and on the relative 
position of the source and the gas with respect to the observer. They therefore 
contain a wealth of information both on the source itself and on the gas 
structures surrounding it.

Several models 
\citep{Sunyaev1998, Murakami2000, Churazov2002,Odaka2011, Marin2015} have been 
developed to simulate 
the reflection of Sgr A$^{*}$ flares by Sgr B2, the brightest of the CMZ XRNe. 
These models considered 
different relative positions of the GMC with respect to the source, different 
total masses of the GMC and different density gradients of its gas. \newline
In this work, we wish to expand on these models to investigate how 
\changeRiII{more realistic models of the substructure of molecular clouds, in 
particular their clumpiness, would affect the reflected X-ray signal.} 
(\changeRiII{From now on} we will refer to both clumps and cores as 
\textit{clumps} for simplicity, as the latter term merely refers to the 
low-mass end of the same population of overdensities.) \\
In particular, we investigate the effects of the following clump population 
model parameters 
on the XRNe's X-ray emission:
\begin{itemize}
\item Slope of the clump mass function ($\alpha$);
\item  Minimum clump mass in the clump mass 
function ($m_{min}$);
\item Fraction of the total cloud's mass found 
in clumps, or dense gas mass fraction ($f_{\text{DGMF}}$);
\item The mass-size relation of individual clumps 
($m = m_{norm} (r/pc)^{\gamma}$);
\end{itemize}

We use a grid-based Monte Carlo radiative transfer code in order to compute the 
reflected energy spectrum and \changeRi{polarisation} of Sgr B2's X-ray 
emission.
In section \ref{MonteCarlo} we discuss the physical processes accounted for and 
the Monte Carlo method used in our
calculations. In section \ref{SgrB2}, we discuss the models for Sgr B2 
considered, 
including the parameters used and the assumptions made in simulating its clump 
population (section \ref{clumpmodels}). In section 
\ref{results} we discuss the results obtained. 

Finally, in section \ref{timeevu}, we discuss the time-variability of the X-ray 
morphology of XRNe in the case of clumps, and suggest that under illumination 
by a non-persistent, flaring source such as Sgr A$^{*}$, the evolution of 
reflected X-ray intensity can reveal information about the location of the 
clumps along their line of sight, and therefore on the 3d distribution of these 
substructures inside GMCs. 

\changeRi{Even though we simulate the particular case of Sgr B2, our results 
are more generally applicable to other GMCs in the Galaxy when illuminated by 
other X-ray sources such as X-ray binaries.}

\section{Monte Carlo simulation of X-ray propagation in inhomogeneous media}
\label{MonteCarlo}
In this section, we describe the physical processes accounted for in simulating
the 
propagation of X-ray photons in neutral gas (section \ref{Xrayphys}) and the 
the Monte Carlo radiative transfer code used (section \ref{WeightingMethod}).

\subsection{X-ray interaction with neutral matter}
\label{Xrayphys}
X-rays interaction with atoms and molecules in the interstellar gas takes place 
through two processes: scattering and absorption through photoionisation. 
\newline
Scattering processes of X-ray photons on bound electrons are classified as 
Rayleigh in the case of elastic scattering, Raman for scattering which results 
in the excitation of electrons in the atom or molecule, and Compton for 
scattering which results in the ionisation of the atom \citep{Sunyaev1996}. In 
our code, we account for these scattering processes on neutral 
HI, H2 and He using the results \changeRi{for the doubly differentiated 
cross-section of 
\cite{Vainshtein1998}}. The contribution 
of 
heavier elements to 
the total scattering cross section, for which such results are not currently 
available, is accounted for by approximating the 
interaction of X-rays with their electrons as if they were unbound. \newline
We further include the effect of polarisation in scattering processes using the 
prescription of \cite{Namito1993}. \newline
Photoionisation, on the other hand, takes place through the ionisation by X-ray 
photons of tightly bound 
electrons in the atoms' innermost shells, which results in the release 
 of a free electron. We use cross sections 
from 
\cite{Verner1995}, and include both K- and L-shell photoabsorption. 
The unstable electron configuration of the 
ionised atom prompts the filling of the K-shell vacancy by an electron in one 
of the higher energy levels, which causes a release of energy. This 
energy can either be released through the emission of a photon in the X-ray 
range (fluorescence) or be transferred to another electron, which is then 
ejected from the atom (Auger effect). The probability of either process taking 
place varies depending on the atomic configuration and on the original energy 
level of the electron that fills the K-shell vacancy. The probability of 
fluorescence is called the \changeRi{radiative} yield ($Y$). In our 
calculation, K-shell 
fluorescence 
yields 
are taken from Krause et al 
1979, and K$_{\beta}$-to-K$_{\alpha}$ ratios are taken from Ertugral et al 
2007. \changeRi{K$_{\alpha 2}$-to-K$_{\alpha 1}$ ratios are decided by the 
degeneracy of $2_{p_{1/2}}$ and $2_{p_{3/2}}$ levels}, \changeRiNo{which we fix 
to 0.5}. The 
energies of 
the fluorescent lines are taken from 
Thompson et al 2001. \newline
We assume a chemical composition of the \changeSu{Sgr B2} cloud given by a factor of 1.5 compared 
to protosolar abundance, as estimated by \cite{Lodders2003}. 

\subsection{Photon weighing method}
\label{WeightingMethod}
We use a Monte Carlo grid-based radiative transfer code to simulate the 
propagation of X-ray photons through a cloud of complex internal structure, 
containing both diffuse and dense regions. We use an 
octree-based approach for gridding the cloud's internal structure, 
\changeRi{which allows 
us to 
have finer grids in high density regions}. \newline 
The Monte Carlo code makes use of the \cite{Pozdnyakov1983} prescription for 
Monte Carlo methods of X-ray propagation. This applies a weight-based approach 
to the radiative transfer, in which photon packages, rather than 
individual 
photons, are followed. Each package is described by a statistical 
weight $w$, which 
reflects the relative probability of photons undergoing different types of 
interactions, a position, $\bf{r}$, a direction of 
travel, $\bf{\Omega}$, and an energy, $hv$. \\
Photon-packages are initially emitted by the source with weight 1. Their 
starting position is the 
source's own position, and their energy is randomly sampled from the source's 
spectrum. Their initial direction is finally randomly 
sampled from an isotropic distribution (assuming the source radiates 
isotropically). \\
In order to compute the propagation of photon packages in our grid, we estimate 
at each step the relative probability of different processes occurring.
A photon package $\bf{P}$, found in a given grid-cell $g$, has in fact a 
probability of:
\begin{itemize}
\item escaping the grid-cell $g$,  $(L)$; 
\item not escaping the grid-cell $g$,  $(1-L$);
\end{itemize}
and, if not escaping the grid-cell $g$, a probability of:
\begin{itemize}
\item being scattered, $(p_{scatt,Z})$; 
\item being absorbed by a K or an L shell, $(p_{abs,Z} = p_{K ion,Z} + p_{L 
ion,Z}$);
\end{itemize}

By considering photon packages rather than single photons, we are able to 
account for all the above events at once by splitting the initial 
photon package weight $w$ as follows:
\begin{itemize}
\item $w_u = w L$ is the probability of $\bf{P}$ crossing the grid-cell without 
undergoing any interaction;
\item $w_s = w (1-L)p_{scatt,Z}$ is the probability of $\bf{P}$ undergoing a 
scattering event inside the grid-cell;
\item $w_{fluo,Z} = w (1-L)p_{K ion,Z} Y$ is the probability of $\bf{P}$ 
photoionising the K shell of element $Z$ and resulting in a fluorescent 
emission;
\end{itemize}
We can account for these events by assigning their weight to 
secondary packages $\bf{P_u}$,$\bf{P_s}$, and $\bf{P_{f,Z}}$s, which will 
represent the relative probability of each one of those physical events taking 
place. Parameters $\bf{r}$, $\bf{\Omega}$ and $hv$ in $\bf{P_u}$,$\bf{P_s}$, 
and $\bf{P_{f,Z}}$s have, of course, to be updated, each according to the 
physical processes relevant for that secondary package.
\\Once all parameters have been updated, the calculation is iterated by taking 
each one of the secondary packages as an initial package $\bf{P}$, and so on 
for the secondary packages then produced. To limit the number of secondary 
photon packages that the code has to follow, we define a statistical threshold 
$\epsilon$ below which secondary packages are discarded. \\
Note that other possible events, for example the emission of an Auger 
electron, not listed above, as well as any secondary processes they may cause, 
will not result in the emission of X-ray radiation, and can therefore be safely 
ignored in the processing of the X-ray radiation field. They will however 
contribute to the deposition of X-ray energy to the interstellar gas.
\changeRiII{The convergence of the code with respect to the threshold $\epsilon$ and 
other parameters was verified, as well as the consistency of our results with those of 
\cite{Odaka2011} (for the energy spectrum) and \cite{Churazov2002} (for the polarisation spectrum).}

\section{Sgr B2 model}
\label{SgrB2}
Containing $\sim$10\% of the molecular 
mass in the CMZ, Sgr B2 is not only one of the most massive, but also one of 
the most complex molecular structures observed in our Galaxy. It is also one of 
the brightest XRNe observed, making it an ideal target for our study. \newline
Located at a projected distance from the GC of $\sim 100$ 
pc, its exact position on the line of sight still remains uncertain.
\changeRi{In a coordinate system in which the GC is located at (0,0,0), the 
observer at the Sun's location (0,-8kpc, 0), and the Galactic longitude is 
defined in the direction of the positive $y$ axis, 
 we assume a fixed position of the GMC at (0,100 pc,0), so that 
the angle between the source, the cloud and the observer is $\sim 90^{\circ}$. }
Discussions on how different geometries should affect the reflected X-ray 
signal can be found in \cite{Churazov2002} and \cite{Odaka2011}. \\
Studies of the large scale morphology of Sgr B2 \citep{Lis1990} show it is 
surrounded by a diffuse gas envelope, extending up to 22.5 pc in radius, with a 
near-uniform density of $n_{H2} \sim 10^3 \text{cm}^{-3}$. 
What is generally referred to as the Sgr B2 cloud, and where most of the 
reflected X-ray signal 
originates, is a dense region within this envelope, of density 
$n_{H2} \sim 10^{4-5}\text{cm}^{-3}$ and extending out to a radius of ~ 10 pc 
\citep{Hasegawa1994}. Within this region, multiple dense clumps are observed. 
The Sgr B2 GMC contains three well known dense clumps, named B2(N), Sgr B2(M) 
and Sgr B2(S). These are known 
to be hosting clusters of compact
HII regions, providing evidence for star-formation activity within this GMC \changeSu{\citep{Gordon1993}}. 
\addII{Two of these cores, SgrB2(N) and SgrB2(M) (with masses of 
3313$M_{\sun}$ 
and 3532 $M_{\sun}$ \citep{Qin2011} and radii 0.47 
pc and 0.62 pc \citep{Etxaluze2013} respectively),
have been resolved at a subarcmin scale in X-rays \citep{Zhang2015} and at a 
subarcsec scale using the Submillimeter 
Array (SMA) by \cite{Qin2011}}. The high 
angular resolution reached by the latter study revealed a remarkable difference 
in the internal structure of the 
two cores, 
with SgrB2(M) appearing to be highly fragmented to 12 sub-cores and SgrB2(N) 
only 
appearing to be divided to only 2 sub-cores, one of which contains most of the 
mass. The very different morphologies of 
the two have been speculated to be evidence in support of the idea that the two 
cores may represent different evolutionary stages of basically the same core, 
given that SgrB2(N) and SgrB2(M) \changeRiII{are} of comparable size and mass. 
\\
In our calculations, we assume Sgr B2 to have a mass of $M_{B2} = 2.5 \times 
10^{6} 
M_{\sun}$ within radius 10 pc, and a diffuse H2 envelope around it as described 
above. In our calculations, 
we approximate this envelope following \cite{Odaka2011}'s prescription, by 
considering an initial 
absorption to the incoming spectrum due to a column density of $N_{H2} = 6 
\times 
10^{22}$ cm$^{-2}$. Note that this initial absorption will only affect the 
incoming spectrum below $\sim 4$ keV energies.\newline
We then consider, given this mass and size, different possible models 
for the clump population inside the cloud, as discussed in the next section.

\subsection{Simulating Sgr B2's clump population}
\label{clumpmodels}
\begin{figure*}
\includegraphics[trim = 0 0 50 0, clip, height = 7cm]{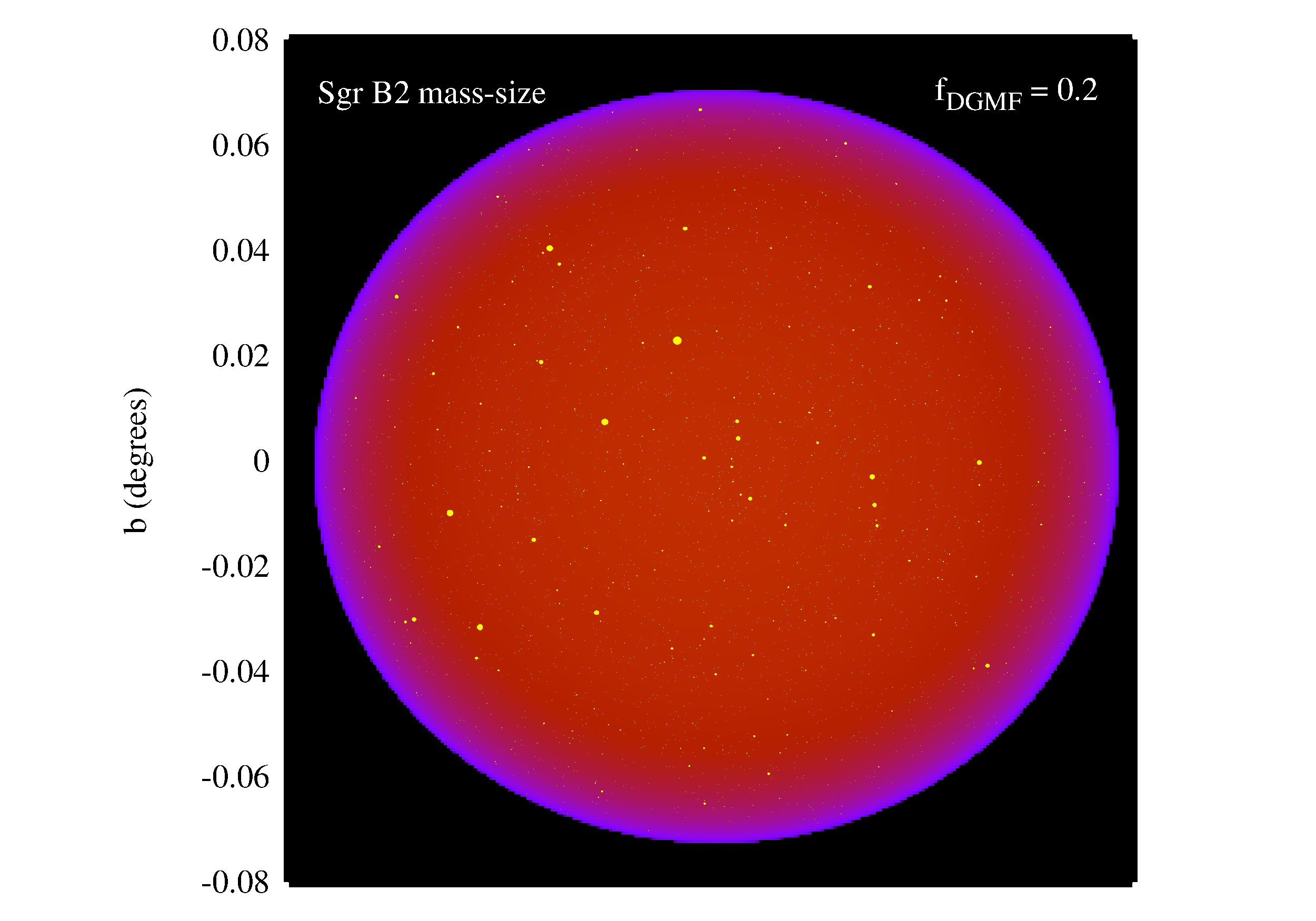} 
\includegraphics[trim = 0 0 50 0, clip, height = 7cm]{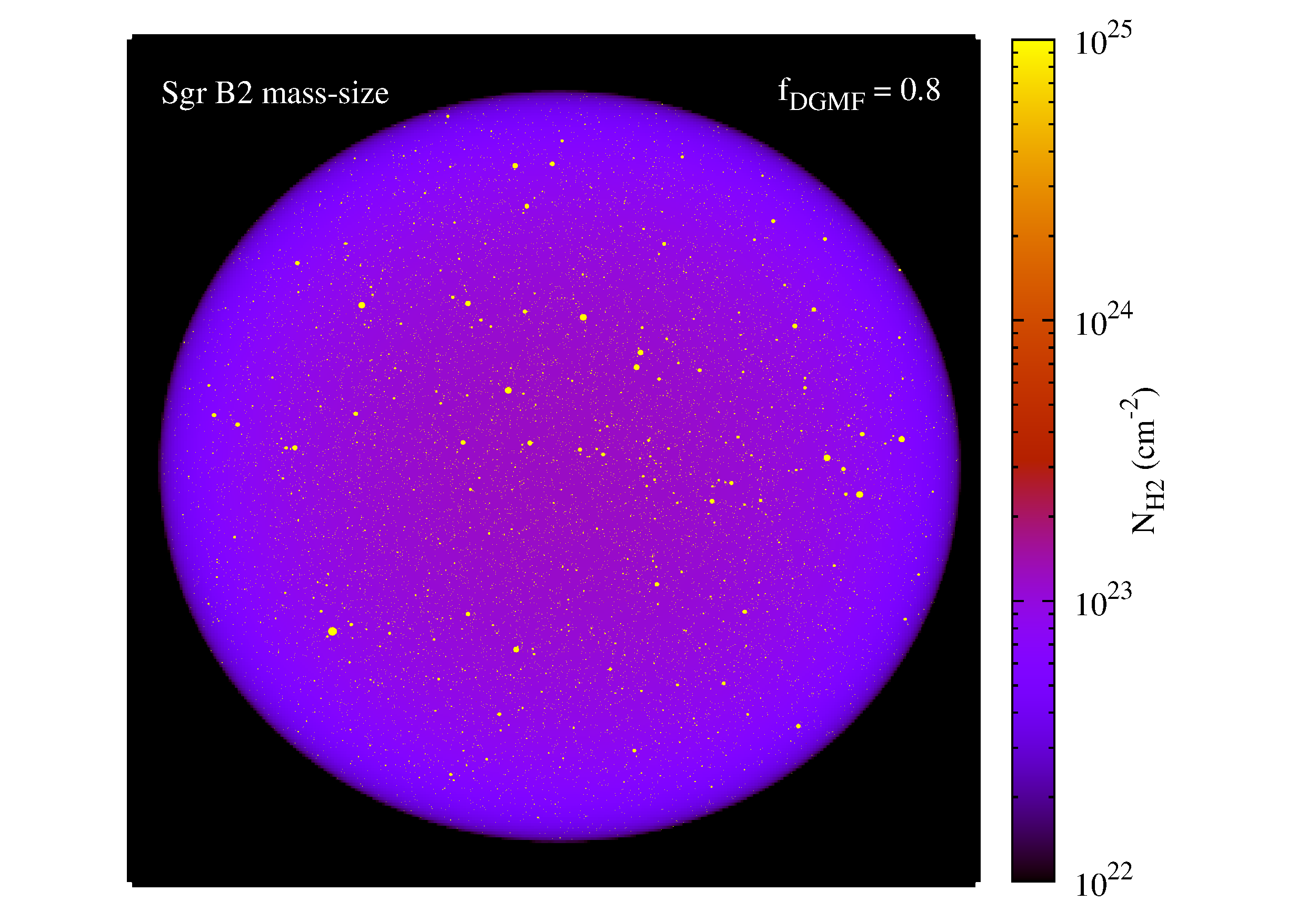} \\
\includegraphics[trim = 0 0 50 0, clip,  height = 7cm]{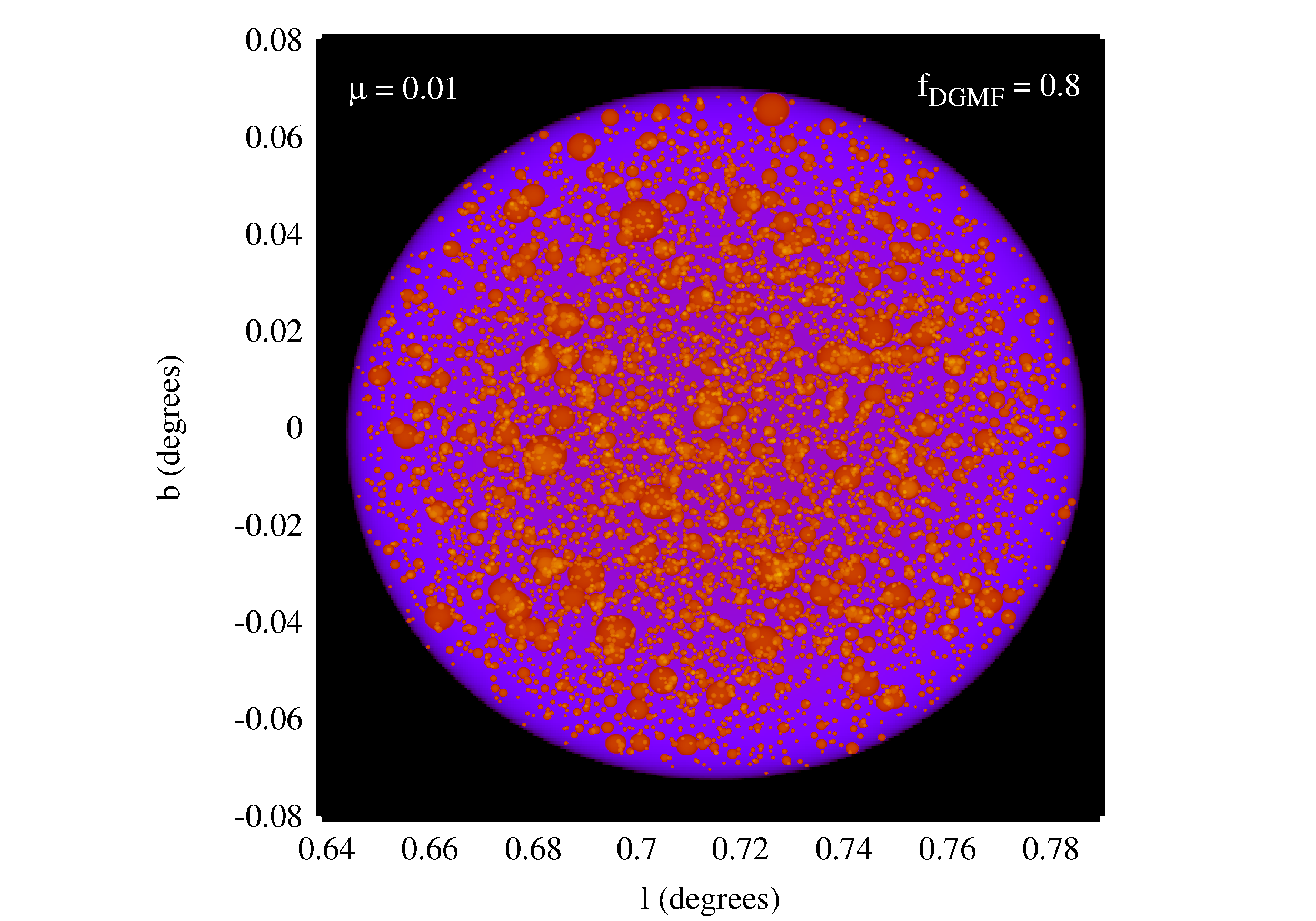}
\includegraphics[trim = 0 0 50 0, clip,  height = 7cm]{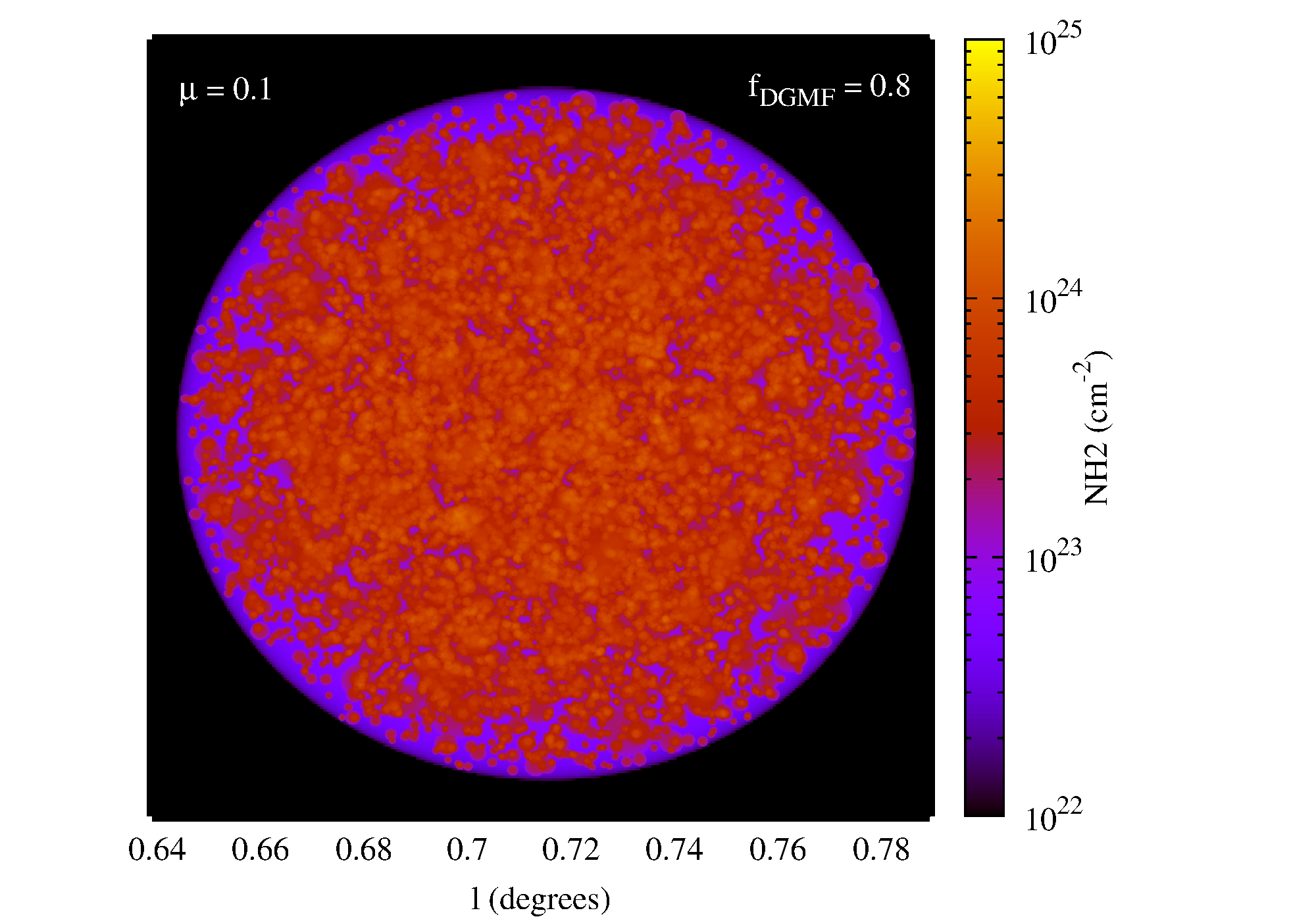}
\caption{Cumulative column density ($N_{\text{H2}}$) along line of sights, 
excluding ISM and diffuse envelope contributions. The maps show Sgr B2 models 
with fixed parameters 
$\alpha = 1.35$, $m_{min} = 10 M_{\sun}$ and Sgr B2 mass-size relation, and 
two values of $f_{\text{DGMF}}$. The concentration of 
mass 
into small dense regions reduces the interclump density, as seen from the 
figures. For the case of $f_{\text{DGMF}} = 0.8$, we also show clump 
populations with \changeRiII{mass-size relations tuned} to obtain a volume 
filling 
fraction of $\mu = 0.01$ and $\mu = 0.1$. 
\label{column_dens}}
 \end{figure*}

Statistical descriptions of the internal structure of GMCs 
have, since the outset of this field, been formulated 
in terms of discrete over-dense regions within GMCs (i.e. clumps 
and cores). More recently, a growing number of studies have 
adopted a different approach and described these density fields
in terms of 
fractals, placing emphasis on the self-similarity of structural 
features at all scales (see review by \cite{Williams1999}). In our work we will 
\changeRi{use} 
the 
former 
approach.\\
The statistical study of such discrete over-dense regions has been applied to a
number of nearby Galactic Plane GMCs, such as, for example, Orion and W51 (eg. 
\citet{Parsons2012}). This has led to 
the formulation of standard population properties and relations, generalised 
from studies of Galactic molecular clouds (eg \citet{Larson1981}): the clump 
mass function (CMF) 
and mass-size relation of individual clumps. Such studies, along with 
simulations, suggest there exist similarities both in 
the CMF and mass-size relations between different GMCs, and therefore 
point towards a 
universal structure of GMCs in the Galactic Plane (see sections \ref{masssize} 
and \ref{CMF}). 
The internal structure of GMCs in the CMZ, on the other hand, is likely to 
greatly differ from that observed in Galactic Plane GMCs, due to the 
extreme environment (both in density \citep{Lis1990} and 
pressure \citep{Kruijssen2014}). Understanding how the clump 
population of these objects differs from the ones found in Galactic Plane GMCs, 
would, therefore, give an important insight into how environmental factors 
affect 
the process of internal structure 
formation. \\
As previously discussed, such studies for GMCs in the CMZ are rather more 
difficult to 
perform, 
due to the large distance and mass of these complexes: catalogs of 
clump and core populations in these GMCs are known to be incomplete, and 
therefore
don't allow for a reliable estimate of their statistical properties. \\ 
In our calculations, we consider different possible shapes of the CMF and 
mass-size relation (see \changeRi{sections} \ref{masssize} and \ref{CMF} 
respectively). 
We also consider different levels of fragmentation of the cloud into clumps by 
considering different possible values of the dense 
gas mass fraction, or the
fraction of the cloud's mass
found in clumps, $f_{\text{DGMF}}$ (see Fig. \ref{column_dens}). 
\newline
The simulation of different clump population models for Sgr B2 is then 
performed as 
follows: given a  $f_{\text{DGMF}}$ and CMF, we sample the 
clumps' masses from the CMF in the mass range $m_{min} - m_{max}$, with 
$m_{max} = 10^4$ M$_{\sun}$, until we have 
obtained a total mass of $M_{DGMF} \sim f_{\text{DGMF}} \times M_{B2}$. We then 
proceed 
to 
assign each clump with a size based on its mass \changeRiII{assuming a 
mass-size relation}. Finally, we
uniformly distribute the resulting clumps inside the 
cloud, and calculate the interclump density using the "left-over" mass, 
$M_{interclump} \sim (1 -f_{\text{DGMF}}) \times M_{B2}$ and the "empty" (i.e. 
not 
occupied by clumps) volume, assuming the interclump density is 
homogeneously distributed.\newline
In the following sections, we discuss the details of this procedure, together 
with the range of parameters considered and the assumptions made in doing so. 
 
\subsubsection{Clump Mass Function}
\label{CMF}
\begin{figure}
\includegraphics[trim = 0 0 0 0, height = 7cm, width = 9cm]{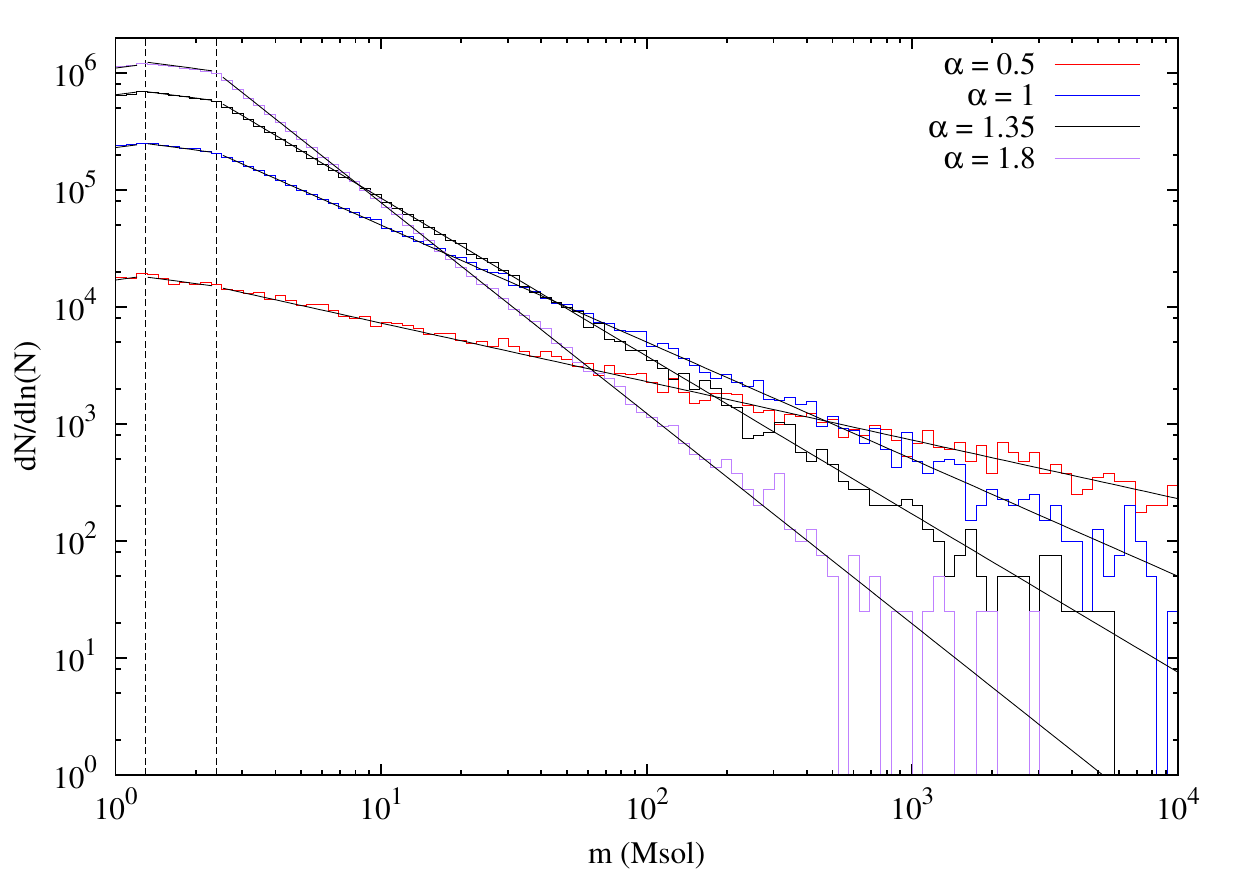} \\
\caption{Clump mass functions for different $\alpha$ with 
$f_{\text{DGMF}} = 0.4$. The grey 
solid lines show the distributions sampled, while the coloured lines show the 
actual realisation. The vertical 
lines show the values at which the CMF changes in slope.
\label{CMFplot}}
\end{figure}

The clump mass function (CMF) of clump populations observed in Galactic plane 
GMCs takes 
the following form \citep{McKee2007}: 
\begin{equation}
\label{CMF}
m\frac{\text{d}N}{\text{d}m} \propto m^{-\alpha}
\end{equation}
The slope of the clump mass function is similar to that for GMCs as a whole, 
possibly because both are
determined by turbulent processes within larger, gravitationally bound systems 
\citep{McKee2007}.
The parameter $\alpha$ takes different values in different mass ranges. The low 
mass end of the function is known as the core mass function. Its shape is 
particularly 
important in relation to the Initial Mass Function (IMF) of stellar populations 
\citep{McKee2007}.
Evidence of similarity in the core mass function with the IMF were first 
studied by \citet{Nutter2007} 
in the Orion complex, which for the first time observed a turnover at $\sim 1.3 
M_{\sun}$, which mimics the turnover seen in the stellar IMF at ~0.1 
$M_{\sun}$. Fitting the data with a three part power law function similar to 
that observed \changeRiII{in the stellar IMF:}
\begin{equation}
\alpha =
  \begin{cases}
   -0.3 & \mbox{if } 0.4 M_{\sun} < m < 1.3 M_{\sun}  \\
     0.3  & \mbox{if } 1.3 M_{\sun} \leq m \leq 2.4 M_{\sun}  \\
    1.35 & \mbox{if } m > 2.4 M_{\sun} 
  \end{cases}
\end{equation}
for clump mass ranges below 100$M_{\sun}$. The 
physical significance of the turnover mass is not clear, as the same work 
highlights that similar studies in the lower-mass and nearer cloud complex 
$\rho$ Ophiuchi showed no turnover in their CMF (\citep{Motte1998}). Later 
studies on the low-mass end of the CMF in the Aquila rift cloud complex also 
confirmed a variation in this 
parameter: \citet{Konyves2010} found a turnover mass of $\sim 0.6 M_{\sun}$ in 
the starless 
core sample, and $\sim 0.9 M_{\sun}$ prestellar (starless and gravitationally 
bound) sub-sample. \\
The similarity of the CMF with the \citet{Salpeter1955} stellar IMF has been 
further confirmed in 
higher mass ranges by \citet{Tsuboi2012} (range > 900 $M_{\sun}$) and 
\citet{Parsons2012} 
(range > 200 $M_{\sun}$).
\newline Despite the fact that the CMF appears to be consistent throughout 
Galactic Plane GMCs, the extreme environment of Sgr B2, as previously discussed,
suggests the CMF for this GMC could be significantly different. Unfortunately, 
because of a lack of exhaustive data, how this function should differ is not 
obvious. 
We therefore adopt the \cite{Nutter2007} three-part power law fitting, but 
consider \changeRiII{different $\alpha$ values in the highest mass interval}. 
In particular, we consider $\alpha = 
1.35$ (i.e. Galactic Plane value)$, 1, 0.5, 1.8$ (see Fig. \ref{CMFplot}). 
We sample this range up to clump masses $10^4 M_{\sun}$, to be consistent with 
the observed massive cores in Sgr B2 (eg \cite{Qin2011}). We maintain the 
low-mass threshold of the CMF as a parameter in the model, $m_{min}$, and 
investigate its effect on the reflected X-ray signal. 

\subsubsection{Clumps mass-size relation}
\label{masssize}
\begin{figure}
\includegraphics[trim = 0 0 0 0, height = 
7cm, width = 9cm]{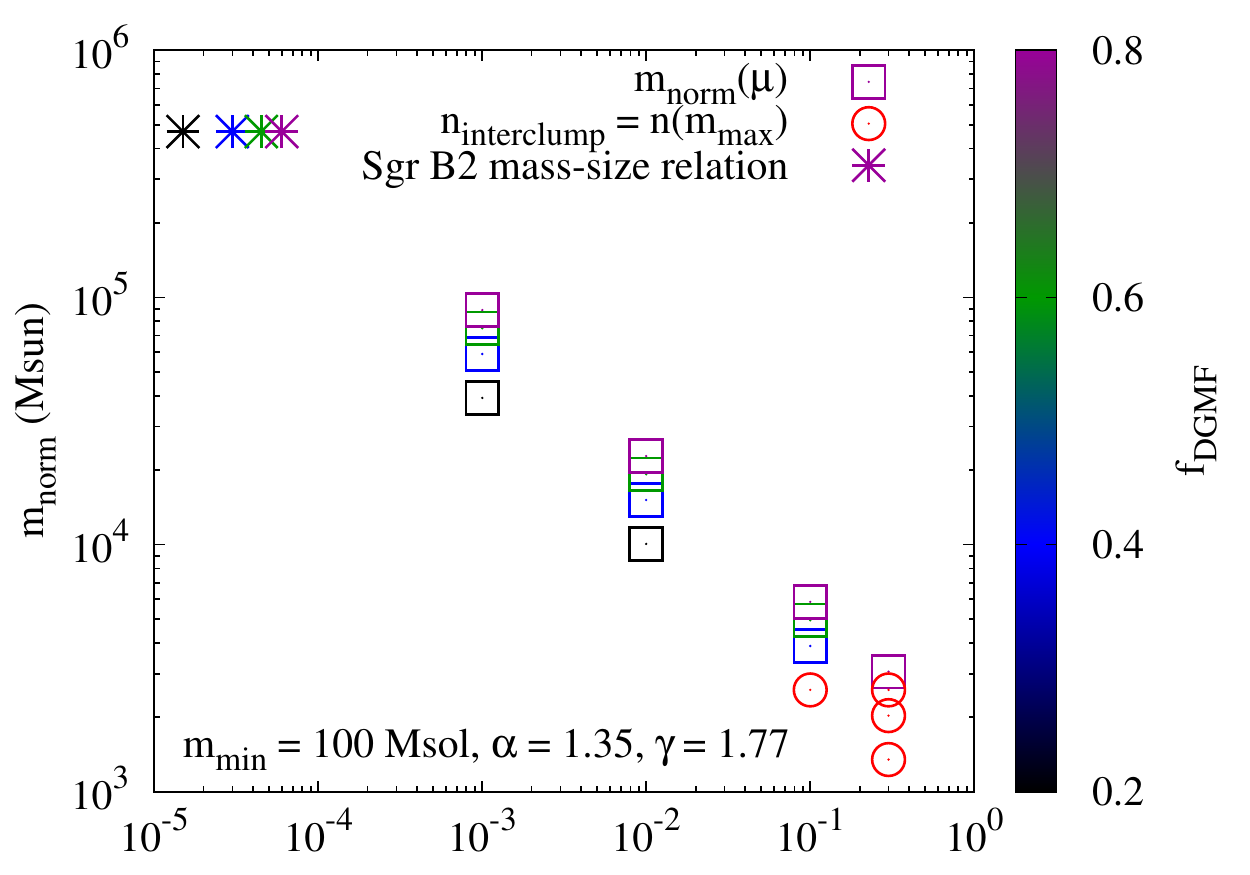} \\
\caption{Normalisation parameter $m_{norm}$ in the mass-size relation as a 
function of the volume filling factor $\mu$, calculated using Eqn. 
\ref{filling_factor}, for different $f_{\text{DGMF}}$ at fixed  $\alpha = 
1.35$, 
$m_{min} = 100 M_{\sun}$ and $\gamma = 1.77$.  We also show the $\mu$ values 
obtained using the fixed Sgr B2 relation. Note that for 
high $\mu$ and low $f_{\text{DGMF}}$ values, we obtain mass-size relations 
yielding 
clump populations where the least dense clump, given by the clump 
with the highest mass ($m_{max}$), is less dense than the interclump 
density $n_{interclump}$, in contradiction with the very definition of clump as 
an overdensity; we clearly mark these models in red in the plot. 
We will use the $f_{\text{DGMF}}$ value that allows us to explore a widest 
possible 
range of $\mu$, $f_{\text{DGMF}} = 0.8$, to investigate the impact of 
increasing $\mu$ 
on the reflected X-ray signal (see Fig. \ref{compare_mu_zoom}).
\label{Sgr_B2_volume_fraction}}
\end{figure}

The mass-size relation of individual clumps is particularly important for our 
study, as it determines the volume filling fraction, $\mu$, of the clump 
population given its $\alpha$, $m_{min}$ and $f_{\text{DGMF}}$ parameters. The 
volume 
filling fraction in return determines how effectively clumps "hide" part of 
the cloud's 
mass, by concentrating it into small volumes which X-rays have a low 
probability 
of intercepting. \\
For clumps formed inside GMCs via turbulent 
fragmentation, one would expect a 
mass-density, or similarly a mass-size, relation \citep{Donkov2011}. A first 
suggestion of this relation based on observations was formulated in the seminal 
paper of \cite{Larson1981}, which claimed a universal power-law mass-size 
relationship should describe all clouds and clumps found in the Galaxy. This is 
commonly referred to as "Larson's third law", and it suggested that 
the relation should go as:
\changeRi{
\begin{equation}
m = m_{norm} r^{\gamma}
\label{larsons}
\end{equation}
with $\gamma = 2$. }
Since being first formulated, this empirical observation has been extensively 
studied 
both in observations (eg \citet{Kauffmann2010, Lombardi2010}) and in 
simulations 
\citep{Shetty2010, Donkov2011}, which all find a deviation from the power of 2 
originally found by \cite{Larson1981}. The latter two find that on scales $< 1 
$pc, the mass size relation studied in eleven different GMC structures in the 
Galactic Plane, can be described as $m(r)=400 M_{\sun}(r/pc)^{1.7}$ and 
$m(r)=380 M_{\sun}(r/pc)^{1.6}$ respectively. By looking at these clouds 
individually, \cite{Lombardi2010} find that these clouds have quite similar 
exponents, but rather different normalisation 
masses (ranging from 170 to 710 $M_{\sun}$).  \newline
Although such studies haven't been performed for the clump population of Sgr 
B2, we can use 
observational data currently available to get a rough estimate of what this 
relation should look like for this GMC. \\
We use the data available from the 14 cores within Sgr B2(N) and SgrB2(M) 
resolved in the subarsec observations of \cite{Qin2011} (see section 
\ref{SgrB2}) to infer a mass-size relation for 
Sgr B2.
\changeRiII{Assuming a homogeneous density distribution inside the clumps, we fit the power law to these observations and find:}
\begin{equation}
\label{SgrB2masssize}
m(r)= 4.68 \times 10^5 M_{\sun}(r/pc)^{1.77} 
\end{equation}
which we refer to here-forth as the "Sgr B2 mass-size 
relation". We see 
that the exponent is consistent with the results of 
\citet{Lombardi2010}, but that the normalisation mass is considerably higher. 
This is somehow expected, considering that the average density in Sgr B2 is 
considerably higher than that found in Galactic plane GMCs.
Note that the range of sizes of these cores is 
considerably below the \cite{Lombardi2010} range. In our model, we assume this 
relation, inferred in a somewhat limited mass range, holds for all possible 
clump masses. \\
As already mentioned, the mass-size relation is particularly important for our 
study as it effectively determines \changeRiII{the volume filling factor} of a 
given 
population. The volume filling factor, $\mu$, can be expressed 
in terms of other clump population parameters as follows:
\begin{equation}
\label{filling_factor}
\mu = \frac{4\pi}{3 
V_{SgrB2}}\frac{C}{m_{norm}^{3/\gamma}}\frac{[m^{-{\alpha}+3/\gamma}]^{m_{max}}_
{m_{min}}}{-{\alpha}+3/\gamma}
\end{equation}
where 
\begin{equation}
C =
  \begin{cases}
   M_{DGMF} (1-\alpha)([m^{-\alpha+1}]^{m_{max}}_{m_{min}})^{-1} & \mbox{if } 
\alpha \neq 1 \\
   M_{DGMF} [\text{ln}m] ^{m_{max}}_{m_{min}}  & \mbox{if } \alpha = 1
  \end{cases}
\end{equation}
Using equation \ref{filling_factor}, and assuming the mass-size relation given 
in Eqn. \ref{SgrB2masssize}, we obtain 
a maximum filling fraction, among all $f_{\text{DGMF}}, \alpha$ and $m_{min}$ 
values 
considered, of only $10^{-4}$ (see Fig. \ref{Sgr_B2_volume_fraction}). Therefore the probability of X-rays intercepting 
the clumps 
will be extremely low. Hence, we don't expect 
parameters related to the mass distribution of the clumps, i.e. $\alpha$ and 
$m_{min}$, to significantly affect the X-ray signal when assuming this 
mass-size relation.
We will also consider how the X-ray signal should be affected by increasingly 
large values of $\mu$, by varying the $m_{norm}$ value in the mass-size 
relation 
accordingly (see section \ref{varymasssize}). \\
The mass-size relation can be alternatively expressed as a column density-mass 
relation \changeRi{to the center of the cloud}. Defining $\tau_{\text{H2}} = r 
n_{\text{H2}} \sigma$, this takes the 
following form:
\begin{equation}
\label{tau}
\tau_{\text{H2}} = m_{norm}^{2/\gamma} \frac{3 \sigma}{m 4\pi} m^{-2/\gamma + 
1} 
\end{equation}
The effect of the mass-size relation on the clumps' contribution to the column 
density of the GMC is shown in Fig. \ref{column_dens}.

\subsubsection{Dense gas mass fraction}
Studies of this parameter in Galactic GMCs, as for 
example in W51 \citep{Battisti2014, Ginsburg2015}, show a variety of 
possible stages of fragmentation in different clouds, reflecting the different 
evolutionary stages GMCs are found in. We therefore span a wide
range of $f_{\text{DGMF}}$ values: $f_{\text{DGMF}} = 0.2,0.4,0.6$ and 0.8.

\subsubsection{Spatial distribution of clumps}
\label{spatial_distr}
In our model we assume clumps are not overlapping and are isotropically 
distributed inside the cloud. \\
The latter assumption is a clear simplification of what is currently being 
observed in CMZ GMCs (eg \cite{Chen2015}). In the 
case of very low volume filling factors, as in the case of clumps 
following the Sgr B2 mass-size relation, this assumption should have a 
negligible effect on the reflected signal. In the case of larger volume filling 
factors, on the other hand, this may become relevant. The 
percolation of the X-ray photons will be dependent on the 
projected area filling on the plane perpendicular to the source-cloud 
direction, rather than on the volume filling factor itself. The maximum 
possible projected area occupied by the clumps, in the case where no line of 
sight from the source intercepts more than one clump, is $\propto \pi 
\sum_{i=1}^N v_i^{2/3}$, where $v_i$ is the volume of the single clump $i$. The 
minimum possible project area on the other hand, in the case where all clumps 
are centered along a single line of sight, is $\propto \pi 
v(r_{max})^{2/3}$, 
where $v(r_{max})$ is the volume of the largest clump in the population. The 
actual value of the projected area occupied by the clumps will be somewhere 
between these two extremes, and will be determined by the distribution of the 
clumps inside the cloud. We postpone an investigation of how the spatial 
distribution of discrete over-dense regions affects 
the 
reflected X-ray signal to a later study. 

\section{Results}
\label{results}
We consider a persistent Sgr A$^{*}$ flare of luminosity $1.3 \times 10^{39}$ 
erg/s, modeled with a power law photon index of 1.8 and assumed to be 
completely unpolarised. For each 
model considered, we plot the energy spectrum and
polarisation fraction  
 of the reflected X-ray emission. The 
polarisation fraction is calculated as the fraction of the Stokes Q parameter 
\changeRi{(with the frame of reference chosen so that U=0)}
intensity over the total intensity reaching the observer.\\
First, we consider the case where the mass-size relation of the clumps is the 
Sgr B2 relation discussed in section \ref{masssize}. We then consider, for 
fixed  
$f_{\text{DGMF}}, \alpha$ and $m_{min}$ parameters, the effect of varying the 
normalisation of the mass-size relation, and therefore the volume 
filling factor of the clump population, on the reflected signal. \newline

\subsection{Fixed Sgr B2 mass-size relation}
\label{res_sgrB2relation}
We consider a fiducial model given by $f_{\text{DGMF}} = 0.4, m_{min} = 10 
M_{\sun}, 
\alpha = 1.35$, and vary each parameter individually around it while 
maintaining the mass-size relationship constant at $m_{norm} = 4.68 \times 
10^{5} M_{\sun}$ and $\gamma = 1.77$ according to Eqn. \ref{SgrB2masssize}.

\begin{figure}
\includegraphics[trim = 0 80 0 80, height = 2.5cm]{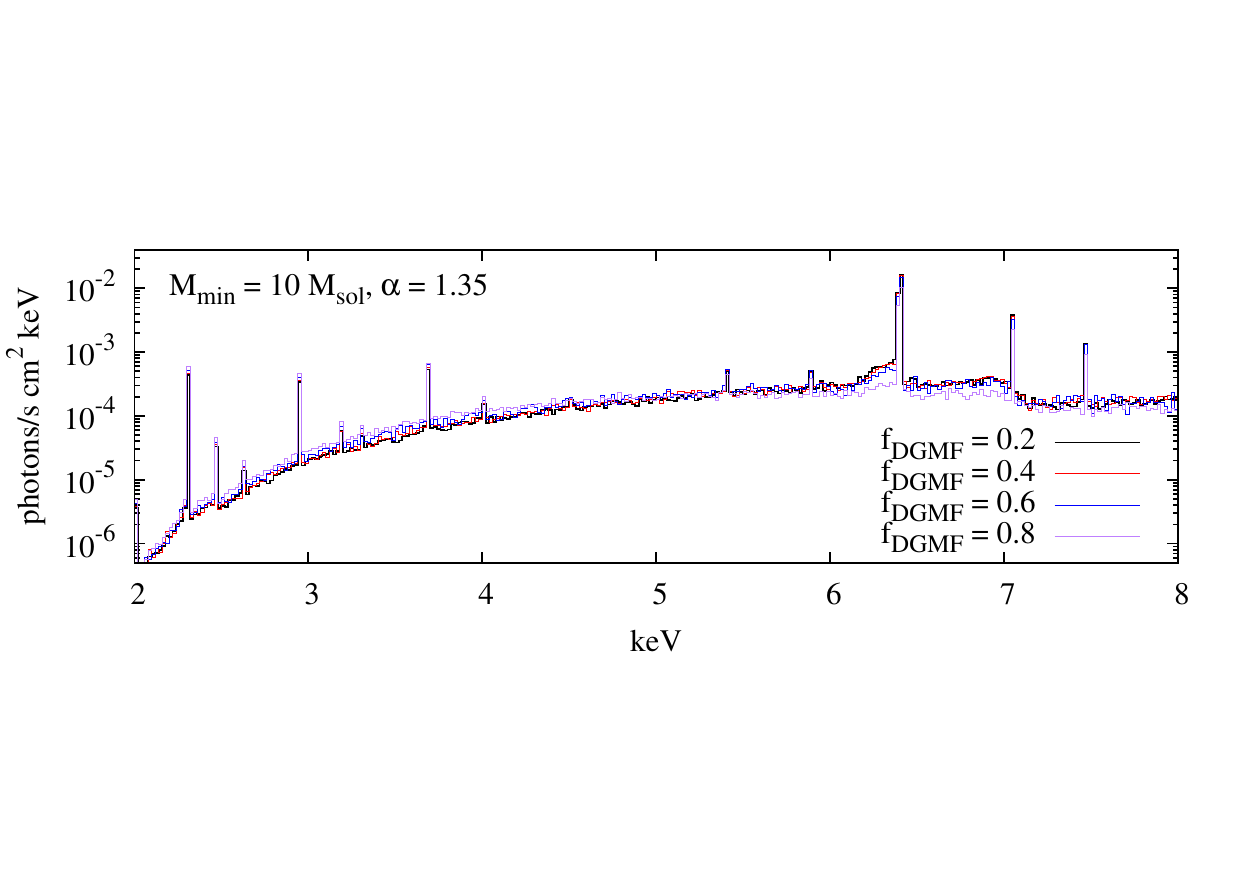} \\
\includegraphics[trim = 0 80 0 80, height = 2.5cm]{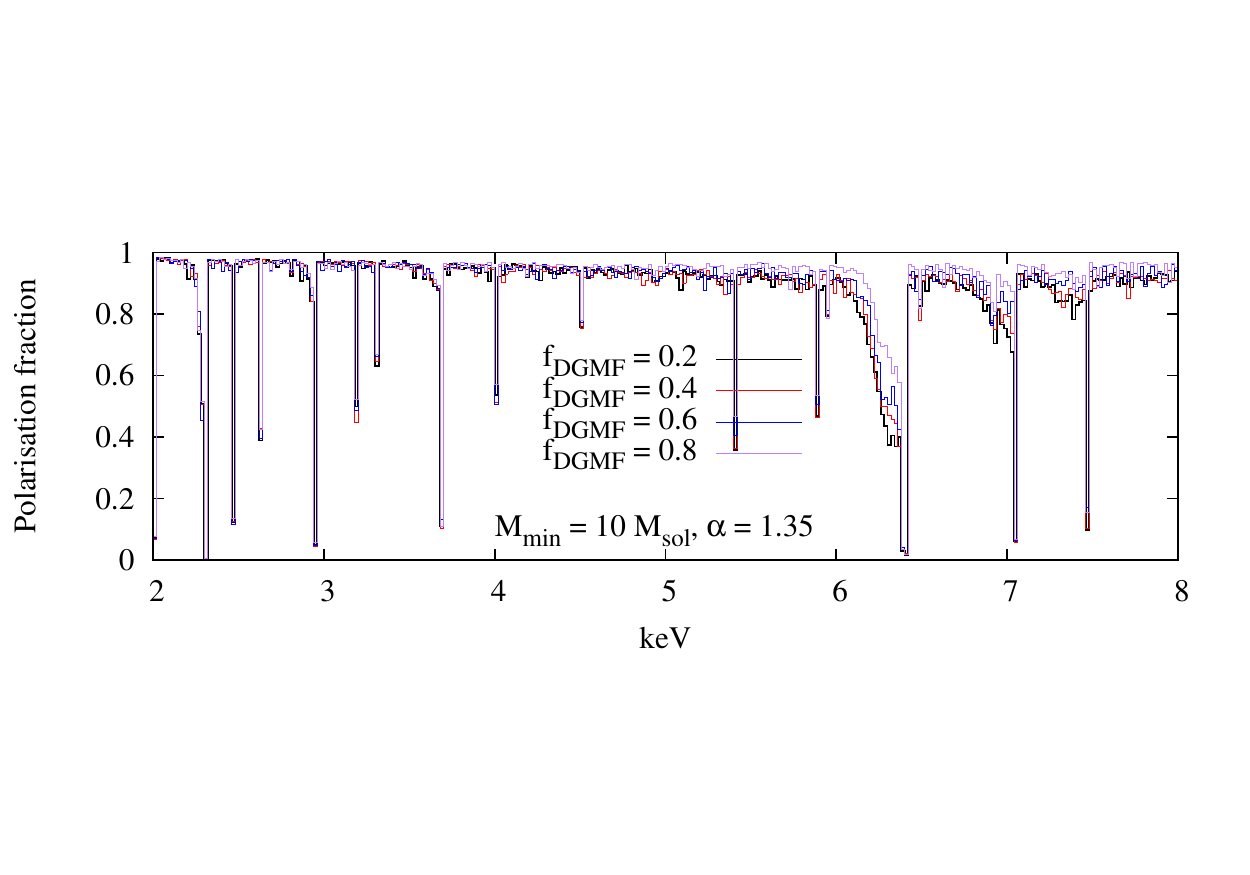} 
\caption{Reflected energy spectrum and polarisation fraction for varying 
$f_{\text{DGMF}}$ cloud models. The energy spectrum 
is shown with resolution of 20 eV.
\label{compare_f_frac}}
\end{figure}

\begin{figure*}
\includegraphics[trim = 50 0 90 20,height = 6cm, width = 
6cm]{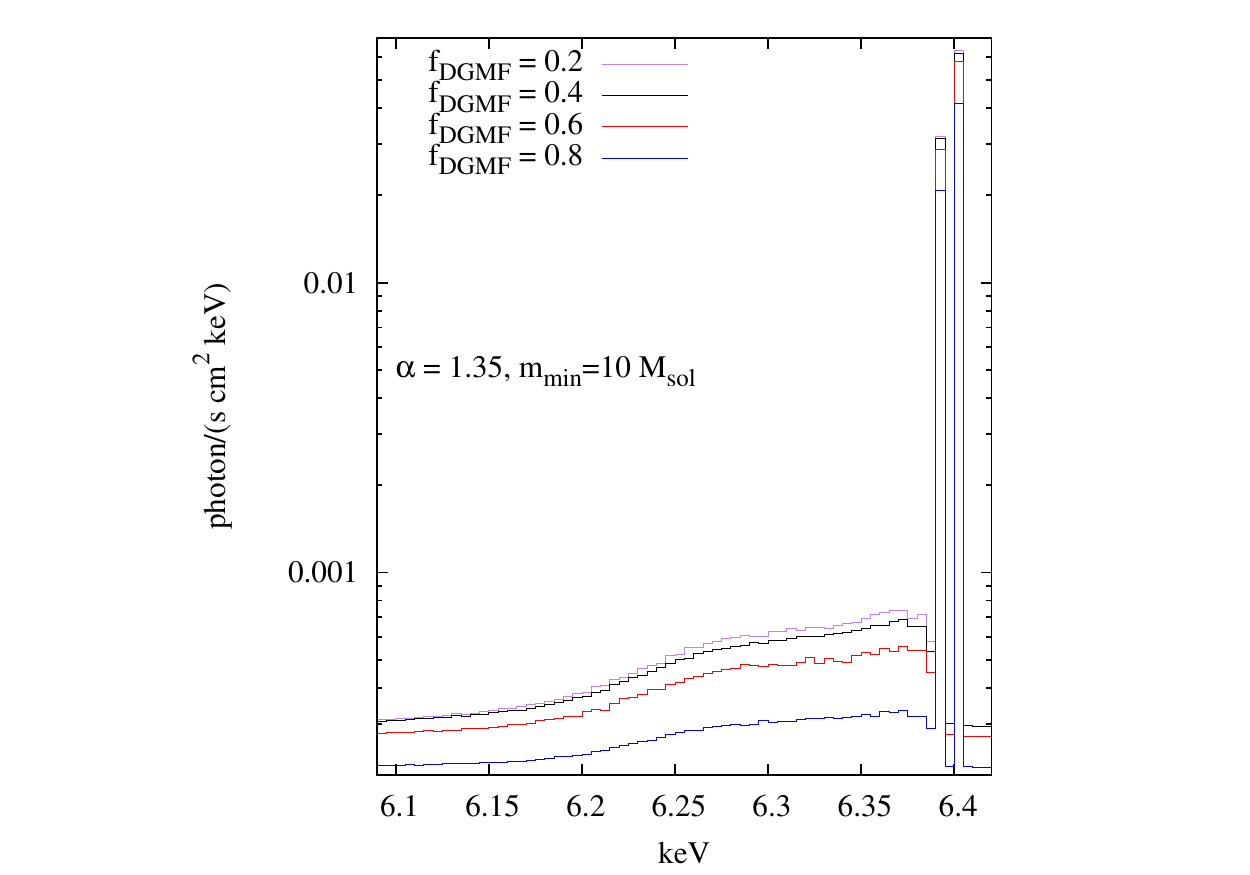} 
\includegraphics[trim = 40 10 70 20,height = 6cm, width = 
6cm]{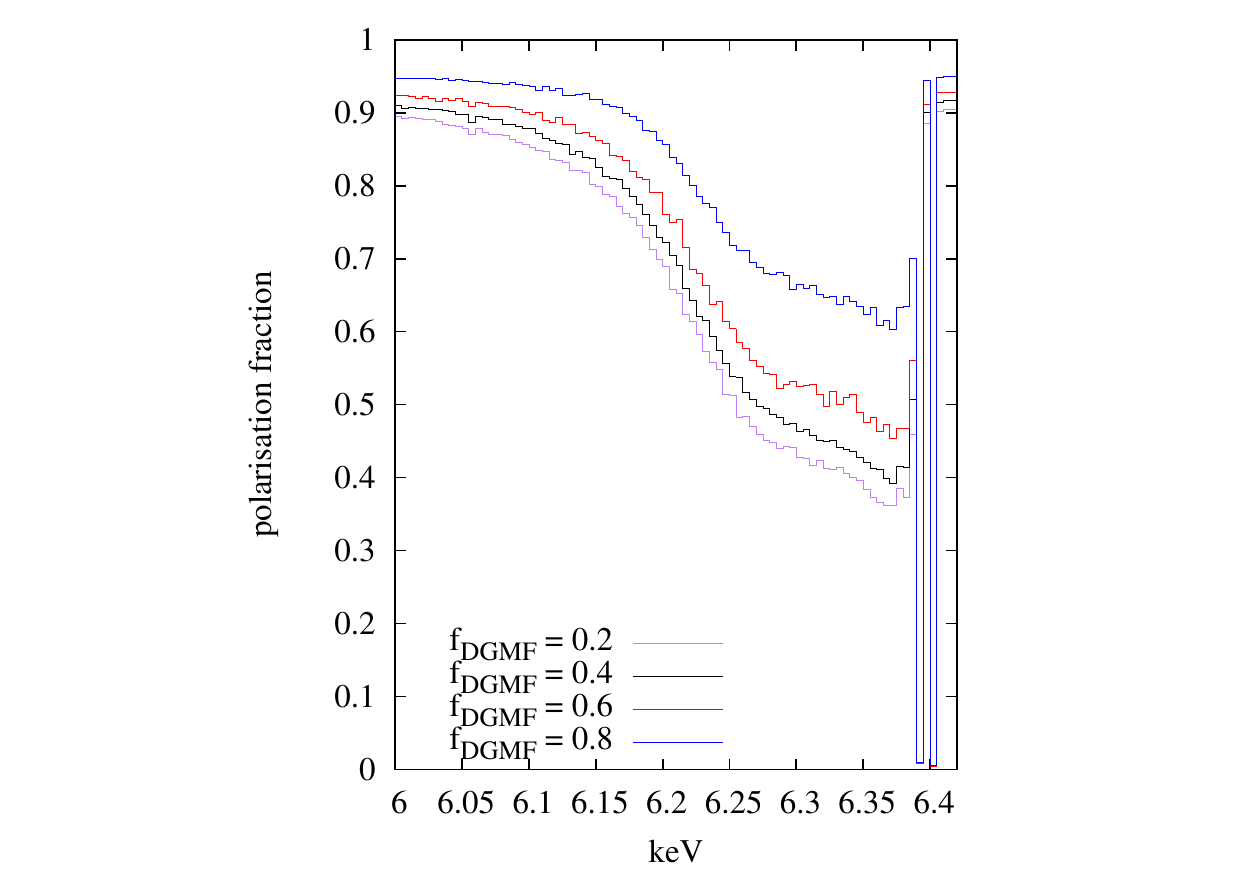} 
\includegraphics[height = 6cm, width = 6.5cm]{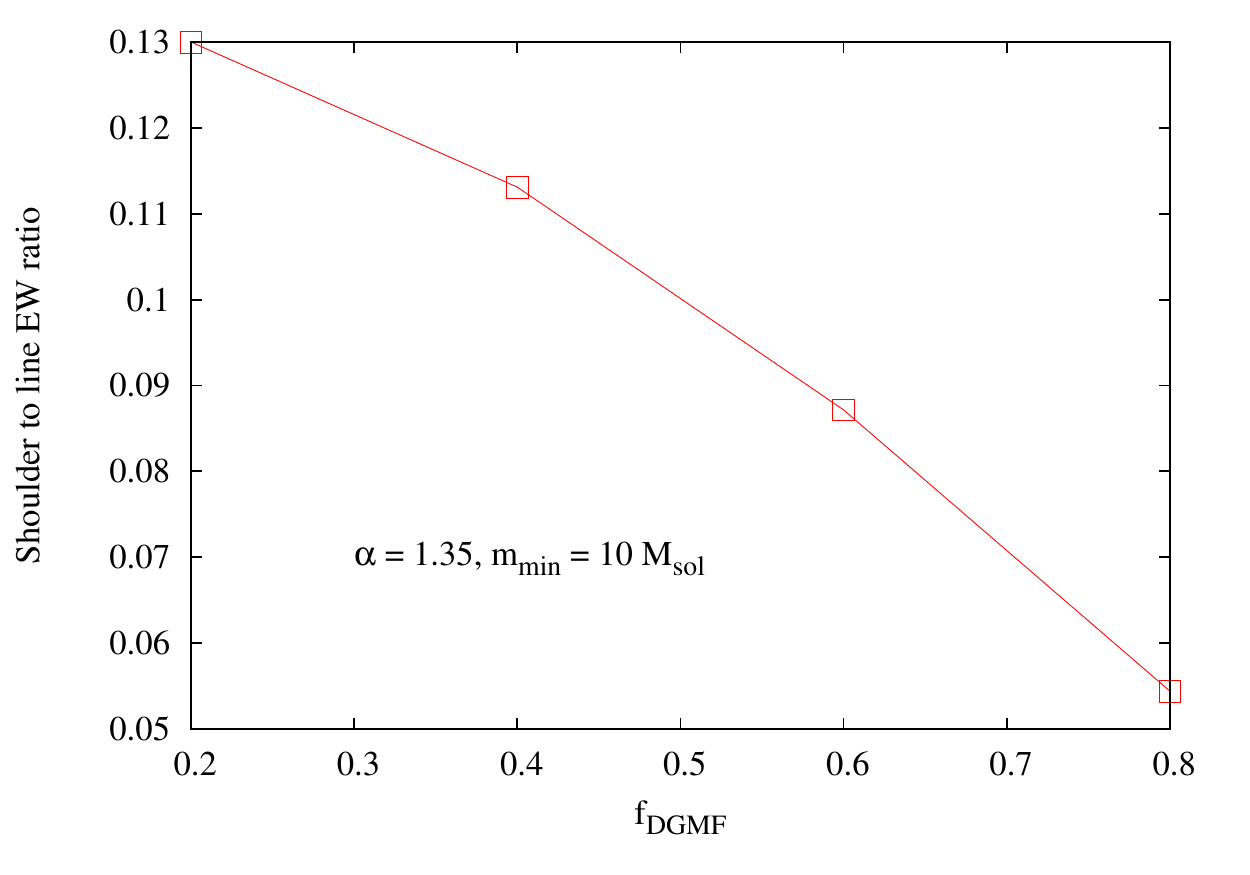} 
\caption{\changeRiII{Reflected energy spectrum and polarisation} around the 6.4 
keV Fe K-$\alpha$ line for varying \changeRiII{$f_{\text{DGMF}}$}. The 
spectral resolution is 5 eV. \changeRiII{The rightmost panel shows the ratio of 
EW for the 6.4 keV shoulder to the line}.
\label{compare_fDGMF_zoom}}
\end{figure*}

In Fig. \ref{compare_f_frac}, we compare the reflected X-ray emission for 
different values of the fragmentation parameter 
$f_{\text{DGMF}}$.  For all models 
considered, we observe that an increase in the fragmentation level 
of 
the cloud into clumps results in: a slight increase in the flux of low 
energy photons, a decrease in the flux of higher energy photons, and a decrease 
in the Fe shoulder's flux (see Fig. \ref{compare_fDGMF_zoom}). \\

\begin{figure}
\includegraphics[trim = 20 0 0 
0, height = 8cm]{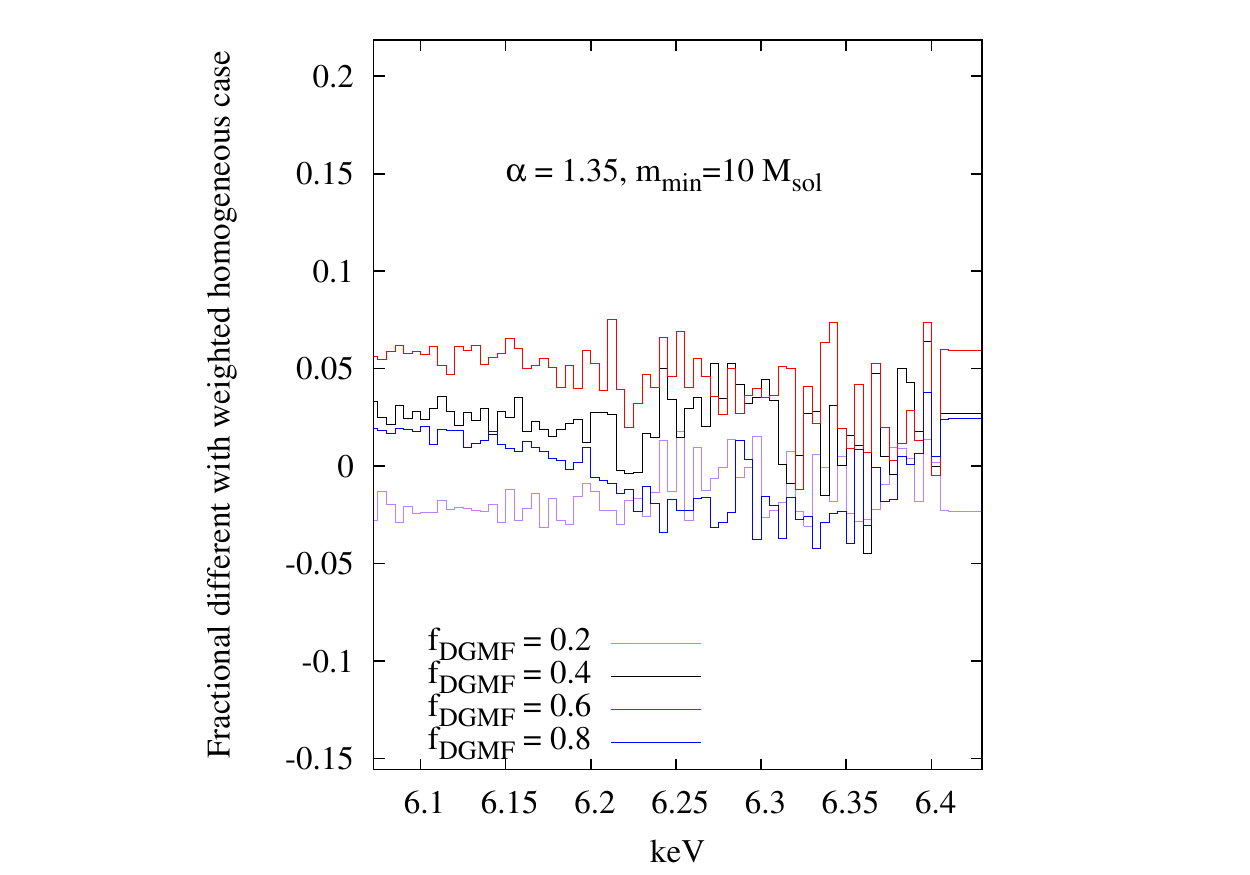} 
\caption{Fractional difference between the reflected energy spectrum for a 
cloud model with mass $M_{B2}$ and a fraction $f_{\text{DGMF}}$ of its mass 
found in 
clumps and a homogeneous cloud model with total mass $M_{B2} \times 
(1-f_{\text{DGMF}})$. The fractional difference is negligible for all energies, 
reinforcing the idea that the clumping of part of the cloud's mass into very 
dense region effectively "hides" that mass from the incoming X-rays.
\label{ratio_fDGMF_homogeneous}}
\end{figure}

These three 
effects can all be accounted for by 
considering percolation: because the probability of intercepting the clumps 
will be extremely low, due to the small ($< 10^{-4}$) volume filling fraction 
of the clumps, the X-ray photons will mainly interact with the atoms and 
molecules in the interclump medium. \changeRi{The resulting reflected spectra 
will therefore be consistent with those resulting from reflection off 
homogeneous \changeRiII{clouds with the same size, and density equal to the 
interclump density}. In Fig. 
\ref{ratio_fDGMF_homogeneous} we compare the X-ray emission obtained 
from cloud models where a fraction $f_{\text{DGMF}}$ of the total cloud's mass 
is 
found in clumps and homogeneous cloud models with a total mass of 
$M_{B2} \times (1-f_{\text{DGMF}})$. We find indeed that the fractional 
difference 
between the two cases is negligible for all energies, and that the 
$f_{\text{DGMF}} = 0.2, 0.4, 0.6$ and 
$0.8$ models can be approximated by homogeneous clouds with $\tau_{\text{HI}} 
\sim 0.32, 0.24, 0.16$ and 0.08 respectively (where $\tau = R \times 
n_{\text{HI}} \times  \sigma_{Thoms}$).}

The decrease in the number of scatterings due to an increase in the dense gas 
mass fraction can also be observed in the plots of the polarisation fraction of 
the reflected spectra (Fig. \ref{compare_f_frac}). 
An analytic approximation to the polarisation fraction of an X-ray photon 
undergoing $n$ scatterings is given by \citep{Churazov2002}:
\begin{equation}
P_{n} = \frac{1-\eta^2}{1+\eta^2+ 
\frac{20}{15}\big(\big(\frac{10}{7})^{n-1}-1\big)}
\end{equation}
where $\eta = cos(\theta)$ and $\theta$ is the average scattering angle. From 
this analytic prescription,   it is indeed for the geometry considered 
in these calculations, the polarisation fraction should be close to unity for 
singly scattered photons, in the case of scattering close to ~90$^{\circ}$ as 
is 
the case here, as and should progressively decrease from unity as the number of 
scatterings increases. With increasing energy the absorption 
\changeRiII{optical} 
depth decreases and the relative contribution to the radiation escaping from 
the cloud from multiple scatterings increases. This is evident in Fig. 
\ref{compare_f_frac} as decrease in degree of polarisation with increasing 
energy.

In Fig. \ref{compare_f_frac}, we can clearly see that the 
polarisation fraction of the shoulder progressively increases 
with the dense gas mass fraction. This means that the higher the fraction of 
the cloud's gas found in dense regions, the lower the number of multiple 
scatterings photons experience, in agreement with a picture of an increasing 
rate of percolation. Fluorescent photons, on the other hand, are 
emitted isotropically by photoionised atoms, and therefore are completely 
unpolarised. 
\changeRi{For varying $\alpha$ and $m_{min}$ parameters we find that, on the 
other hand, these have no effect on the overall reflected signal. This 
reinforces the idea that the dominant effect of clumping 
 within the XRN (in the case of the Sgr B2 mass-size relation) is percolation, 
and that the mass concentrated in clumps is 
effectively "hidden" from incoming X-ray photons because of the small volume it occupies.}

\subsection{Variable mass-size relation}
\label{varymasssize}
\begin{figure}[h]
\includegraphics[trim = 40 0 140 
20,height = 6cm, width = 6cm]{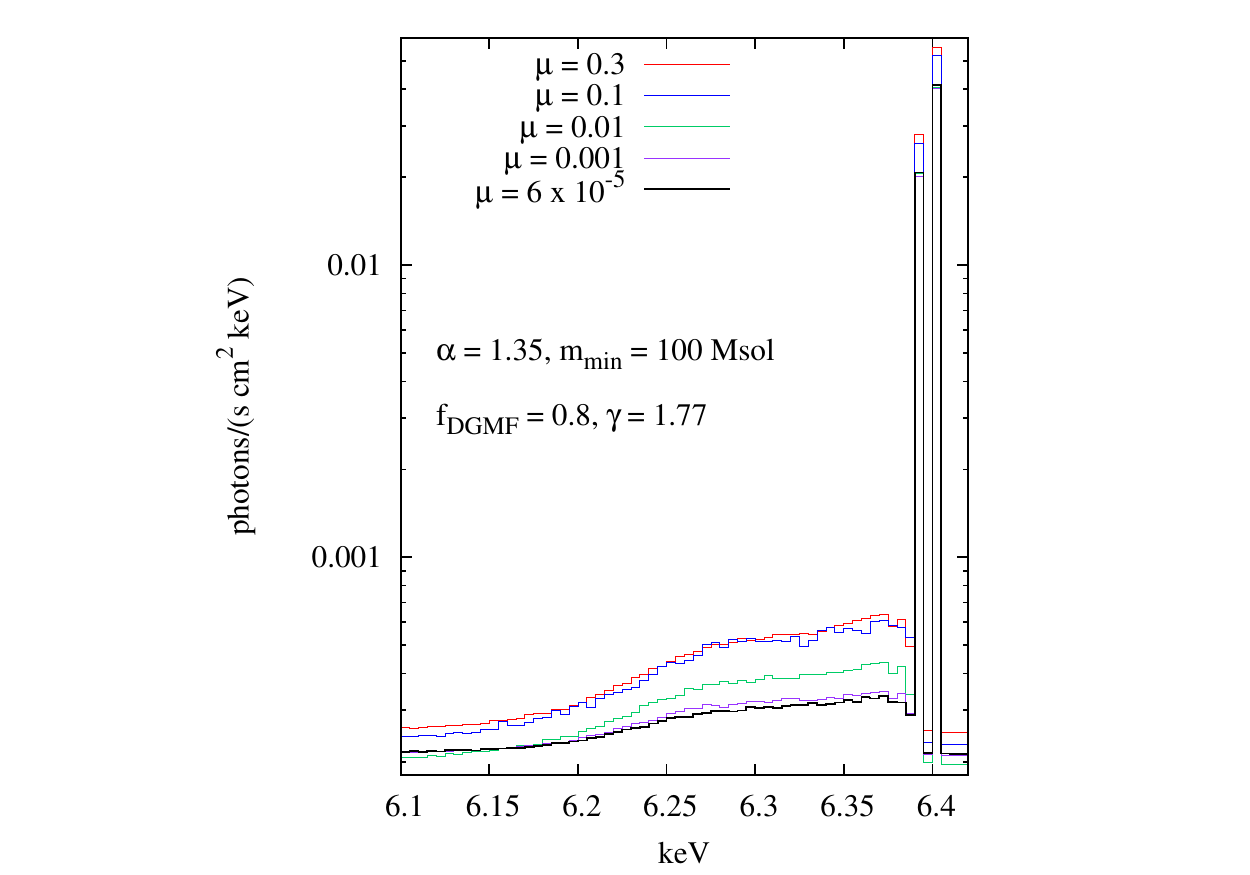}  \\
\includegraphics[trim = 40 0 140 
20,height = 6cm, width = 6cm]{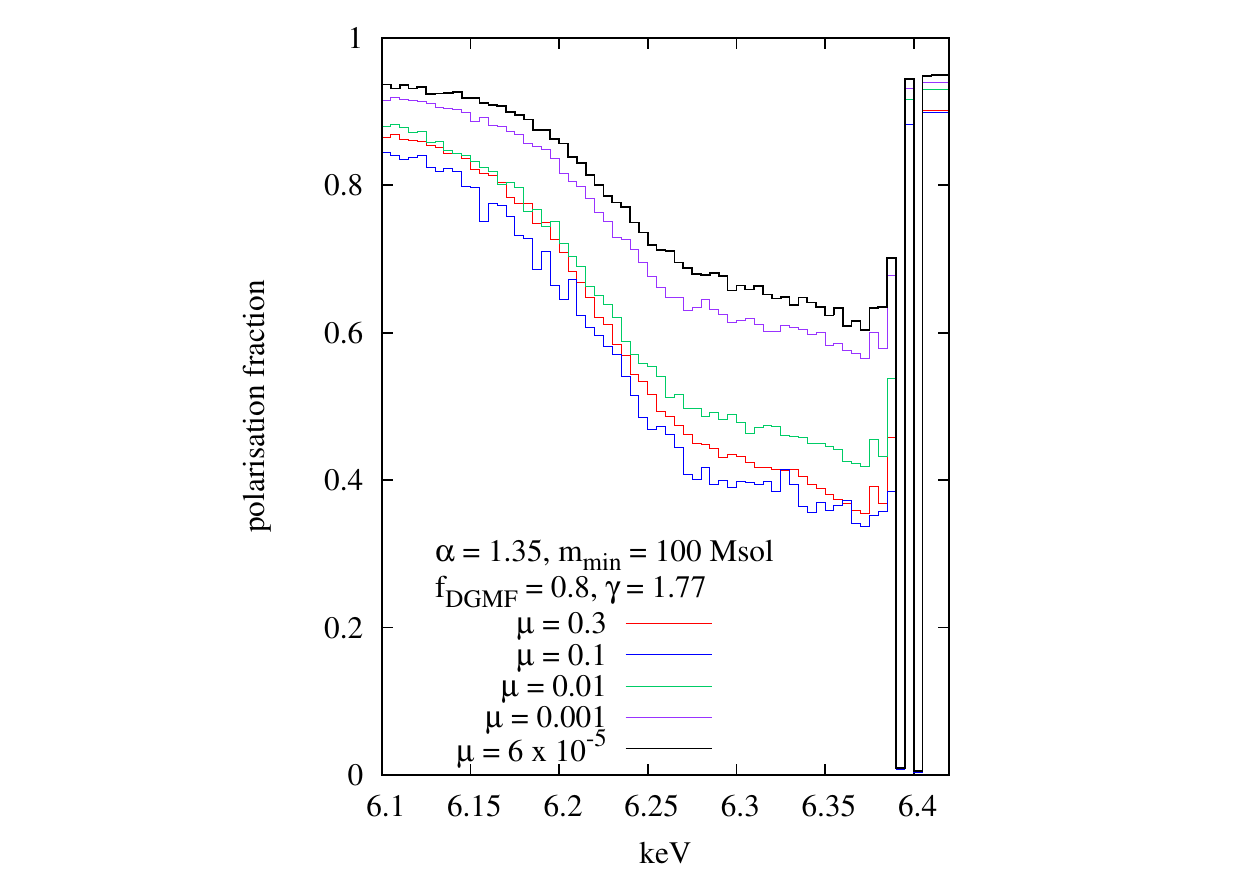} 
\caption{Reflected energy \textit{(top plot)} and polarisation \textit{(bottom 
plot)} around the 6.4 
keV Fe K-$\alpha$ line for varying $\mu$, shown with resolution of 5 eV. Values 
of $\mu$ illustrated were obtained by adjusting the 
$m_{norm}$ parameter in the mass-relation of clumps using Eqn. 
\ref{filling_factor}. As expected, for increasingly large volume filling 
factors, the probability of fluorescent photons intercepting the clumps 
increases, resulting in an increase in the shoulder's flux. Note that, 
because the X-ray signal is really dependent on the projected area (see section 
\ref{spatial_distr}), which roughly goes as $\mu^{2/3}$, we would expect a 
volume filling factor of $\mu \sim 0.001$ (project area $\sim 0.01$) to already 
produce a visible signature in the X-ray spectrum. Indeed, from the 
polarisation fraction plot, it is clearly visible that at $\mu = 0.001$ the 
signal \changeRiII{starts} deviating from the virtually homogeneous case. 
\label{compare_mu_zoom}}
\end{figure}

\begin{figure*}
\includegraphics[trim = 10 0 120 0,height = 6cm, width = 
6cm]{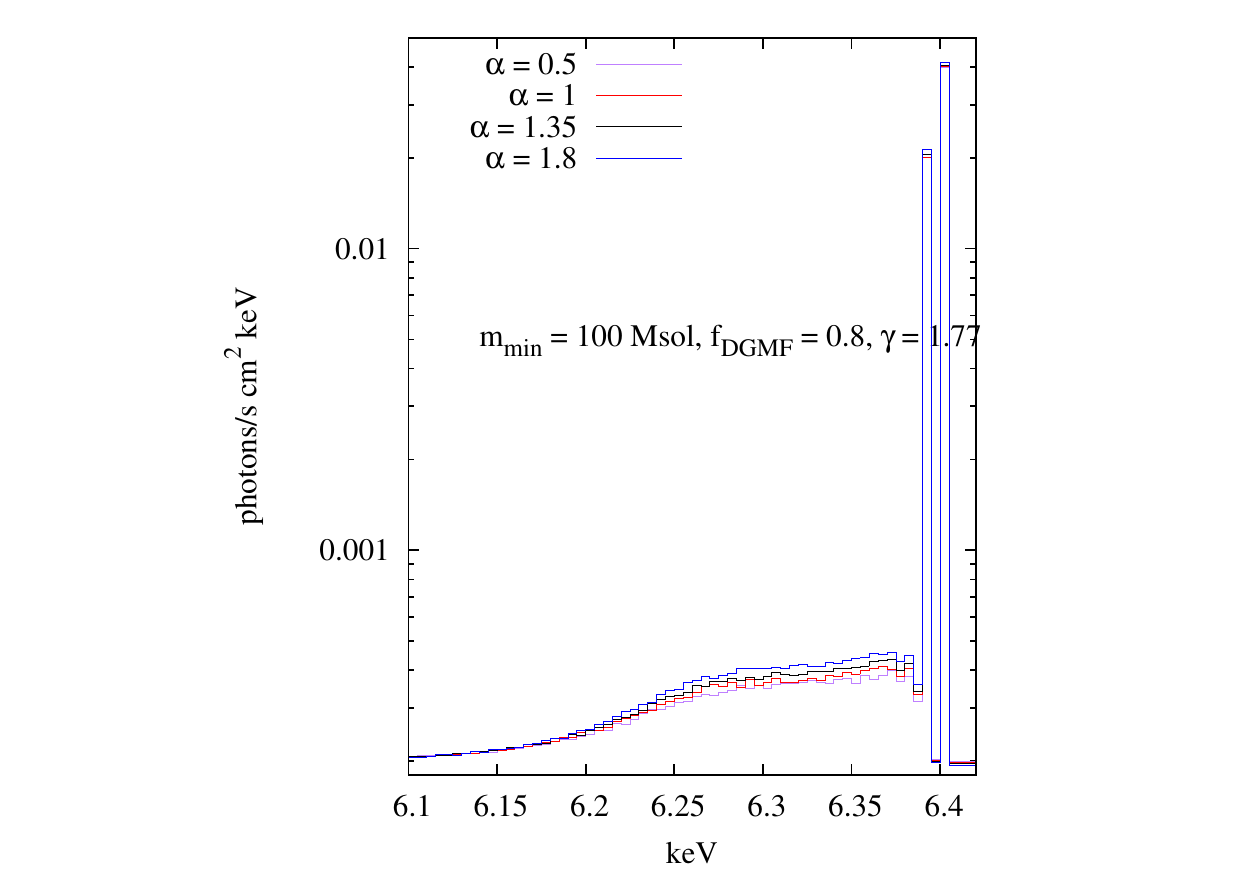}
\includegraphics[trim = 10 0 70 0,height = 6cm, width = 
6cm]{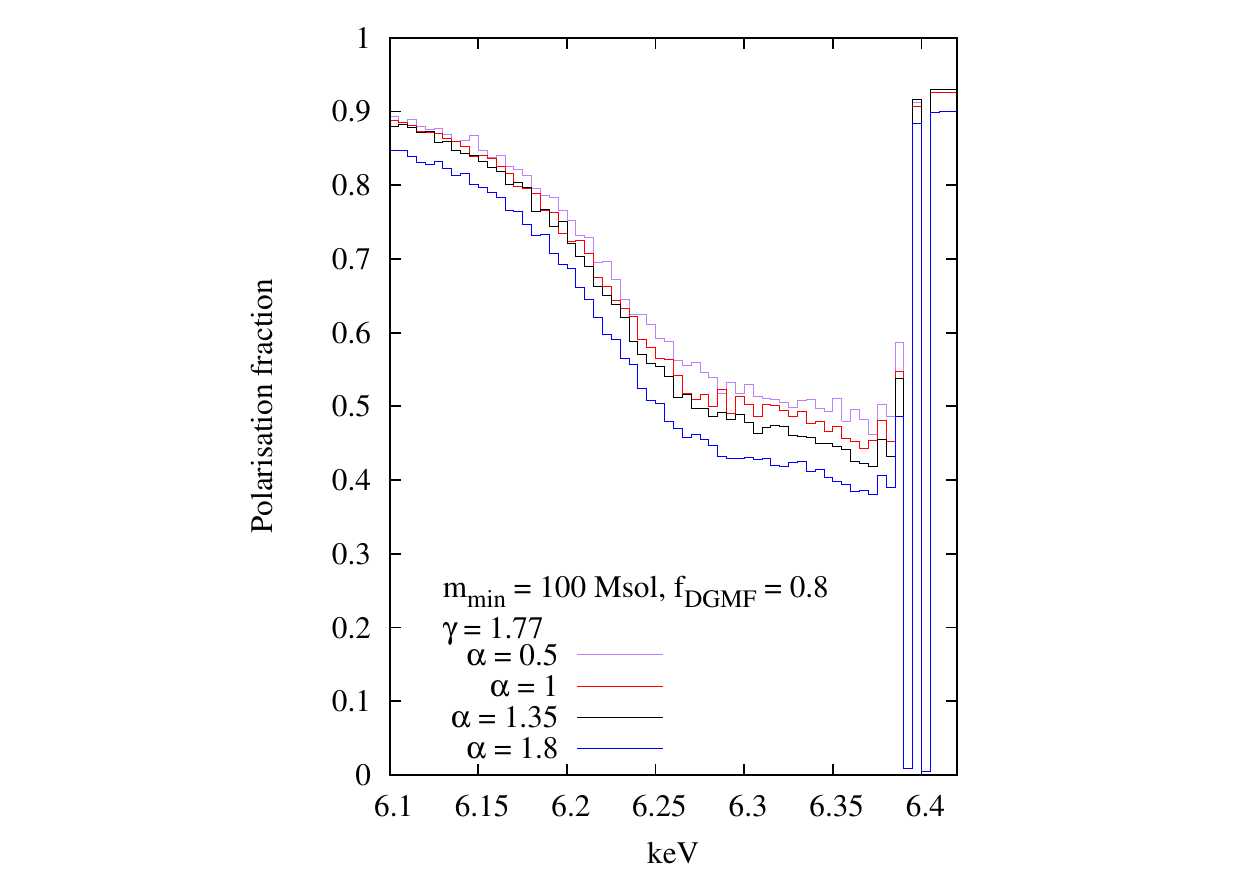} 
\includegraphics[trim = 10 0 14 0,height = 6cm, width = 
6cm]{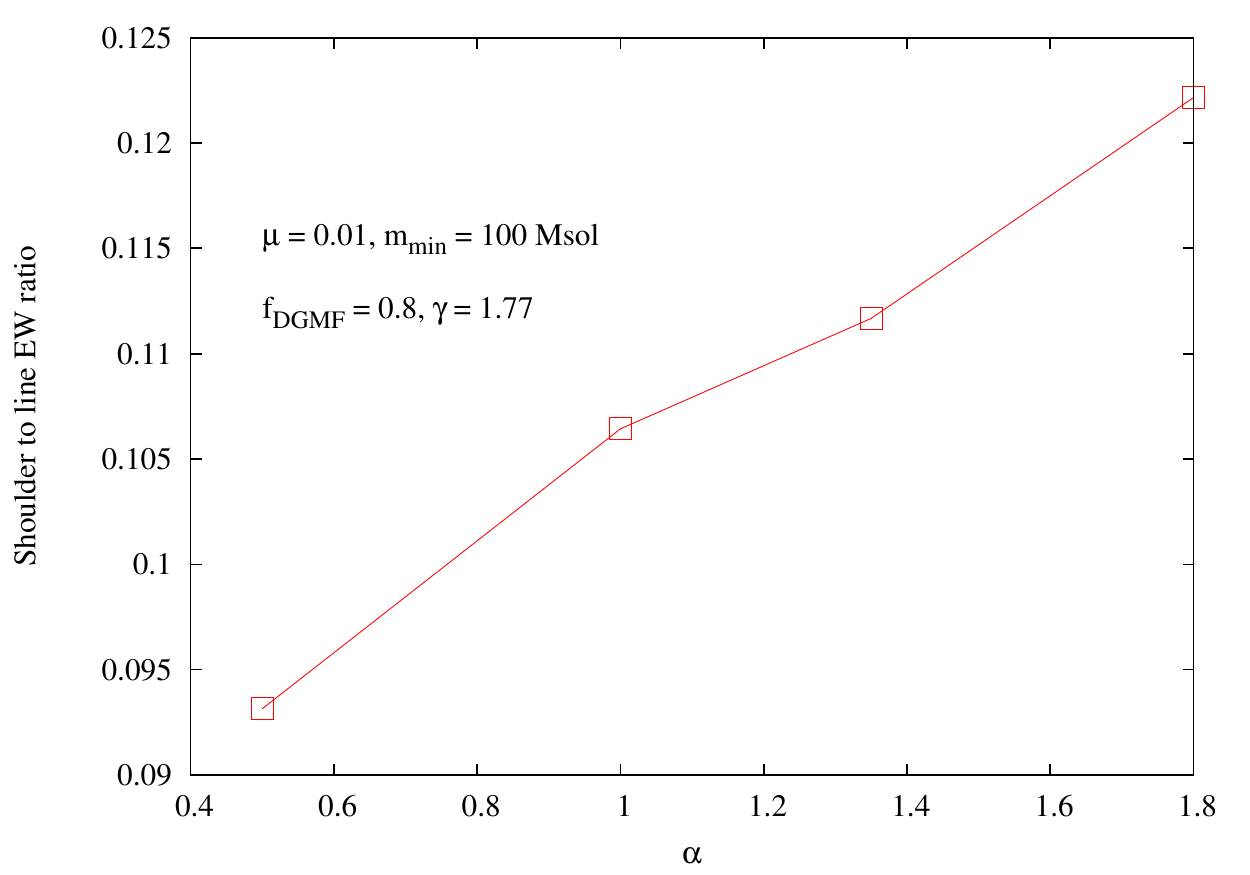} \\
\includegraphics[trim = 10 0 120 0,height = 6cm, width = 
6cm]{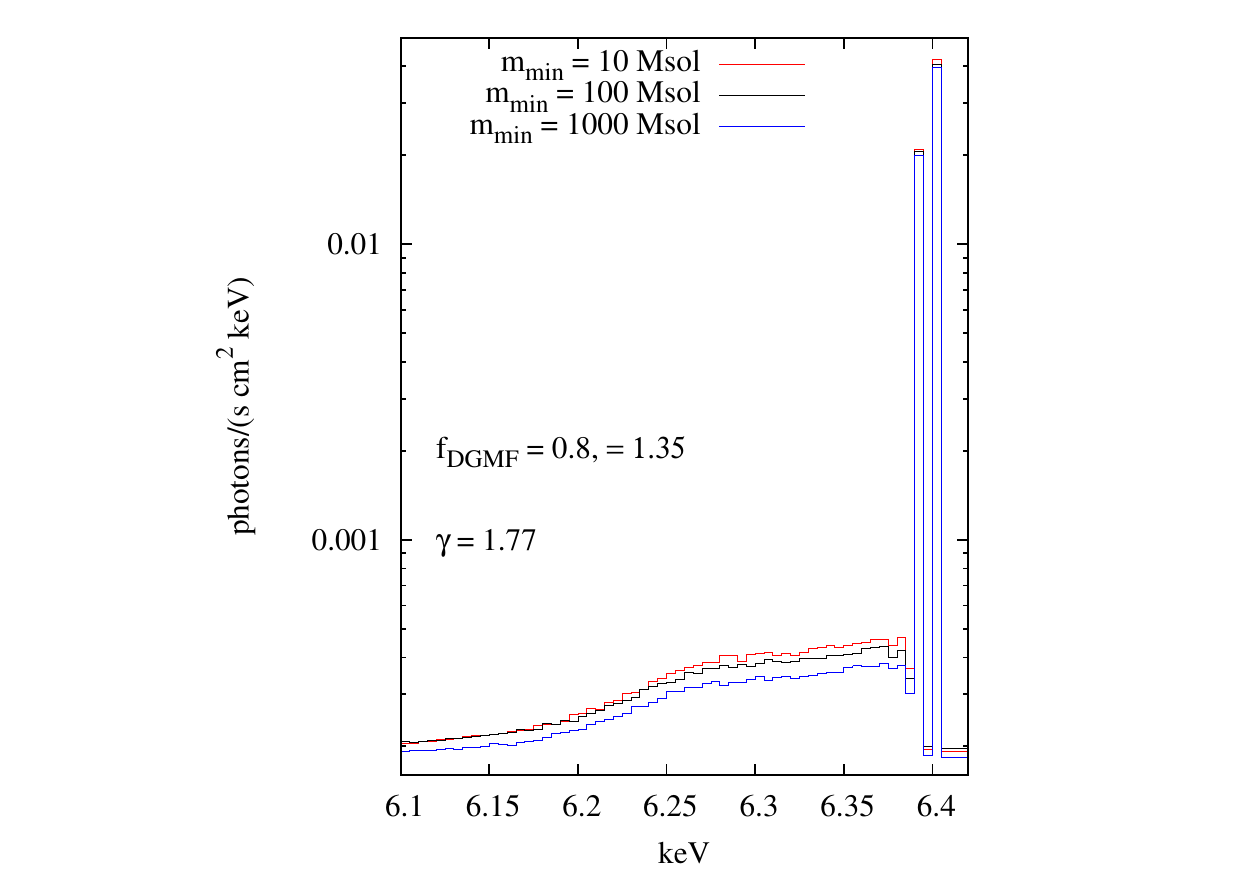} 
\includegraphics[trim = 10 0 70 0,height = 6cm, width = 
6cm]{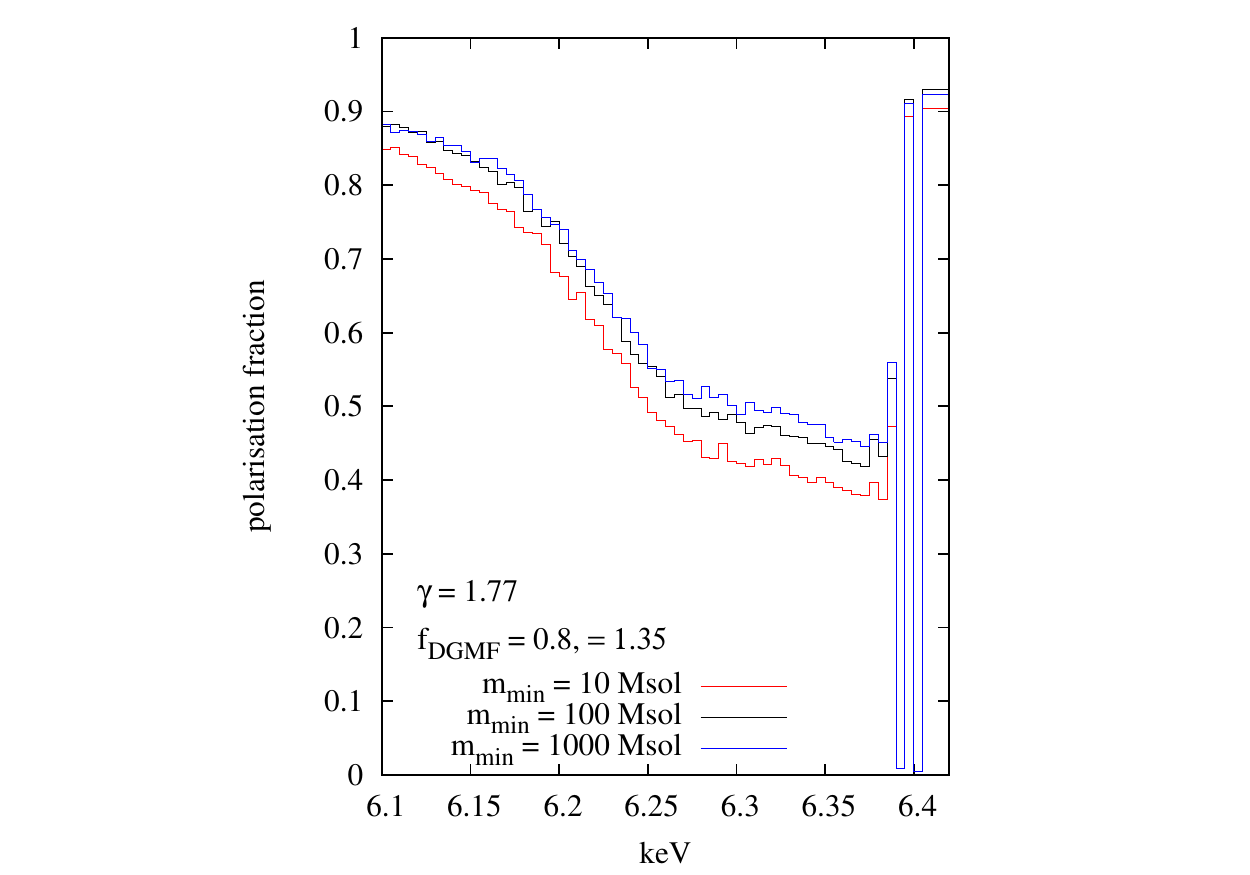} 
\includegraphics[trim = 10 0 14 0,height = 6cm, width = 
6cm]{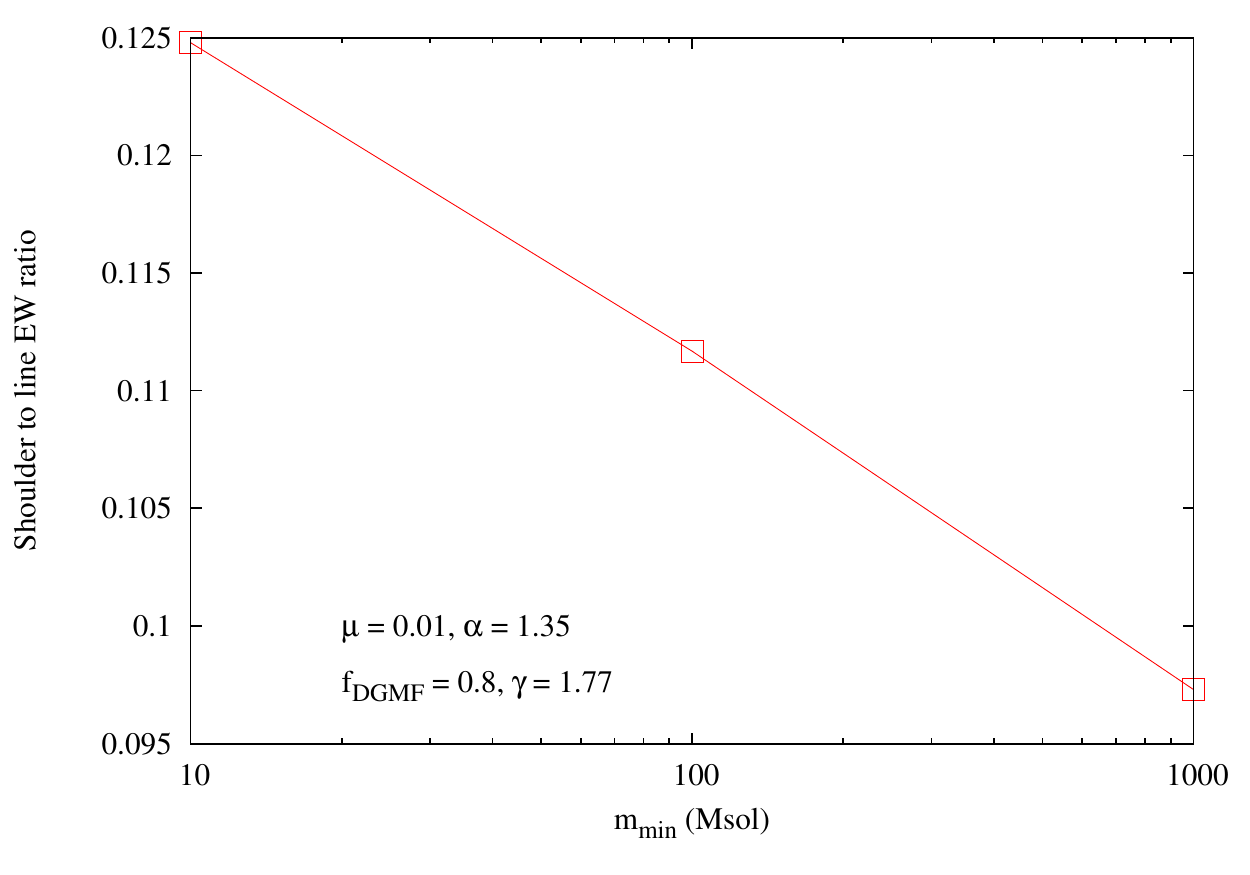} 
\caption{\changeRi{Reflected energy \textit{(left plots)} and polarisation 
\textit{(centre 
plots)} around the 6.4 
keV Fe K-$\alpha$ line (shown 
with resolution of 5 eV) and shoulder to line ratio \textit{(right 
plots)} for varying $\alpha$ \textit{(top plots)} and $m_{min}$ 
\textit{(bottom plots)} parameters at fixed $\mu = 0.01$. Values 
of $\mu$ illustrated were obtained by adjusting the 
$m_{norm}$ parameter in the mass-relation of clumps using Eqn. 
\ref{filling_factor}. For increasing \changeRiII{$\alpha$}, we 
expect an increase 
in the number of clumps being sampled from the higher mass 
range, resulting in a greater average size of the clumps in the population. In 
return, this results in an increase in the probability of X-rays intercepting 
them. Indeed, an increase in the Fe shoulder for increasing $\alpha$ is 
observed both in the energy and polarisation spectrum. The fragmentation of the 
clumps to lower and lower $m_{min}$, on the other hand, results in a larger 
projected area of the clump population, which increases the probability of 
interaction with incoming X-rays. This effect in also observed in the plots.
\label{compare_mu_0.01}}}
\end{figure*}

The picture painted in section \ref{res_sgrB2relation} of course only holds if 
the volume filling fraction of the clump population is low enough to 
effectively reduce the probability of interaction between photons and 
overdensities to a 
negligible value. Should the volume filling fraction increase, as would result 
from a variation in the mass-size relation of the clumps (see section 
\ref{masssize}), then X-rays should start intercepting the clumps at a more 
significant rate, with 
consequences to the reflected spectrum. \\
In particular, an increase in the absorption probability should result in an 
increase in the fluorescent lines, while an increase in the scattering 
probability should result in an increase in the fraction of fluorescent photons 
scattered, and therefore of the fluorescent line's shoulder's flux.
In Fig. \ref{compare_mu_zoom}, we show the energy and polarisation spectrum 
around the 6.4 keV K-$\alpha$ line for cases of increasing high volume filling 
factor. We find that, as expected, both the line and the shoulder's flux 
increase with increasing $\mu$. \newline
Once the probability of intercepting clumps increases, we expect properties of 
the clump populations such as $m_{min}$ and $\alpha$ to play a more significant 
role in shaping the reflected X-ray signal. In Fig. \ref{compare_mu_0.01}, 
we compare, for fixed $\mu = 0.01$, the reflected signal in the case of varying 
$\alpha$ and $m_{min}$ parameters respectively. Indeed, we observe that already 
for this volume filling fraction the slope shows a dependence on the two 
population parameters: a higher CMF slope will 
result in a higher number of clumps with larger masses (and hence radii) being 
selected. Because these are more likely to \changeRi{intercept the incident 
X-rays}, we expect 
an increase in the Fe shoulder's flux in correspondence with increasing 
$\alpha$, as it is indeed observed in the figure. 
\changeRi{A decrease in $m_{min}$ puts more mass in smaller clumps, resulting 
in a decrease in the fluorescent lines and shoulders in Fig. 
\ref{compare_mu_0.01}. However, a photon which is emitted inside a denser clump 
is also more likely to be scattered before it escapes, and therefore the ratio 
of the shoulder to line increases with decreasing $m_{min}$, as seen in the EW 
ratio plot in the same figure.}

\section{Time-evolution of the XRN morphology as a probe of the 3d 
distribution of substructures}
\label{timeevu}

\begin{figure}[ht!]
\includegraphics[trim = 100 20 100 60, clip,height = 
6cm]{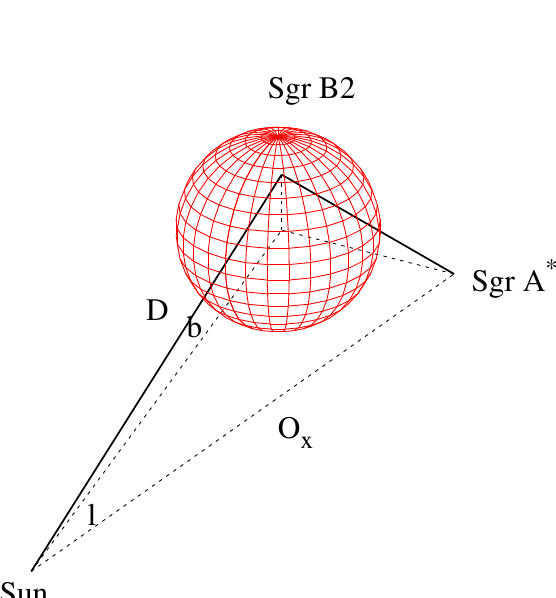} \\
\caption{
\changeRiII{Schematic representation (not to scale) of parameters used in Eqn. \ref{Dfunc}}
\label{diagram}
}
\end{figure}

\begin{figure}[ht!]
\includegraphics[trim = 40 0 50 0, clip, height = 8cm]{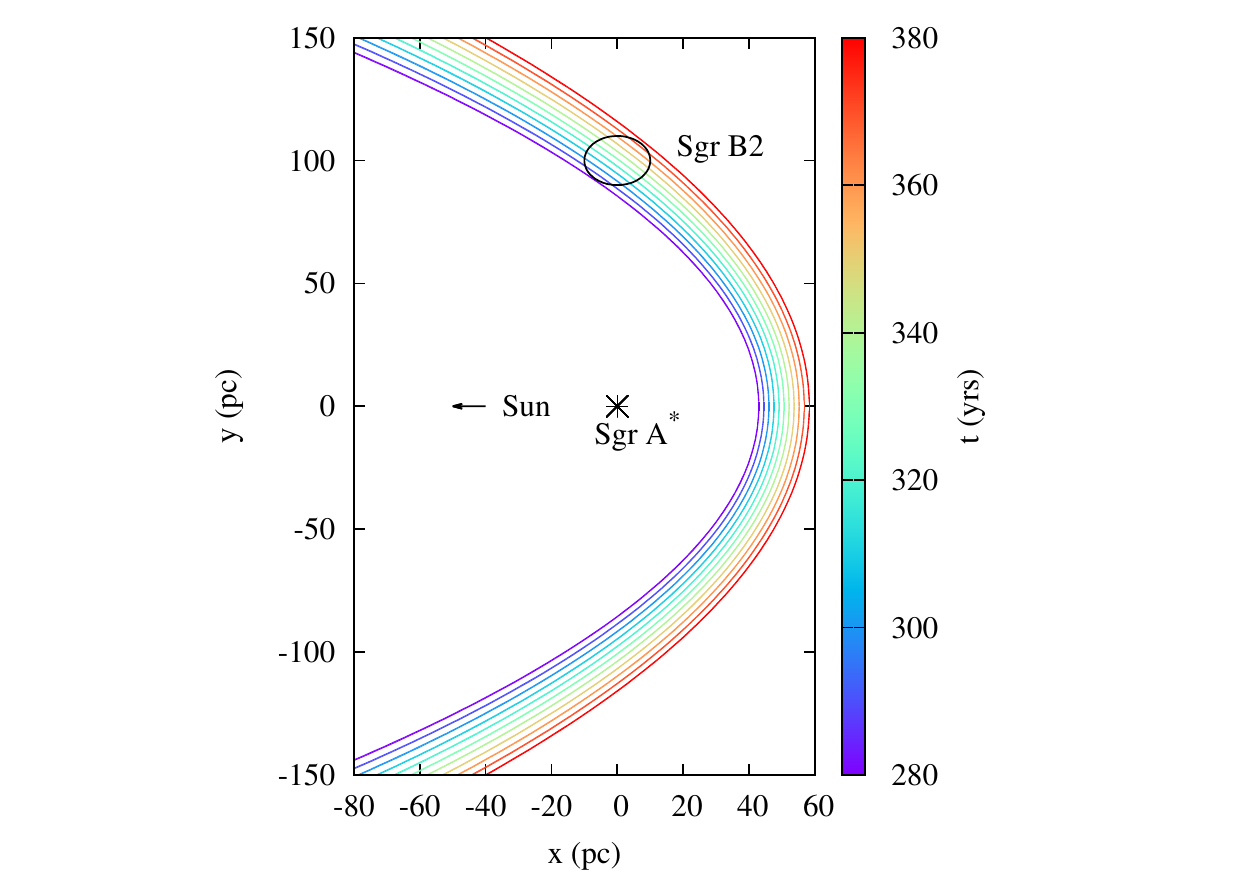} \\
\includegraphics[trim = 40 0 50 0, clip, height = 8cm]{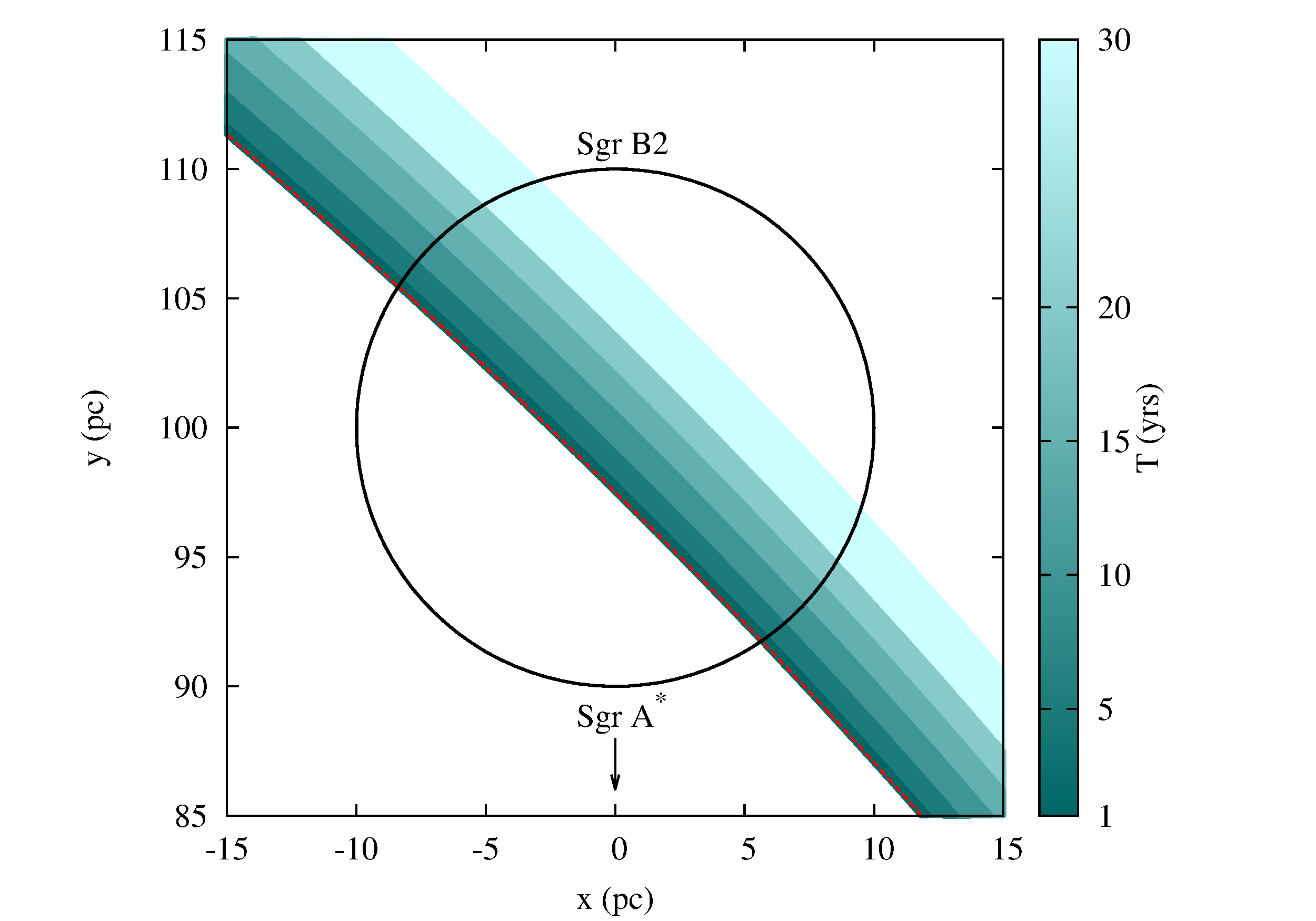} 
\caption{\textit{Top plot:} Regions of the x-y plane visible to the observer at 
different times
through scattered X-ray photons, in the case of photons originally emitted by 
an instantaneous flare of Sgr A$^{*}$ (i.e. all photons emitted at $t=0$). 
\textit{Bottom plot:} Regions of the x-y plane visible to the observer through 
scattered X-rays at 
time $t = 320$ yrs in the case of an instantaneous flare (dashed line) and in 
the case of flares of duration $T$ (coloured maps). The longer the duration of 
the flare, the 
thicker the region of the sky observable simultaneously. 
\label{propagation_wavefront}
}
\end{figure}

\begin{figure*}[ht!]
\includegraphics[trim = 10 0 30 0, clip,  height = 7cm]{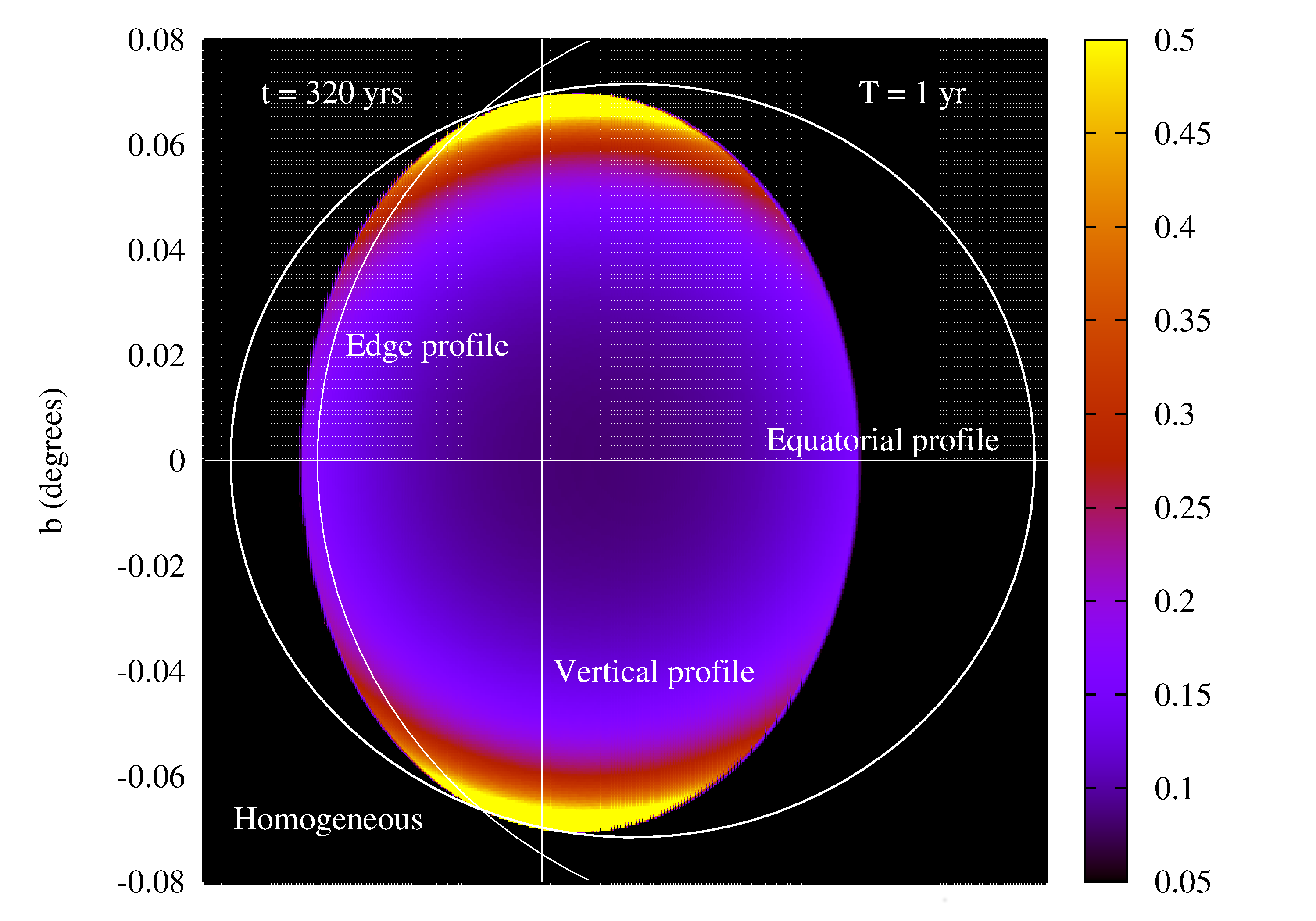} 
\includegraphics[trim = 20 0 30 0, clip,  height = 7cm]{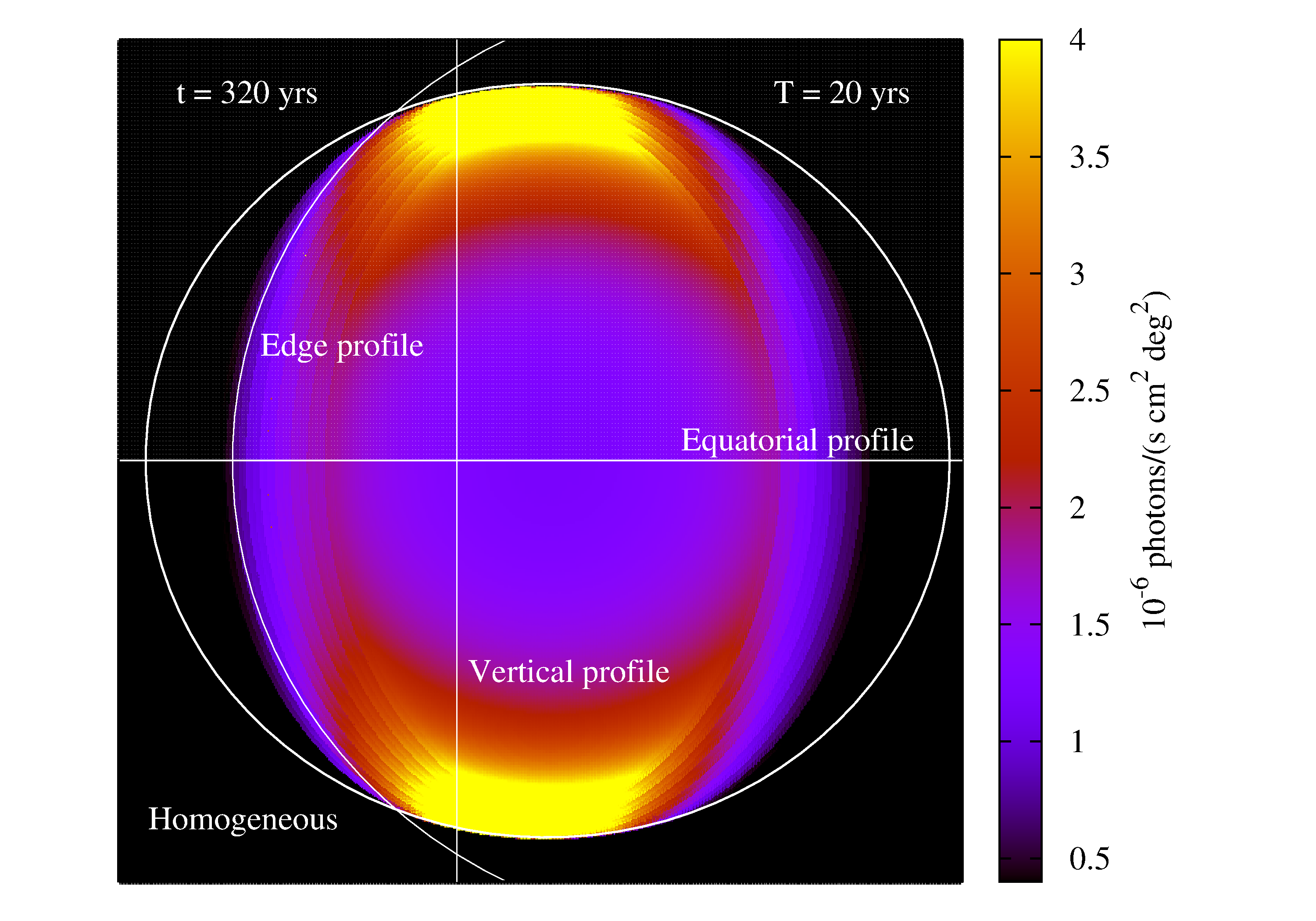} \\
\includegraphics[trim = 10 0 30 0, clip,  height = 7cm]{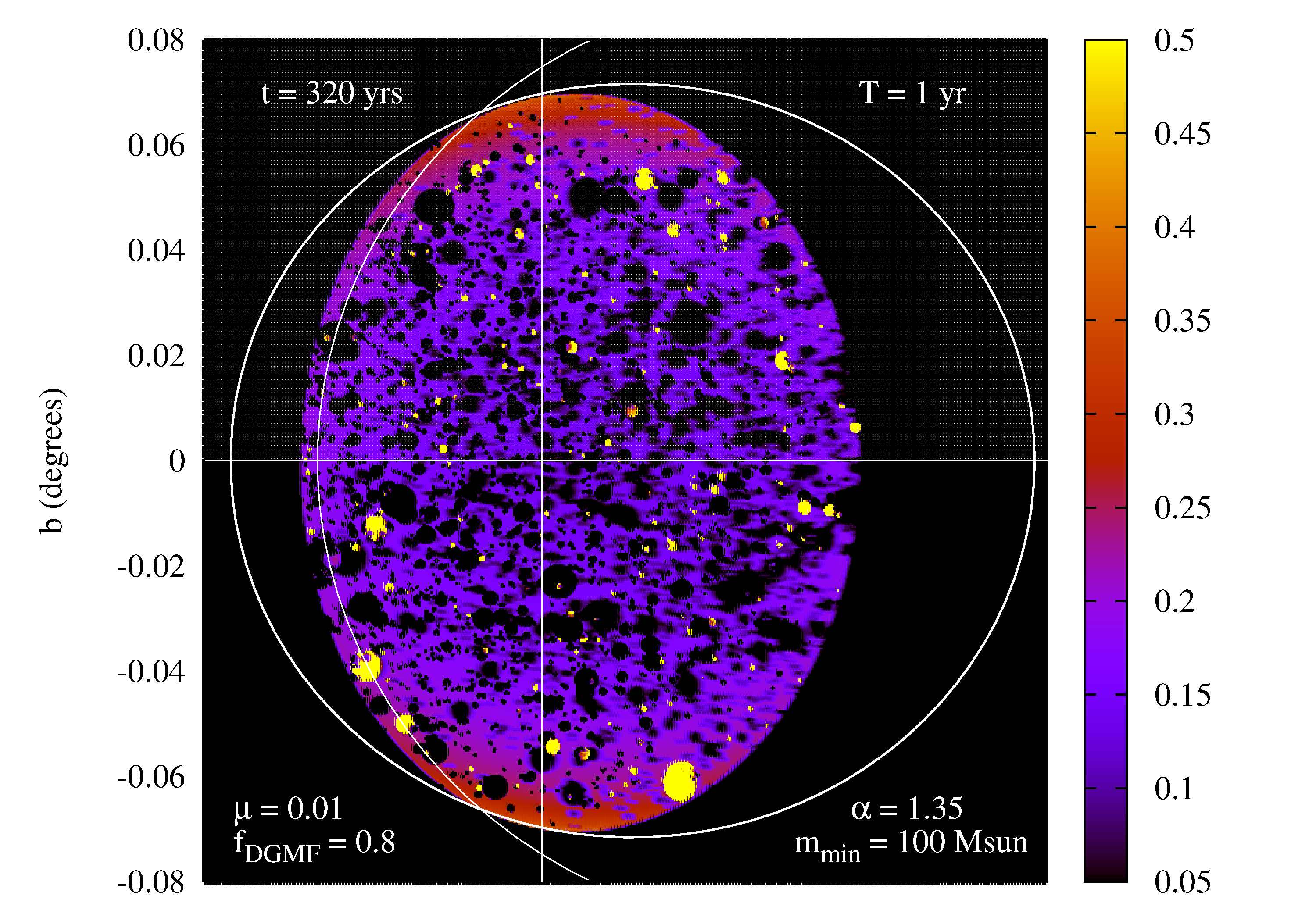} 
\includegraphics[trim = 20 0 30 0, clip,  height = 7cm]{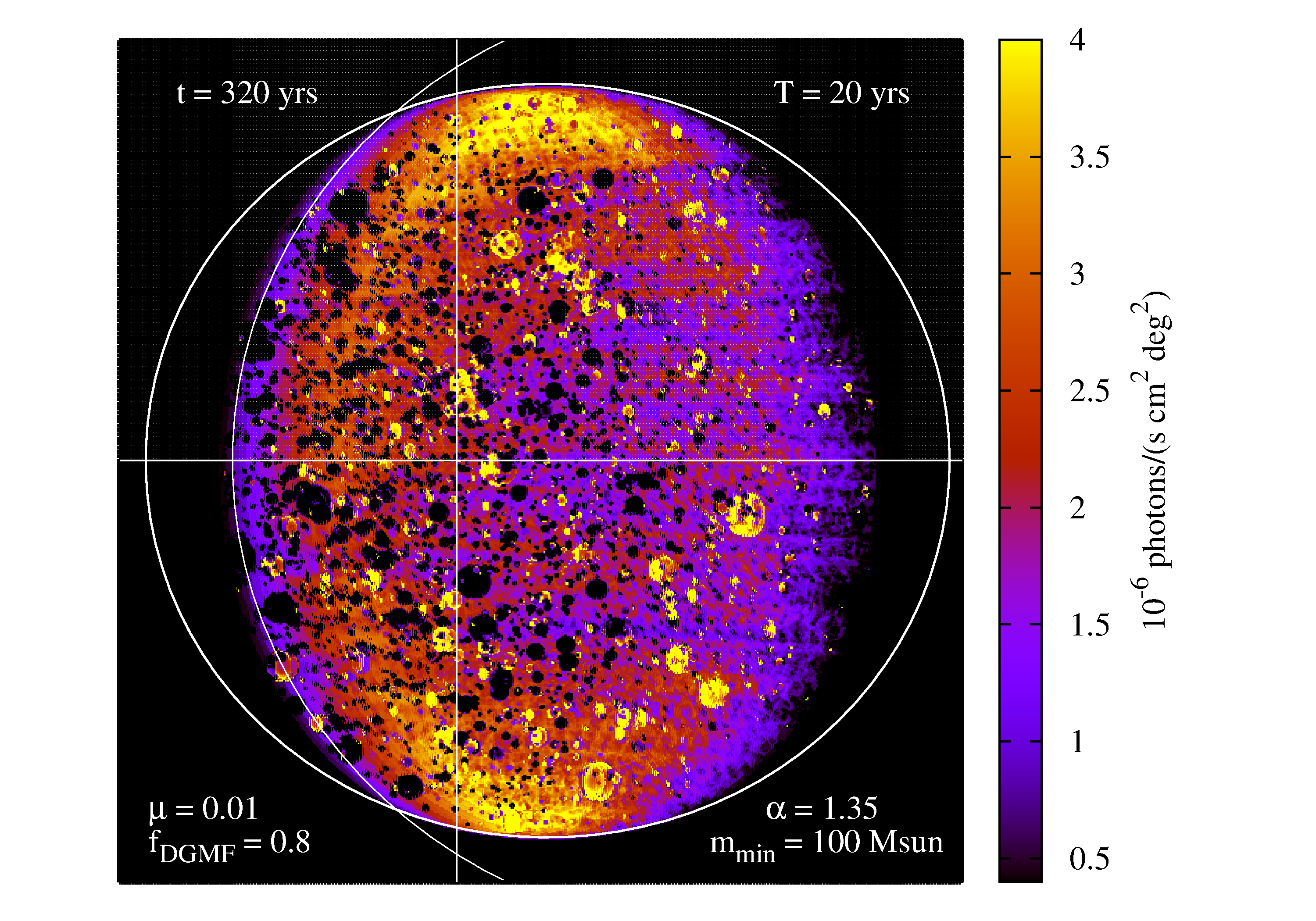} \\
\includegraphics[trim = 20 0 40 0, clip,  height = 7.5cm]{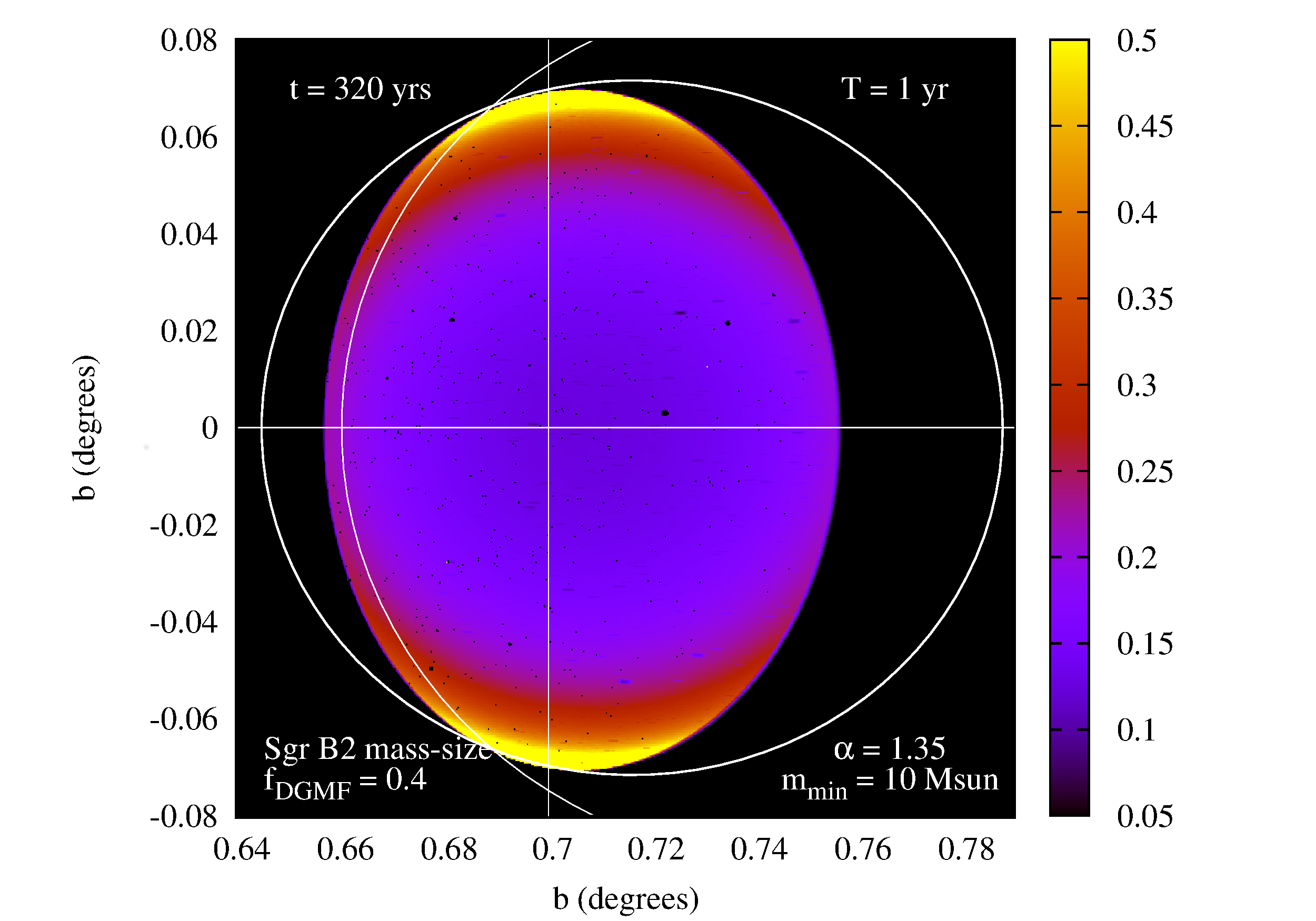} 
\includegraphics[trim = 20 0 40 0, clip,  height = 7.5cm]{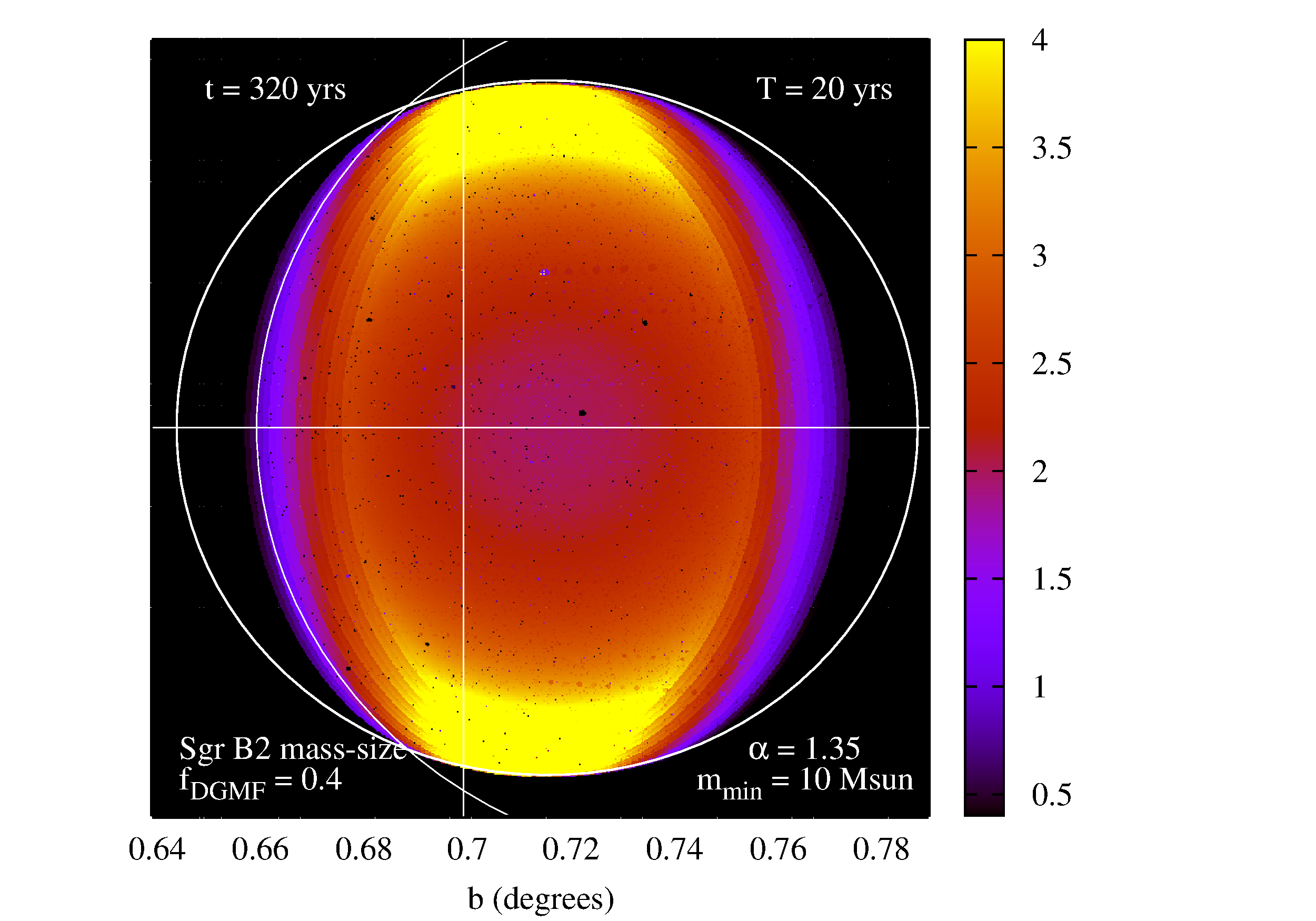} 
\caption{Analytic\add{, single scattering} approximation of the 3-20keV 
reflected X-ray intensity 
observed at time $t = 320$ yrs for the three cloud models discussed in the 
text, 
and for the case of a short (1 yr) and long (20 yr) flare. The angular 
resolution used is of 1 arcsec. 
The intensity and column density profiles for the equatorial, vertical 
and edge profiles indicated in the maps are shown 
in Figs. \ref{intensity_prof} and \ref{profiles_NHI} respectively.
\label{analytic_calculations}}
\end{figure*}

\begin{figure}[ht!]
\includegraphics[trim = 0 0 0 0, clip, width = 9cm]{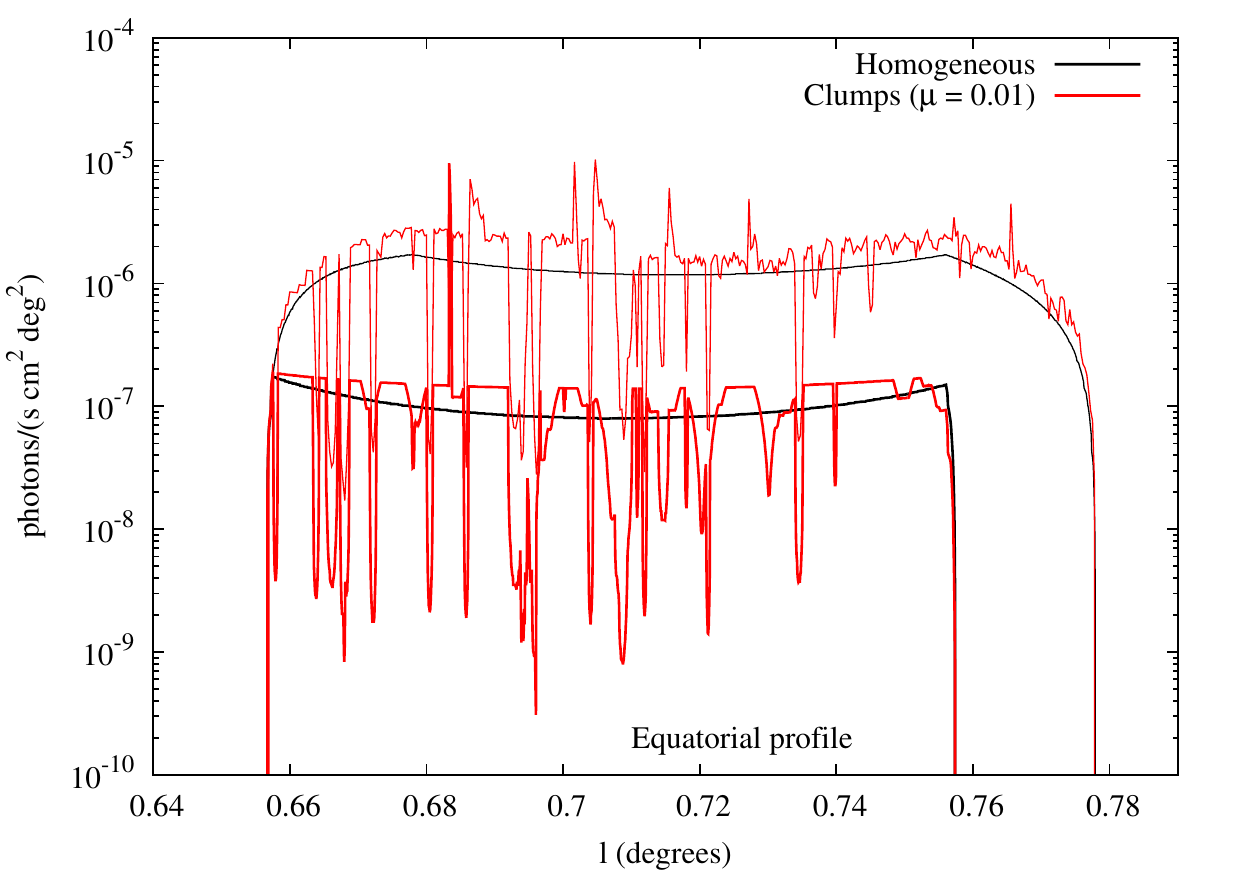} 
\includegraphics[trim = 0 0 0 0, clip, width = 9cm]{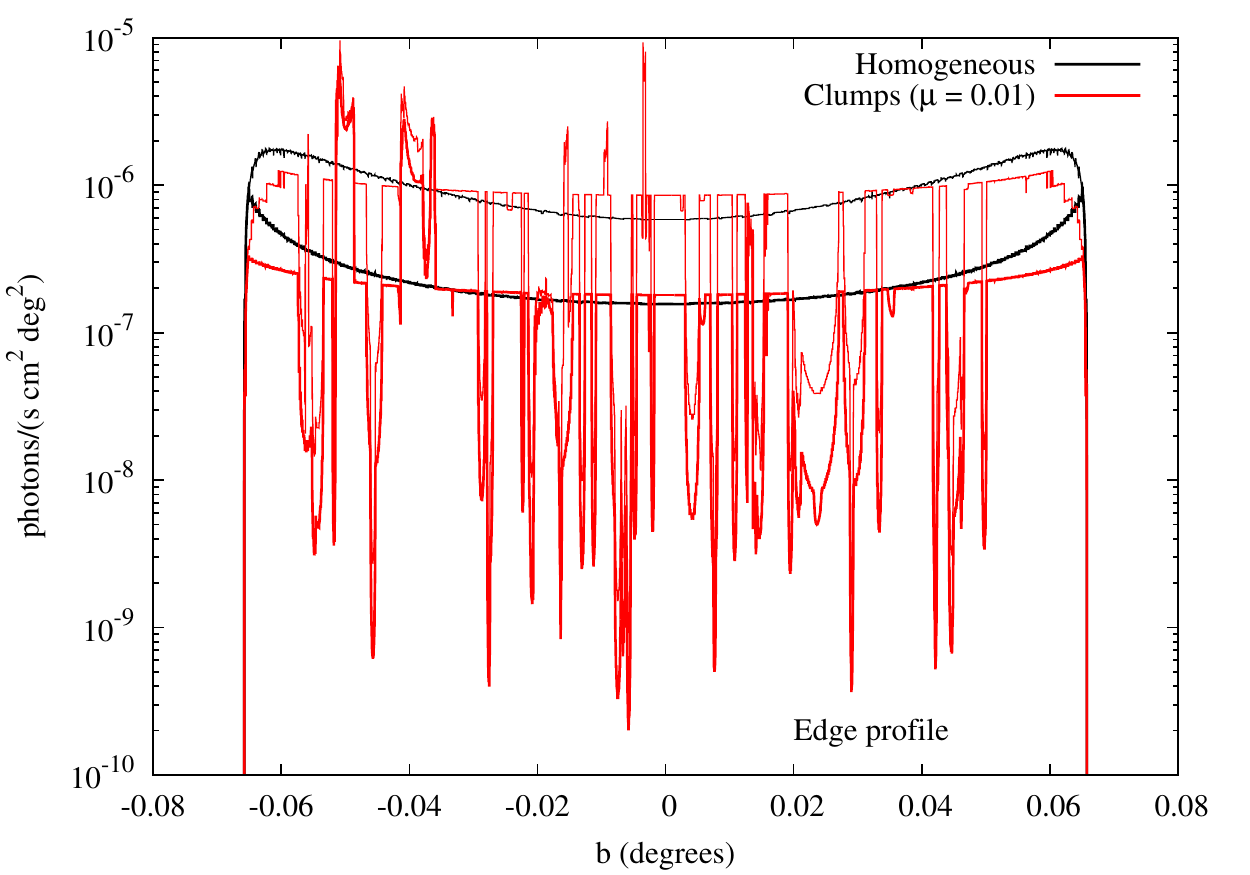} 
\includegraphics[trim = 0 0 0 0, clip, width = 9cm]{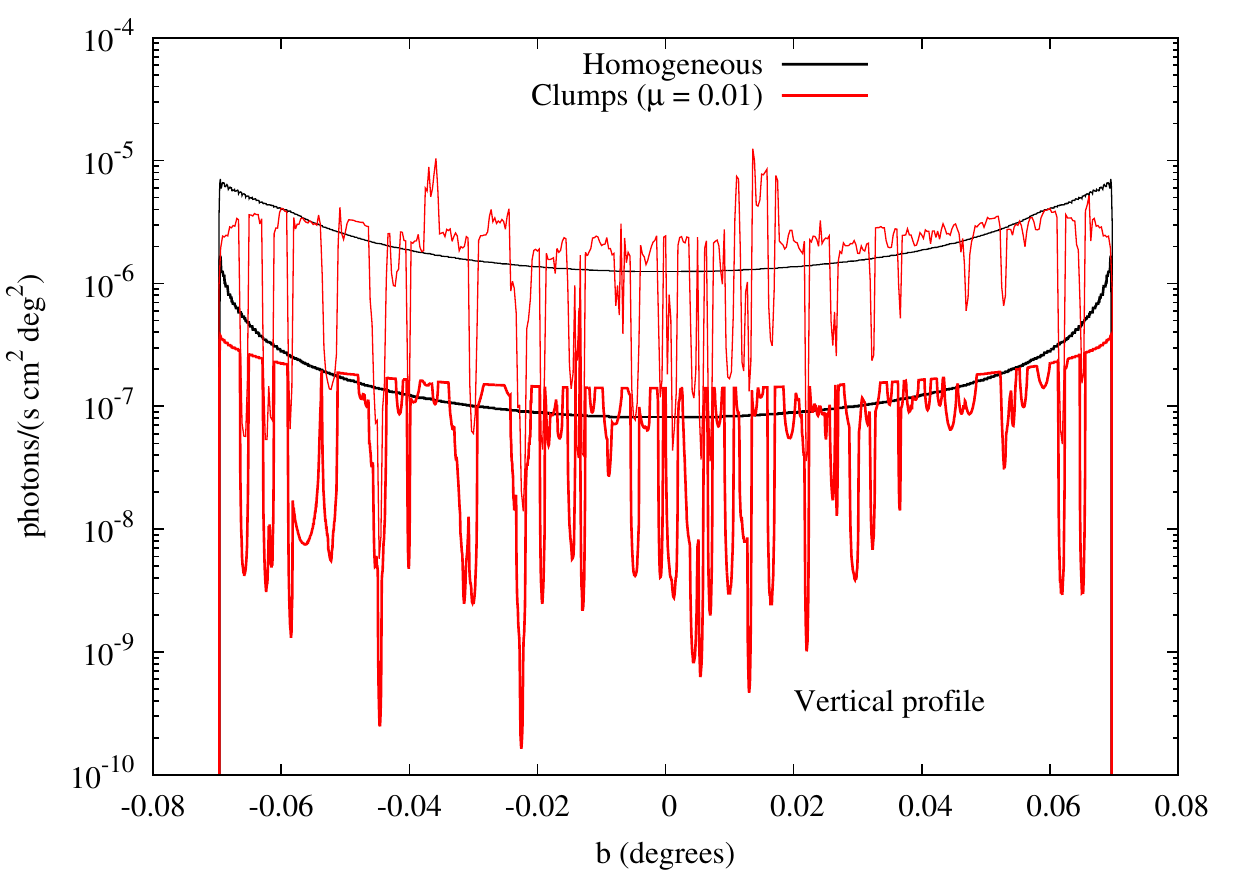} 
\caption{Scattered intensity profiles for the homogeneous and $\mu = 0.01$ 
model for the $T = 1$ yr (thick lines) and $T = 20$ yrs (thin lines).   
\label{intensity_prof}}
\end{figure}

\begin{figure}[ht!]
\includegraphics[trim = 0 0 0 0, clip, width = 9cm]{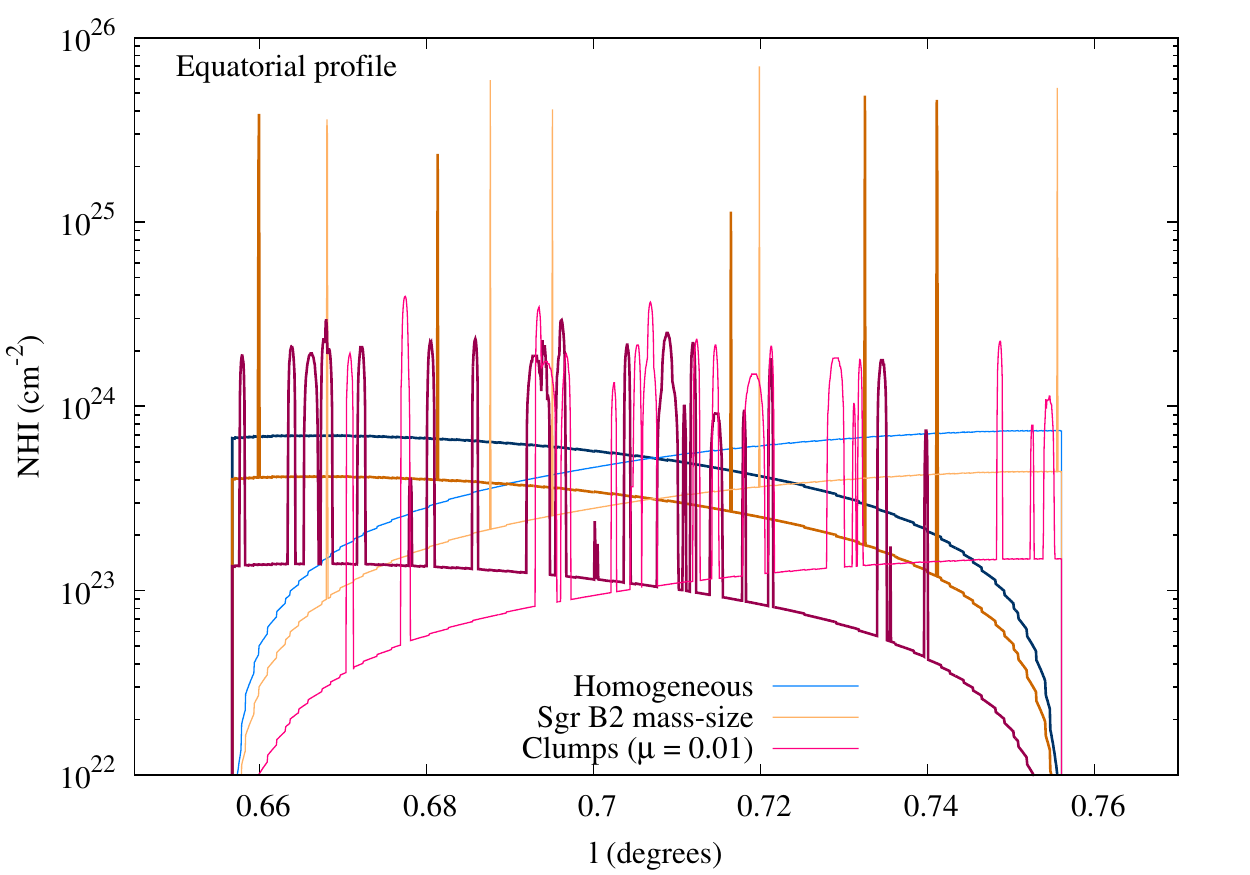} 
\includegraphics[trim = 0 0 0 0, clip, width = 9cm]{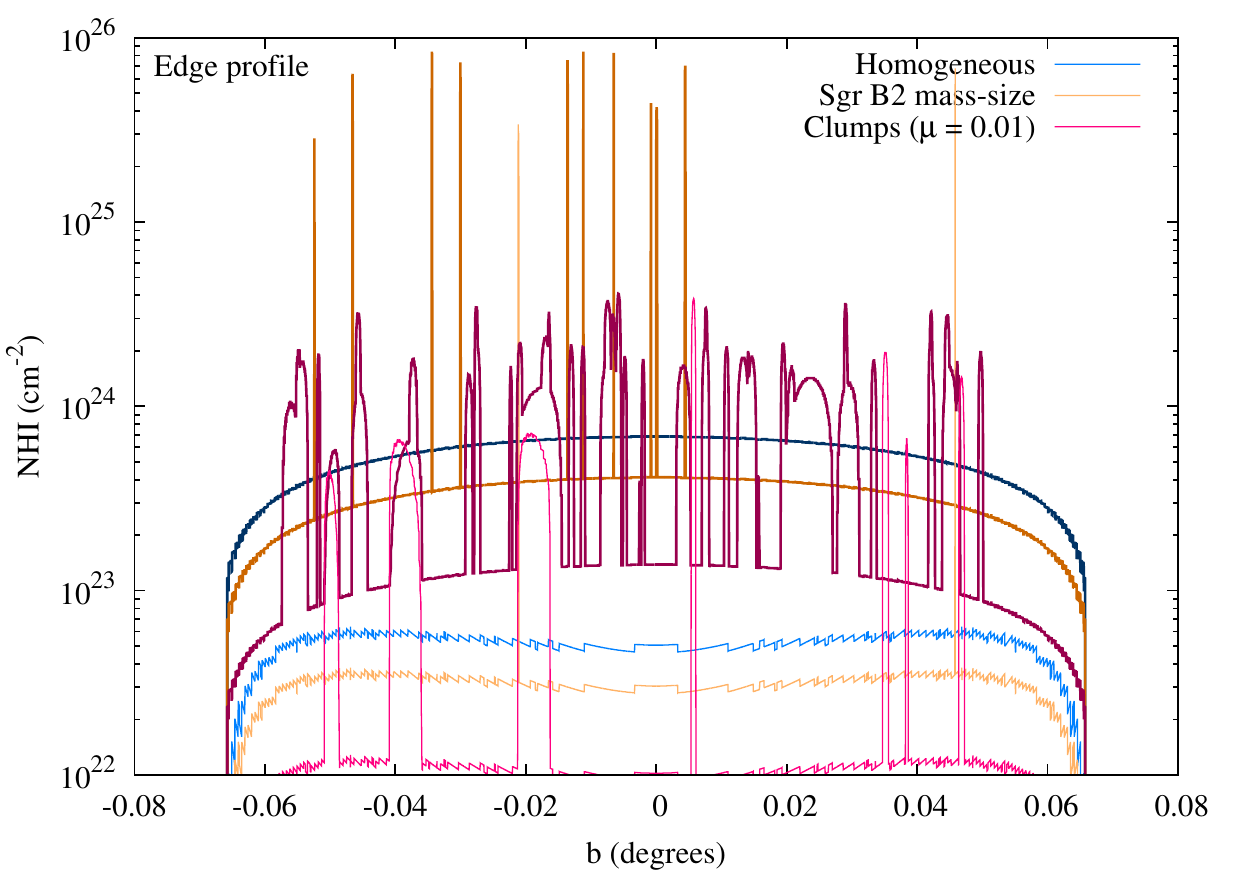} 
\includegraphics[trim = 0 0 0 0, clip, width = 9cm]{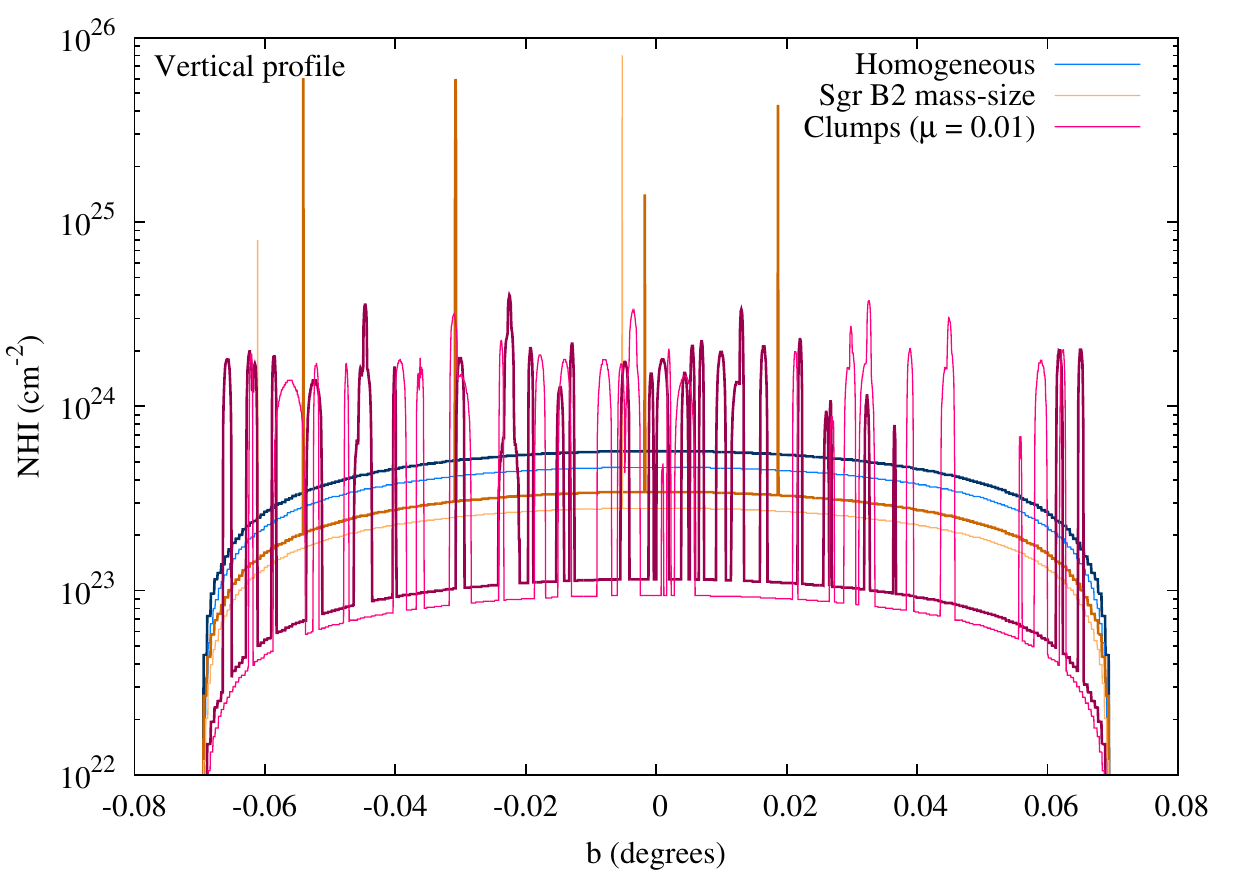} 
\caption{Column density from the source to the point of scattering (darker 
colours) and from the point of scattering to the observer (lighter colours) 
along profiles indicated in Fig. \ref{analytic_calculations}, excluding the 
contribution of the diffuse envelope. The plots show the case of minimum column 
density ($t' = 0$) at time $t = 320$ yrs. The different mass-size relations 
assumed in clump models "Sgr B2 mass-size" and "$\mu = 0.01$" result in the 
latter case presenting: a higher probability of intercepting clumps, broader 
absorption features in the intensity profiles, and lower column density peaks.  
\label{profiles_NHI}}
\end{figure}

\begin{figure}[ht!]
\includegraphics[height = 6cm]{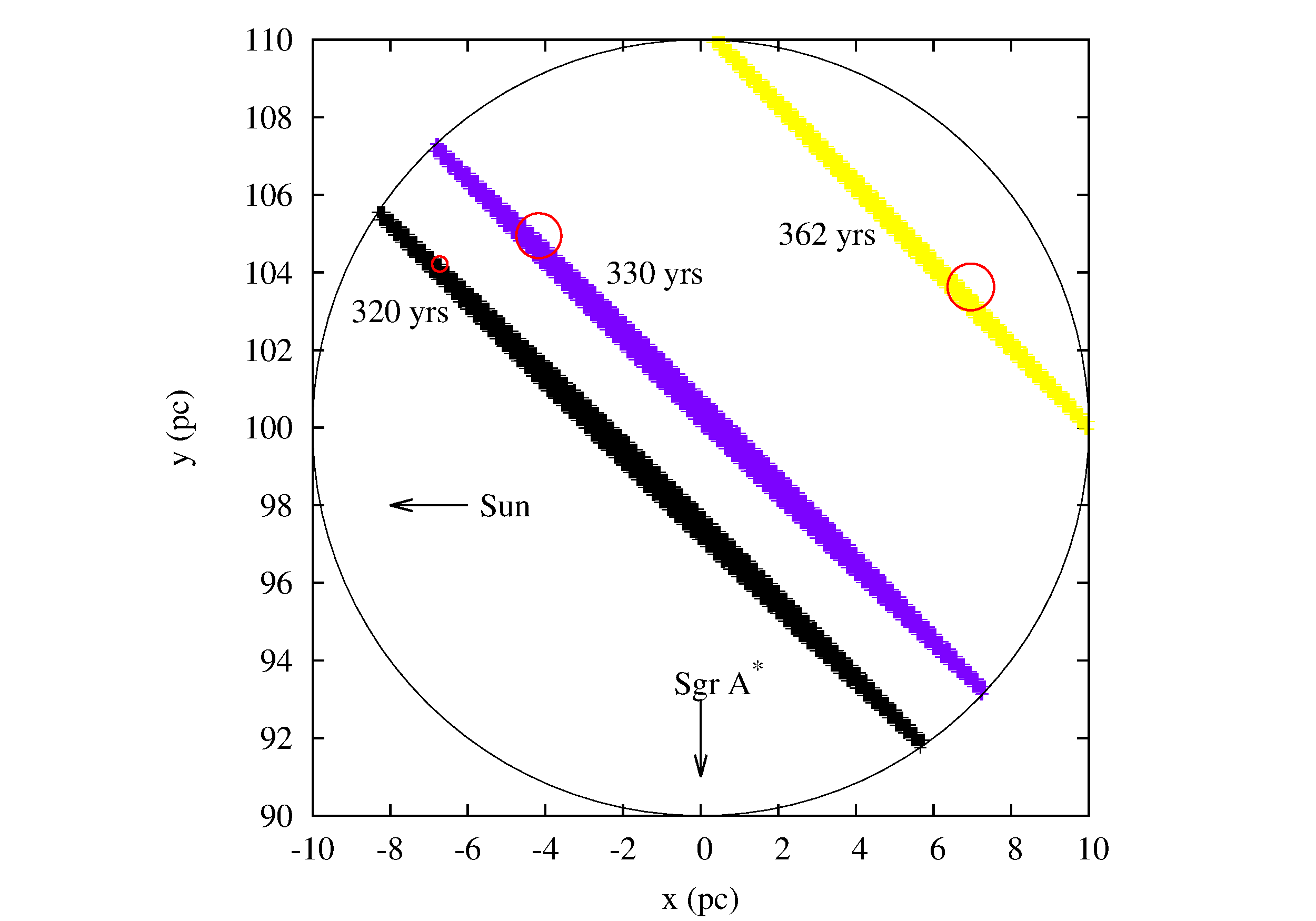} \\
\includegraphics[trim = 0 70 0 50, clip, width = 9cm]{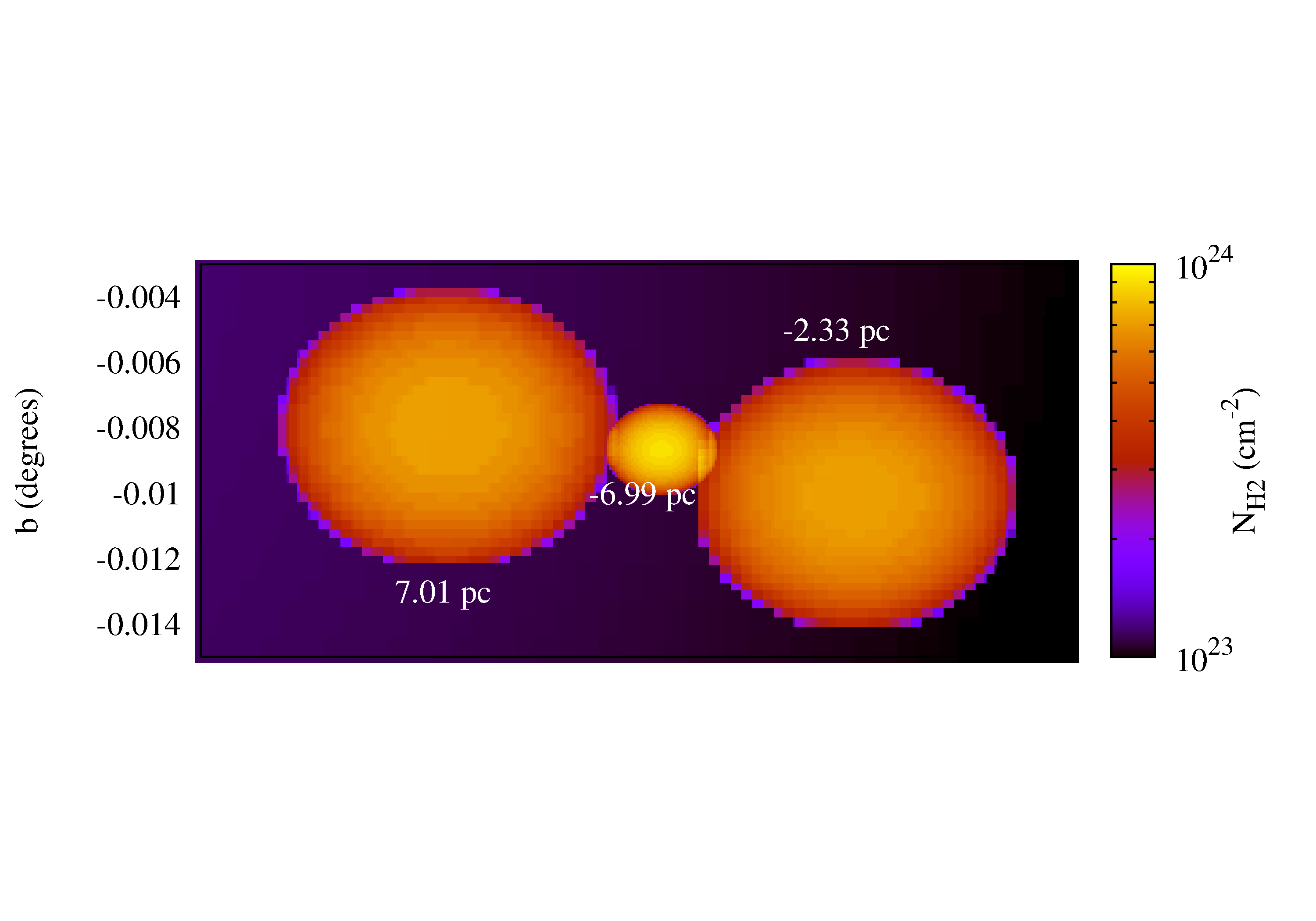} \\
\includegraphics[trim = 0 70 0 50, clip,  width = 9cm]{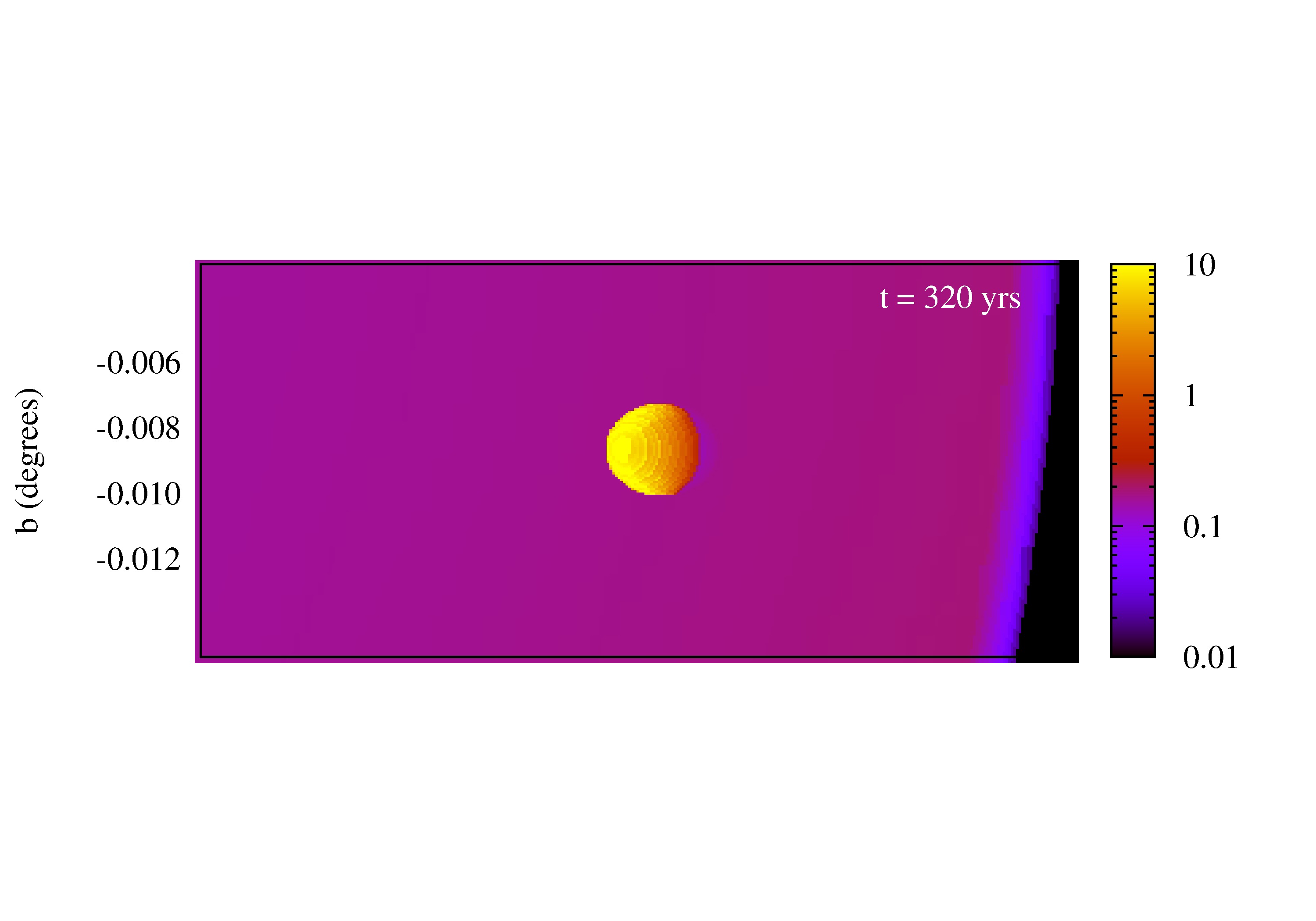}\\
\includegraphics[trim = 0 70 0 50, clip,  width = 9cm]{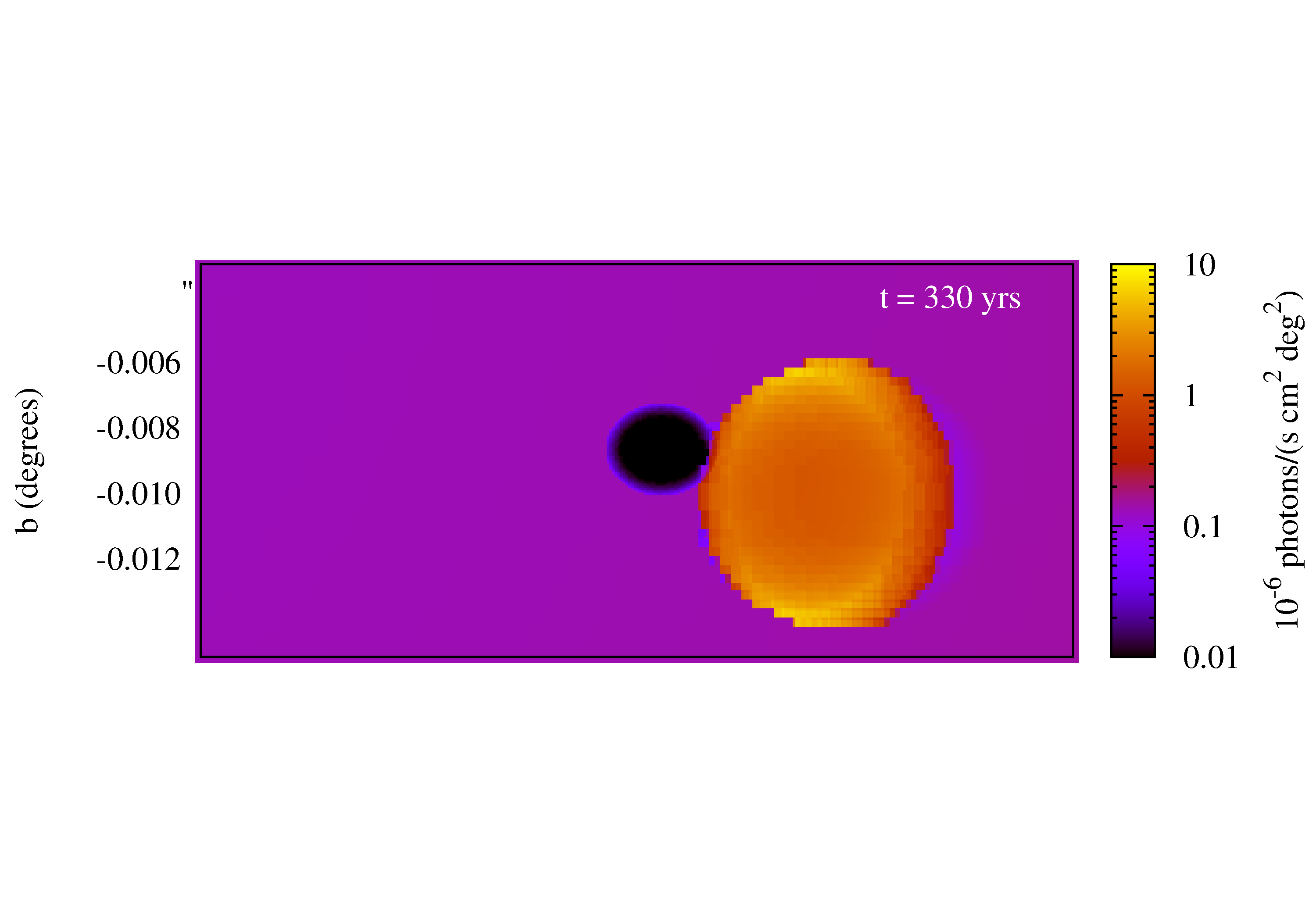}\\
\includegraphics[trim = 0 50 0 40, clip,  width = 9cm]{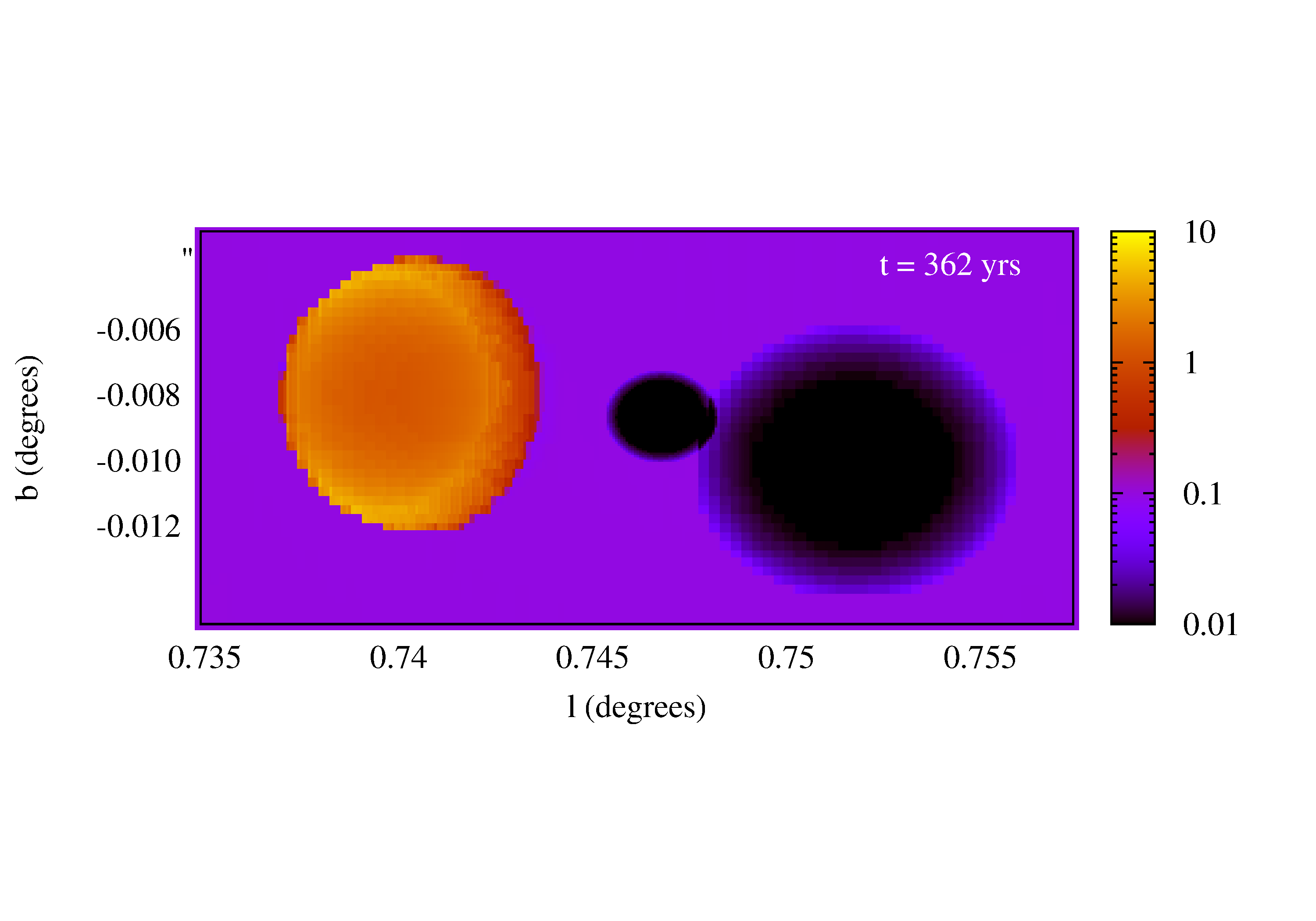}
\caption{
\textit{Top plot}: Column density map of three simulated clumps within Sgr B2, 
found in a 
small region of the sky (labels in plot indicate the clumps' distance from the 
center of the cloud, approximated as $D - D_{\text{SgrB2}}$). \textit{Second 
plot}: projected location of the 
clumps inside Sgr B2 and regions inside Sgr B2 visible to the observer at times 
when the substructures should become visible (assuming illumination by an 
instantaneous flare). \textit{Third, fourth and fifth plot}: reflected 
X-ray intensity reaching the observer at different times. Despite the apparent 
proximity of the clumps on the sky, because of the different distance at which 
they are located along the line of sight, the clumps are visible through 
their reflected X-ray emission at 
different times, so that when one is visible, the others are not.
\label{clumps_illumination}
}
\end{figure}

\begin{figure}[ht!]
\includegraphics[trim = 40 0 50 0, height = 
6cm]{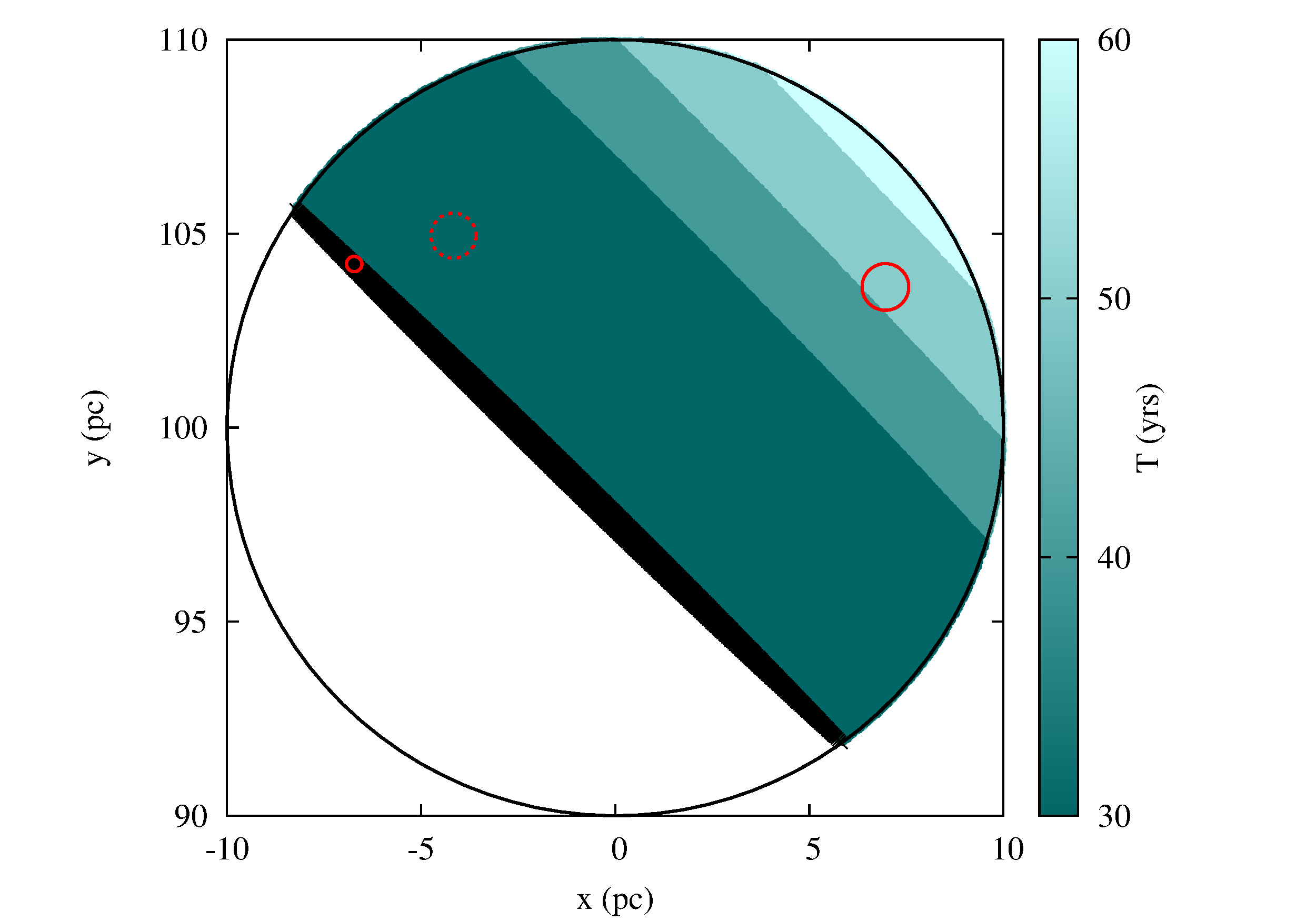} \\
\caption{
Projected regions inside the cloud observed simultaneously at time 320 yrs by 
flares of different durations $T$ (where the black region corresponds to 
illumination by an instantaneous flare). In the case of the three clumps 
considered in Fig. \ref{clumps_illumination}, the minimum flare duration 
required for the 
two 
most further apart clumps (shown with solid red lines in the plot) to both be 
visible to 
the observer at the same time is $T \sim 40$ yrs. In the case in which both 
substructures are visible simultaneously in the reflected X-ray intensity maps, 
one can infer that the source's flare must have lasted at least $T \sim 40$ yrs.
\label{clumps_illumination_longflare}
}
\end{figure}

Due to the finite speed of light, illumination by a flare of duration shorter 
than the light-crossing time of the cloud  
results in different regions of the GMC being visible 
to the observer, in the form of reflected X-ray emission, at different times. 
The evolution of the reflected X-ray intensity, therefore, acts as a scan of 
the density structure of the cloud as 
the wavefront propagates through it \citep{Sunyaev1998}. In this section we 
discuss the importance of this effect in the context of the study of the GMC's 
clumps properties and distribution.

For these calculations we focus on three of the Sgr B2 models considered in the 
previous sections: 
\begin{itemize}
\item the 
fiducial model, which assumes parameters $f_{\text{DGMF}} = 0.4$, $m_{min} = 10 
M_{\sun}$, $\alpha = 1.35$ and the Sgr B2 mass-size relation consistent 
with observations of real Sgr B2 clumps (see section \ref{masssize}). We refer 
to this model as the "Sgr B2 mass-size" model;
\item the homogeneous model, where we assume no clumps at all;
\item a more ``visible'' clump population model, which considers a case in 
which most of the gas ($f_{\text{DGMF}} = 0.8$) is contained in relatively 
massive ($m_{min} = 100 
M_{\sun}$) clumps, which are described by a mass-size relation constrained to 
obtain a volume filling fraction as large as $\mu = 0.01$ (see section 
\ref{masssize}). This case ensures clumps will be numerous and voluminous 
enough to be easily recognisable in our calculations. We refer to this model as 
the "$\mu = 0.01$" model; 
\end{itemize}
While these calculations were being performed, NuSTAR was 
able to resolve the Sgr B2 clumps Sgr B2(N) and Sgr B2(M) in 
X-rays for the first time
\citep{Zhang2015}. This new result shows the feasibility and potential that 
high-resolution studies of the X-ray 
morphology of GMCs in the CMZ have in the study of the internal structure of 
these XRNe. We stress, however, that the mass-size relation assumed in our Sgr 
B2 mass-size model makes use of the \cite{Qin2011} observations of the clumps, 
which were able to resolve the Sgr B2(N) and Sgr B2(M) clumps into distinct and 
independent substructures. The clumps for this model obtained in our simulation 
will therefore be more compact than the region of gas considered by the 
\cite{Zhang2015} observations.

In a single scattering approximation, the distance $D$ along the line of sight 
$(l,b)$ at which light has to be scattered in order to reach the observer at 
time $t$, defined such that $t = 0$ is the time at which the flare was last 
observed directly, is given by:
\begin{equation}  
     D(t,t') = \frac{ct^2-O_x^2 + 2 ct (|O_x|-t'c) + 
(|Ox|-t'c)^2}{2(ct+(|O_x|-t'c)+O_x \text{cos}(b)\text{cos}(l))}
\label{Dfunc}
\end{equation}
where $O_x = - 8$ kpc is the Sun's location with respect to the emitting 
source (assuming $O_y = O_z = 0 $), and $-T \leq t' \leq 0$ is the time during 
the flare of duration $T$ at 
which the photon was 
emitted, \changeRiII{as illustrated in Fig. \ref{diagram}}. The region illuminated at a given time is therefore an ellipsoid, with 
its focus at the observer's position. For the case of an observer located at 
the 
Sun's position, this can be approximated, in the proximity of Sgr B2, by 
a paraboloid 
\citep{Cramphorn2002}. The propagation of the section of the ellipsoid on the 
x-y plane is 
illustrated, 
for the case of an instantaneous flare ($T = 0$), in Fig. 
\ref{propagation_wavefront}.  

In the case of a flare with finite duration, that is $T > 0$, the duration of 
the flare determines the ``thickness'' of the ellipsoid, or in other words the 
thickness of the region simultaneously visible to the observer, as illustrated 
in Fig. 
\ref{propagation_wavefront}.
The surface brightness observed along a line of sight at a
given moment, $I(l,b,t)$, will therefore be determined not by the total optical 
depth of 
the
cloud in that direction, but rather by the surface density in the  
section of the cloud delimited by the thick paraboloid \citep{Sunyaev1998}, 
whose boundaries are determined by the beginning and end of the flare. 
\changeRi{The 
reflected intensity can therefore be described, under a single scattering 
approximation, as:
\begin{equation}
I(l,b,t,\nu) = \int_{-T}^{0} \sum_Z \frac{\rho(\nu)}{4 \pi R^2} n_Z 
\frac{\text{d} \sigma_Z}{\text{d}\Omega} \text{exp}(- \tau_Z) c \text{d}t'
\end{equation}
where $\rho(\nu)$ is the number of photons/(s keV)} emitted by the source, $R$ 
is 
the distance from the source to the point of scattering, $n_z$ is the density 
at the point of scattering, $\frac{\text{d} \sigma_Z}{\text{d}\Omega}$ is the 
singly-differentiated cross section, computed using the public library xraylib 
\citep{xraylib} and:
\begin{equation}
\tau_Z = \text{NHI}(\sigma_{abs,Z} + \sigma_{scatt,Z})
\end{equation}
is the total optical depth, from the point of emission to the point of 
observation. In  $\sigma_{scatt}$ we only consider the Raman and Compton 
scattering cross section, since Rayleigh scattering mainly contributes towards 
scattering at small angles, and therefore will have a negligible effect towards 
scattering photons out of the path traveled.

Due to 
their higher 
average density, clumps are able to contribute significantly towards $I$, by 
scattering more X-ray flux towards the 
observer 
compared to 
the interclump medium, and hence should be clearly recognisable in the 
morphology of 
the XRN at times when they are intercepted by the propagating paraboloid, as 
first 
suggested by \cite{Sunyaev1998}. 
Once the paraboloid has passed them, the clumps should significantly contribute 
towards the intervening column density NHI, and therefore still be visible in 
the  
morphology of the 
XRN as regions of absorption. 

In Fig. \ref{analytic_calculations}, we illustrate this in the case of the 
three 
clump models described at the beginning of the section, both in the case of a 
short ($T = 1$ yr) and longer ($T = 20$ yr) flare, in a snapshot at time $t = 
320$ yr. The intensity and column density  for the equatorial, vertical 
and edge profiles indicated on the maps are shown in Figs. 
\ref{intensity_prof} and \ref{profiles_NHI} respectively.

In the maps, the main visible effects are the following:
\begin{itemize}
\item the contribution of clumps towards the scattered intensity is 
indeed clearly visible in the form of bright spots, \changeRiII{consistently} 
with the 
findings of 
\cite{Zhang2015};
\item intervening clumps located between the source and the point of 
scattering, or between the point of scattering and the observer (see Fig. 
\ref{profiles_NHI}) considerably 
contribute to the absorption of the X-ray radiation, and are therefore 
observable as regions of absorption in the maps;
\item The effect of the duration of the flare is also recognisable: 
the longer the duration of the flare, the thicker the region of the cloud 
probed by the ellipsoid at the same time, hence the higher the number of clumps 
probed by the 
paraboloid simultaneously, as clearly seen in the $\mu = 0.01$ model maps.
\item In the case of clumps with Sgr B2 mss-size relation, 
the very small volume they occupy means the probability of a short flare 
intercepting them is very low, and in fact very few clumps are intercepted at 
all for 
this particular distribution, and most clumps are therefore only seen in 
absoprtion;
\item the intensity of the interclump regions in the Sgr B2 mass-size model 
(average density $n_{H2} \sim 1 \times 10^{3}$ cm$^{-3}$), is higher than that 
of the homogeneous model (average density $n_{H2} \sim 1 \times 10^{4}$ 
cm$^{-3}$) in the central region of the cloud. This is due to the fact that, 
although contributing a larger surface density within the thick paraboloid, the 
homogeneous cloud also results in a larger absorbing column density, as shown 
in Fig. \ref{profiles_NHI};
\end{itemize}

The reflected intensity at a given time can therefore reveal information on the 
column density of both the clump and the interclump medium. But it also 
contains information on the distribution of the clumps inside the cloud.

In the case of \changeRiII{a short flare}, the distance along 
the 
line of sight observed at a given time is uniquely defined by the time of 
observation, since $D(t,t') \sim D(t,t'=0)$. In this case, by comparing the 
time at which a clump becomes visible on the reflected 
intensity map with its location on the sky, it is possible to constrain its 
position $D$ along the line of sight, as illustrated in Fig. 
\ref{clumps_illumination}.
 This kind of analysis could 
therefore prove to be important in the study of the 3d distribution of 
substructures within 
GMCs. \\

For flares of 
finite duration, on the other hand, there will be a range of distances 
observable at the same time along a given line of sight, as photons are 
emitted by the source over a period of time $T$, which determines the 
``thickness'' of the 
region observed at a given time. In the case of known duration of flares, as is 
the case for many X-ray sources in the Galaxy, the time-evolution of the 
reflected intensity could still be used in this kind of analysis.
In the case 
of illumination by a flare of unknown duration, on the other hand, it will be 
impossible 
to constrain the distance $D$ of each clump along its line of sight, since the  
range of possible values will be proportional to the duration of the flare 
itself. On the other hand, if the position of at least two clumps is known, it 
would
be possible to reverse the problem and infer a lower-bound to the duration of 
the flare, as illustrated in Fig. \ref{clumps_illumination_longflare}.

The visibility of clumps in the reflected X-ray intensity, their localised 
nature, and the non-persistent illumination from external flaring sources such 
as Sgr A$^{*}$, 
make the time-evolution of the X-ray morphology of Sgr B2 and similar XRNe an 
ideal 
target in the study of the spatial distribution of clumps within them. 
\changeRiII{We leave the study of the intensity light curve of individual 
clumps, as a function of photon energy, for different clump sizes and optical 
depths, for future work.}

\section{Conclusions}
We studied the effect of clumps on the 
X-ray 
emission of GMCs that act as XRNe by modeling Sgr B2, one of the brightest 
and 
most massive XRNe in our Galaxy. \\
\changeRi{We studied the effect of the internal structure of GMCs on the 
properties of X-ray spectrum, polarisation and morphology reflected from them. 
We have considered both persistent sources and transients, in particular giant 
flares, as the source of incident X-rays. We use Sgr B2 as a case study, but 
most of our results are generally applicable to any GMC in the Galaxy.
We defined a simple clump model for simplicity.} We investigated the effect of 
different clump 
population model parameters on the reflected X-ray energy and polarisation 
spectrum. The parameters investigated included the fraction of the total mass 
of 
the cloud 
contained in clumps ($f_{\text{DGMF}}$), the slope of the clump mass 
function ($\alpha$), the minimum mass of clumps found in the population 
($m_{min}$) and the mass-size relation of individual clumps ($m = 
m_{norm}(r/pc)^{\gamma}$). 
We first considered a fixed mass-size relation 
consistent with the clumps observed in Sgr B2, and varied each of the remaining 
parameters around a fiducial model given by $_{DGMF} 
= 0.4$, $m_{min} = 10 M_{\sun}$ and $\alpha = 1.35$, assessing their effect 
on 
the overall reflected X-ray spectrum. \\
In this case, the volume filling fraction of the clumps, and therefore 
the 
relative probability of X-rays being scattered by gas in clumps 
\changeRiII{compared to} 
the interclump medium, \changeRiII{is} negligible. The cloud therefore appears 
in X-rays as having a mass smaller than the total mass by the amount 
\changeRiII{that} is clumped. The extremely low volume filling 
fraction obtained when assuming the mass-size relation 
observed in Sgr B2 allows these clumps to effectively "hide" a fraction 
$f_{\text{DGMF}}$ of the cloud's
mass in an extremely small fraction of the cloud's volume. \changeRi{We 
explicitly check this hypothesis by considering the case of homogeneous clouds 
containing (1-$f_{\text{DGMF}}$) of the cloud's original mass and no clumps at 
all.}
In cases where the mass-size relation of clumps means these occupy a 
much higher volume filling fraction, we find that clumps do 
\changeRiII{contribute} towards reflection, and that the reflected X-rays 
contain information about the internal structure of the cloud. \changeRiII{The 
parameters of the clumpying model could therefore be constrained by X-ray 
observations}. 

We also investigated how the time evolution of the spatially-resolved 
\changeRiII{images} of the reflected X-ray intensity can be used to probe the 
location of individual substructures along the line of sight in the case where 
the incident X-rays have a transient origin, such as a short-duration flare from 
a X-ray binary or the supermassive black hole at the centre of our Galaxy. We 
have shown that in the case of transient sources, the timing information, 
retreivable both in emission and in absorption, can be used to probe the third 
dimension along the line of sight, opening up the possibility of 3d tomography 
of the cloud.
Future X-ray observatories such as Astro-H \citep{Takahashi2010}, Athena 
\citep{Barcons2012} and the X-ray Surveyor \citep{Weisskopf2015} could 
therefore open up a new probe of the internal structure of GMCs.

\begin{acknowledgements}
\changeRi{We would like to thank Eugene Churazov and Diederik Kruijssen for 
discussions.}
\end{acknowledgements}

\bibliography{references}

\begin{thebibliography}{54}
\expandafter\ifx\csname natexlab\endcsname\relax\def\natexlab#1{#1}\fi

\bibitem[{{Barcons} {et~al.}(2012){Barcons}, {Barret}, {Decourchelle}, {den
  Herder}, {Dotani}, {Fabian}, {Fraga-Encinas}, {Kunieda}, {Lumb}, {Matt},
  {Nandra}, {Piro}, {Rando}, {Sciortino}, {Smith}, {Str{\"u}der}, {Watson},
  {White}, \& {Willingale}}]{Barcons2012}
{Barcons}, X., {Barret}, D., {Decourchelle}, A., {et~al.} 2012, ArXiv e-prints

\bibitem[{{Battisti} \& {Heyer}(2014)}]{Battisti2014}
{Battisti}, A.~J. \& {Heyer}, M.~H. 2014, \apj, 780, 173

\bibitem[{{Capelli} {et~al.}(2012){Capelli}, {Warwick}, {Porquet}, {Gillessen},
  \& {Predehl}}]{sgr10}
{Capelli}, R., {Warwick}, R.~S., {Porquet}, D., {Gillessen}, S., \& {Predehl},
  P. 2012, \aap, 545, A35

\bibitem[{{Chen} \& {Ostriker}(2015)}]{Chen2015}
{Chen}, C.-Y. \& {Ostriker}, E.~C. 2015, ArXiv e-prints

\bibitem[{{Churazov} {et~al.}(2002){Churazov}, {Sunyaev}, \&
  {Sazonov}}]{Churazov2002}
{Churazov}, E., {Sunyaev}, R., \& {Sazonov}, S. 2002, \mnras, 330, 817

\bibitem[{{Clavel} {et~al.}(2013){Clavel}, {Terrier}, {Goldwurm}, {Morris},
  {Ponti}, {Soldi}, \& {Trap}}]{sgr13}
{Clavel}, M., {Terrier}, R., {Goldwurm}, A., {et~al.} 2013, ArXiv e-prints

\bibitem[{{Cramphorn} \& {Sunyaev}(2002)}]{Cramphorn2002}
{Cramphorn}, C.~K. \& {Sunyaev}, R.~A. 2002, \aap, 389, 252

\bibitem[{{Donkov} {et~al.}(2011){Donkov}, {Veltchev}, \&
  {Klessen}}]{Donkov2011}
{Donkov}, S., {Veltchev}, T.~V., \& {Klessen}, R.~S. 2011, \mnras, 418, 916

\bibitem[{{Draine}(2011)}]{Draine}
{Draine}, B.~T. 2011, {Physics of the Interstellar and Intergalactic Medium}
  (Princeton University Press)

\bibitem[{{Etxaluze} {et~al.}(2013){Etxaluze}, {Goicoechea}, {Cernicharo},
  {Polehampton}, {Noriega-Crespo}, {Molinari}, {Swinyard}, {Wu}, \&
  {Bally}}]{Etxaluze2013}
{Etxaluze}, M., {Goicoechea}, J.~R., {Cernicharo}, J., {et~al.} 2013, \aap,
  556, A137

\bibitem[{{Gando Ryu} {et~al.}(2012){Gando Ryu}, {Nobukawa}, {Nakashima},
  {Tsuru}, {Koyama}, \& {Uchiyama}}]{sgr12}
{Gando Ryu}, S., {Nobukawa}, M., {Nakashima}, S., {et~al.} 2012, ArXiv e-prints

\bibitem[{{Ginsburg} {et~al.}(2015){Ginsburg}, {Bally}, {Battersby},
  {Youngblood}, {Darling}, {Rosolowsky}, {Arce}, \& {Lebr{\'o}n
  Santos}}]{Ginsburg2015}
{Ginsburg}, A., {Bally}, J., {Battersby}, C., {et~al.} 2015, \aap, 573, A106

\bibitem[{{Gordon} {et~al.}(1993){Gordon}, {Berkermann}, {Mezger}, {Zylka},
  {Haslam}, {Kreysa}, {Sievers}, \& {Lemke}}]{Gordon1993}
{Gordon}, M.~A., {Berkermann}, U., {Mezger}, P.~G., {et~al.} 1993, \aap, 280,
  208

\bibitem[{{Hasegawa} {et~al.}(1994){Hasegawa}, {Sato}, {Whiteoak}, \&
  {Miyawaki}}]{Hasegawa1994}
{Hasegawa}, T., {Sato}, F., {Whiteoak}, J.~B., \& {Miyawaki}, R. 1994, \apjl,
  429, L77

\bibitem[{{Inui} {et~al.}(2009){Inui}, {Koyama}, {Matsumoto}, \&
  {Tsuru}}]{sgr6}
{Inui}, T., {Koyama}, K., {Matsumoto}, H., \& {Tsuru}, T.~G. 2009, \pasj, 61,
  241

\bibitem[{{Kauffmann} {et~al.}(2010){Kauffmann}, {Pillai}, {Shetty}, {Myers},
  \& {Goodman}}]{Kauffmann2010}
{Kauffmann}, J., {Pillai}, T., {Shetty}, R., {Myers}, P.~C., \& {Goodman},
  A.~A. 2010, \apj, 716, 433

\bibitem[{{K{\"o}nyves} {et~al.}(2010){K{\"o}nyves}, {Andr{\'e}},
  {Men'shchikov}, {Schneider}, {Arzoumanian}, {Bontemps}, {Attard}, {Motte},
  {Didelon}, {Maury}, {Abergel}, {Ali}, {Baluteau}, {Bernard}, {Cambr{\'e}sy},
  {Cox}, {di Francesco}, {di Giorgio}, {Griffin}, {Hargrave}, {Huang}, {Kirk},
  {Li}, {Martin}, {Minier}, {Molinari}, {Olofsson}, {Pezzuto}, {Russeil},
  {Roussel}, {Saraceno}, {Sauvage}, {Sibthorpe}, {Spinoglio}, {Testi},
  {Ward-Thompson}, {White}, {Wilson}, {Woodcraft}, \& {Zavagno}}]{Konyves2010}
{K{\"o}nyves}, V., {Andr{\'e}}, P., {Men'shchikov}, A., {et~al.} 2010, \aap,
  518, L106

\bibitem[{{Kruijssen} {et~al.}(2014){Kruijssen}, {Longmore}, {Elmegreen},
  {Murray}, {Bally}, {Testi}, \& {Kennicutt}}]{Kruijssen2014}
{Kruijssen}, J.~M.~D., {Longmore}, S.~N., {Elmegreen}, B.~G., {et~al.} 2014,
  \mnras, 440, 3370

\bibitem[{{Larson}(1981)}]{Larson1981}
{Larson}, R.~B. 1981, \mnras, 194, 809

\bibitem[{{Lis} \& {Goldsmith}(1990)}]{Lis1990}
{Lis}, D.~C. \& {Goldsmith}, P.~F. 1990, \apj, 356, 195

\bibitem[{{Lodders}(2003)}]{Lodders2003}
{Lodders}, K. 2003, \apj, 591, 1220

\bibitem[{{Lombardi} {et~al.}(2010){Lombardi}, {Alves}, \&
  {Lada}}]{Lombardi2010}
{Lombardi}, M., {Alves}, J., \& {Lada}, C.~J. 2010, \aap, 519, L7

\bibitem[{{Marin} {et~al.}(2015){Marin}, {Muleri}, {Soffitta}, {Karas}, \&
  {Kunneriath}}]{Marin2015}
{Marin}, F., {Muleri}, F., {Soffitta}, P., {Karas}, V., \& {Kunneriath}, D.
  2015, \aap, 576, A19

\bibitem[{{McKee} \& {Ostriker}(2007)}]{McKee2007}
{McKee}, C.~F. \& {Ostriker}, E.~C. 2007, \araa, 45, 565

\bibitem[{{Motte} {et~al.}(1998){Motte}, {Andre}, \& {Neri}}]{Motte1998}
{Motte}, F., {Andre}, P., \& {Neri}, R. 1998, \aap, 336, 150

\bibitem[{{Muno} {et~al.}(2007){Muno}, {Baganoff}, {Brandt}, {Park}, \&
  {Morris}}]{sgr5}
{Muno}, M.~P., {Baganoff}, F.~K., {Brandt}, W.~N., {Park}, S., \& {Morris},
  M.~R. 2007, \apjl, 656, L69

\bibitem[{{Murakami} {et~al.}(2000{\natexlab{a}}){Murakami}, {Koyama},
  {Sakano}, {Tsujimoto}, \& {Maeda}}]{sgr3}
{Murakami}, H., {Koyama}, K., {Sakano}, M., {Tsujimoto}, M., \& {Maeda}, Y.
  2000{\natexlab{a}}, \apj, 534, 283

\bibitem[{{Murakami} {et~al.}(2000{\natexlab{b}}){Murakami}, {Koyama},
  {Sakano}, {Tsujimoto}, \& {Maeda}}]{Murakami2000}
{Murakami}, H., {Koyama}, K., {Sakano}, M., {Tsujimoto}, M., \& {Maeda}, Y.
  2000{\natexlab{b}}, \apj, 534, 283

\bibitem[{{Murakami} {et~al.}(2001){Murakami}, {Koyama}, {Tsujimoto}, {Maeda},
  \& {Sakano}}]{Murakami2001}
{Murakami}, H., {Koyama}, K., {Tsujimoto}, M., {Maeda}, Y., \& {Sakano}, M.
  2001, \apj, 550, 297

\bibitem[{{Namito} {et~al.}(1993){Namito}, {Ban}, \& {Hirayama}}]{Namito1993}
{Namito}, Y., {Ban}, S., \& {Hirayama}, H. 1993, Nuclear Instruments and
  Methods in Physics Research A, 332, 277

\bibitem[{{Nobukawa} {et~al.}(2011){Nobukawa}, {Ryu}, {Tsuru}, \&
  {Koyama}}]{sgr11}
{Nobukawa}, M., {Ryu}, S.~G., {Tsuru}, T.~G., \& {Koyama}, K. 2011, \apjl, 739,
  L52

\bibitem[{{Nutter} \& {Ward-Thompson}(2007)}]{Nutter2007}
{Nutter}, D. \& {Ward-Thompson}, D. 2007, \mnras, 374, 1413

\bibitem[{{Odaka} {et~al.}(2011){Odaka}, {Aharonian}, {Watanabe}, {Tanaka},
  {Khangulyan}, \& {Takahashi}}]{Odaka2011}
{Odaka}, H., {Aharonian}, F., {Watanabe}, S., {et~al.} 2011, \apj, 740, 103

\bibitem[{{Parsons} {et~al.}(2012){Parsons}, {Thompson}, {Clark}, \&
  {Chrysostomou}}]{Parsons2012}
{Parsons}, H., {Thompson}, M.~A., {Clark}, J.~S., \& {Chrysostomou}, A. 2012,
  \mnras, 424, 1658

\bibitem[{{Ponti} {et~al.}(2013){Ponti}, {Morris}, {Terrier}, \&
  {Goldwurm}}]{Ponti2013}
{Ponti}, G., {Morris}, M.~R., {Terrier}, R., \& {Goldwurm}, A. 2013, in
  Advances in Solid State Physics, Vol.~34, Cosmic Rays in Star-Forming
  Environments, ed. D.~F. {Torres} \& O.~{Reimer}, 331

\bibitem[{{Ponti} {et~al.}(2010){Ponti}, {Terrier}, {Goldwurm}, {Belanger}, \&
  {Trap}}]{sgr7}
{Ponti}, G., {Terrier}, R., {Goldwurm}, A., {Belanger}, G., \& {Trap}, G. 2010,
  \apj, 714, 732

\bibitem[{{Pozdnyakov} {et~al.}(1983){Pozdnyakov}, {Sobol}, \&
  {Syunyaev}}]{Pozdnyakov1983}
{Pozdnyakov}, L.~A., {Sobol}, I.~M., \& {Syunyaev}, R.~A. 1983, Astrophysics
  and Space Physics Reviews, 2, 189

\bibitem[{{Qin} {et~al.}(2011){Qin}, {Schilke}, {Rolffs}, {Comito}, {Lis}, \&
  {Zhang}}]{Qin2011}
{Qin}, S.-L., {Schilke}, P., {Rolffs}, R., {et~al.} 2011, \aap, 530, L9

\bibitem[{{Revnivtsev} {et~al.}(2004){Revnivtsev}, {Churazov}, {Sazonov},
  {Sunyaev}, {Lutovinov}, {Gilfanov}, {Vikhlinin}, {Shtykovsky}, \&
  {Pavlinsky}}]{Revnivtsev2004}
{Revnivtsev}, M.~G., {Churazov}, E.~M., {Sazonov}, S.~Y., {et~al.} 2004, \aap,
  425, L49

\bibitem[{{Salpeter}(1955)}]{Salpeter1955}
{Salpeter}, E.~E. 1955, \apj, 121, 161

\bibitem[{Schoonjans {et~al.}(2011)Schoonjans, Brunetti, Golosio, del Rio,
  Sole, \& C.~Ferrero}]{xraylib}
Schoonjans, T., Brunetti, A., Golosio, B., {et~al.} 2011, Spectromchimica Acta
  Part B: Atomic Spectroscopy, 66, 776

\bibitem[{{Shetty} {et~al.}(2010){Shetty}, {Collins}, {Kauffmann}, {Goodman},
  {Rosolowsky}, \& {Norman}}]{Shetty2010}
{Shetty}, R., {Collins}, D.~C., {Kauffmann}, J., {et~al.} 2010, \apj, 712, 1049

\bibitem[{{Sunyaev} \& {Churazov}(1998)}]{Sunyaev1998}
{Sunyaev}, R. \& {Churazov}, E. 1998, \mnras, 297, 1279

\bibitem[{{Sunyaev} \& {Churazov}(1996)}]{Sunyaev1996}
{Sunyaev}, R.~A. \& {Churazov}, E.~M. 1996, Astronomy Letters, 22, 648

\bibitem[{{Sunyaev} {et~al.}(1993){Sunyaev}, {Markevitch}, \&
  {Pavlinsky}}]{Sunyaev1993}
{Sunyaev}, R.~A., {Markevitch}, M., \& {Pavlinsky}, M. 1993, \apj, 407, 606

\bibitem[{{Takahashi} {et~al.}(2010){Takahashi}, {Mitsuda}, {Kelley},
  {Aharonian}, {Akimoto}, {Allen}, {Anabuki}, {Angelini}, {Arnaud}, {Awaki},
  {Bamba}, {Bando}, {Bautz}, {Blandford}, {Boyce}, {Brown}, {Chernyakova},
  {Coppi}, {Costantini}, {Cottam}, {Crow}, {de Plaa}, {de Vries}, {den Herder},
  {Dipirro}, {Done}, {Dotani}, {Ebisawa}, {Enoto}, {Ezoe}, {Fabian},
  {Fujimoto}, {Fukazawa}, {Funk}, {Furuzawa}, {Galeazzi}, {Gandhi}, {Gendreau},
  {Gilmore}, {Haba}, {Hamaguchi}, {Hatsukade}, {Hayashida}, {Hiraga}, {Hirose},
  {Hornschemeier}, {Hughes}, {Hwang}, {Iizuka}, {Ishibashi}, {Ishida},
  {Ishimura}, {Ishisaki}, {Isobe}, {Ito}, {Iwata}, {Kaastra}, {Kallman},
  {Kamae}, {Katagiri}, {Kataoka}, {Katsuda}, {Kawaharada}, {Kawai}, {Kawasaki},
  {Khangaluyan}, {Kilbourne}, {Kinugasa}, {Kitamoto}, {Kitayama}, {Kohmura},
  {Kokubun}, {Kosaka}, {Kotani}, {Koyama}, {Kubota}, {Kunieda}, {Laurent},
  {Lebrun}, {Limousin}, {Loewenstein}, {Long}, {Madejski}, {Maeda},
  {Makishima}, {Markevitch}, {Matsumoto}, {Matsushita}, {McCammon}, {Miller},
  {Mineshige}, {Minesugi}, {Miyazawa}, {Mizuno}, {Mori}, {Mori}, {Mukai},
  {Murakami}, {Murakami}, {Mushotzky}, {Nakagawa}, {Nakagawa}, {Nakajima},
  {Nakamori}, {Nakazawa}, {Namba}, {Nomachi}, {O'Dell}, {Ogawa}, {Ogawa},
  {Ogi}, {Ohashi}, {Ohno}, {Ohta}, {Okajima}, {Ota}, {Ozaki}, {Paerels},
  {Paltani}, {Parmar}, {Petre}, {Pohl}, {Porter}, {Ramsey}, {Reynolds},
  {Sakai}, {Sambruna}, {Sato}, {Sato}, {Serlemitsos}, {Shida}, {Shimada},
  {Shinozaki}, {Shirron}, {Smith}, {Sneiderman}, {Soong}, {Stawarz}, {Sugita},
  {Szymkowiak}, {Tajima}, {Takahashi}, {Takei}, {Tamagawa}, {Tamura}, {Tamura},
  {Tanaka}, {Tanaka}, {Tanaka}, {Tashiro}, {Tawara}, {Terada}, {Terashima},
  {Tombesi}, {Tomida}, {Tozuka}, {Tsuboi}, {Tsujimoto}, {Tsunemi}, {Tsuru},
  {Uchida}, {Uchiyama}, {Uchiyama}, {Ueda}, {Uno}, {Urry}, {Watanabe}, {White},
  {Yamada}, {Yamaguchi}, {Yamaoka}, {Yamasaki}, {Yamauchi}, {Yamauchi},
  {Yatsu}, {Yonetoku}, \& {Yoshida}}]{Takahashi2010}
{Takahashi}, T., {Mitsuda}, K., {Kelley}, R., {et~al.} 2010, in Society of
  Photo-Optical Instrumentation Engineers (SPIE) Conference Series, Vol. 7732,
  Society of Photo-Optical Instrumentation Engineers (SPIE) Conference Series,
  0

\bibitem[{{Terrier} {et~al.}(2010){Terrier}, {Ponti}, {B{\'e}langer},
  {Decourchelle}, {Tatischeff}, {Goldwurm}, {Trap}, {Morris}, \&
  {Warwick}}]{sgr8}
{Terrier}, R., {Ponti}, G., {B{\'e}langer}, G., {et~al.} 2010, \apj, 719, 143

\bibitem[{{Tsuboi} \& {Miyazaki}(2012)}]{Tsuboi2012}
{Tsuboi}, M. \& {Miyazaki}, A. 2012, \pasj, 64, 111

\bibitem[{{Vainshtein} {et~al.}(1998){Vainshtein}, {Syunyaev}, \&
  {Churazov}}]{Vainshtein1998}
{Vainshtein}, L.~A., {Syunyaev}, R.~A., \& {Churazov}, E.~M. 1998, Astronomy
  Letters, 24, 271

\bibitem[{{Verner} \& {Yakovlev}(1995)}]{Verner1995}
{Verner}, D.~A. \& {Yakovlev}, D.~G. 1995, \aaps, 109, 125

\bibitem[{{Weisskopf} {et~al.}(2015){Weisskopf}, {Gaskin}, {Tananbaum}, \&
  {Vikhlinin}}]{Weisskopf2015}
{Weisskopf}, M.~C., {Gaskin}, J., {Tananbaum}, H., \& {Vikhlinin}, A. 2015, in
  Society of Photo-Optical Instrumentation Engineers (SPIE) Conference Series,
  Vol. 9510, Society of Photo-Optical Instrumentation Engineers (SPIE)
  Conference Series, 2

\bibitem[{{Williams}(1999)}]{Williams1999}
{Williams}, J. 1999, in Interstellar Turbulence, ed. J.~{Franco} \&
  A.~{Carraminana}, 190

\bibitem[{{Williams} {et~al.}(2000){Williams}, {Blitz}, \&
  {McKee}}]{Williams2000}
{Williams}, J.~P., {Blitz}, L., \& {McKee}, C.~F. 2000, Protostars and Planets
  IV, 97

\bibitem[{{Zhang} {et~al.}(2015){Zhang}, {Hailey}, {Mori}, {Clavel}, {Terrier},
  {Ponti}, {Goldwurm}, {Bauer}, {Boggs}, {Craig}, {Christensen}, {Harrison},
  {Hong}, {Nynka}, {Stern}, {Soldi}, {Tomsick}, \& {Zhang}}]{Zhang2015}
{Zhang}, S., {Hailey}, C.~J., {Mori}, K., {et~al.} 2015, ArXiv e-prints

\end{thebibliography}
\end{document}